\documentclass[prd,showpacs,preprintnumbers,amsmath,amssymb,twocolumn,superscriptaddress,notitlepage,nofootinbib,longbibliography,reprint]{revtex4-2}
\usepackage{graphicx,color}
\usepackage{graphicx}
\usepackage{amsmath,amssymb,amsfonts}
\usepackage{stackrel}
\usepackage{enumitem}
\usepackage[english]{babel}
\usepackage{verbatim}
\usepackage{xcolor}
\usepackage{dcolumn}
\usepackage{physics}
\usepackage{verbatim}
\usepackage{float}
\usepackage{color}
\usepackage{multirow}
\usepackage{graphicx}
\usepackage{epstopdf}
\usepackage{epsfig}
\usepackage{slashed}
\usepackage{physics}
\usepackage{multirow}
\usepackage{makecell}
\usepackage{placeins}
\usepackage{setspace}
\usepackage{tabularx}
\usepackage{colortbl}
\usepackage{float}
\usepackage{placeins}
\usepackage{ragged2e}
\usepackage{listings}
\usepackage{gensymb}
\usepackage{siunitx}
\usepackage{aas_macros}
\usepackage{bm}
\usepackage{topcapt}
\usepackage{tabularx}
\usepackage{makecell}

\definecolor{darkred}{rgb}{.7,.1,.1}
\usepackage{hyperref}
\hypersetup{colorlinks=true,
	breaklinks=true,
	pdfstartview=Fit,
	linkcolor=blue,
	citecolor=blue,
	urlcolor=blue}
\usepackage{cleveref}

\definecolor{dark-green}{rgb}{0.1,0.7,0.3}

\newcommand{\gev}{~\text{GeV}}

\newcommand{\beq}{\begin{equation}}
\newcommand{\eeq}{\end{equation}}
\newcommand{\bal}{\begin{align}}
\newcommand{\eal}{\end{align}}
\newcommand{\bit}{\begin{itemize}}
\newcommand{\eit}{\end{itemize}}
\newcommand{\ben}{\begin{enumerate}}
\newcommand{\een}{\end{enumerate}}

\renewcommand{\eqref}[1]{Eq.~(\ref{eq:#1})}

\newcommand{\figref}[1]{Fig.~\ref{fig:#1}}

\newcommand{\tabref}[1]{Table~\ref{tab:#1}}

\begin{document}

\preprint{}
\title{The Origin of the Fermi Halo-like Emission: A New Model of the Fermi Bubbles}

\author{Ilias Cholis}
\thanks{cholis@oakland.edu, ORCID: orcid.org/0000-0002-3805-6478}
\affiliation{Department of Physics, Oakland University, Rochester, Michigan 48309, USA}

\author{Leo Qiyuan Hu}
\thanks{leo.hu@my.cityu.edu.hk, ORCID:  orcid.org/0009-0008-0980-4482}
\affiliation{Department of Physics, City University of Hong Kong, Kowloon, Hong Kong SAR, China}

\author{Yi-Ming Zhong}
\thanks{yiming.zhong@cityu.edu.hk, ORCID: orcid.org/0000-0001-9922-6162}
\affiliation{Department of Physics, City University of Hong Kong, Kowloon, Hong Kong SAR, China}

\date{\today}

\begin{abstract}
We study the gamma-ray sky at high latitudes for energies between 0.3 and 900 GeV. We search for a  contribution in the high-latitude sky and in the region of the Galactic Stellar Halo that would be beyond known astrophysical emissions, i.e., the galactic diffuse emission, the isotropic mostly extragalactic gamma-ray background, the \textit{Fermi} Bubbles, Loop I, the contribution from the Galactic Center Excess, and the known point and extended gamma-ray sources detected by the \textit{Fermi}-LAT Collaboration.  We perform a sequence of template fits to the \textit{Fermi} gamma-ray data, testing different regions of interest of the galactic sky, utilizing a large set of alternative galactic diffuse emission models, and implementing alternative fitting schemes to assess the impact of increased modeling complexity of the gamma-ray sky. We find that the halo-like emission recently reported by Totani (2025) is real, but it is likely due to emission from the \textit{Fermi} Bubbles rather than dark matter annihilation in the Galactic halo. We provide new templates for the \textit{Fermi} Bubbles, including an updated model for the Flat Bubbles and a detailed model of their substructures, the Cocoons and the Inner Bubbles. In combination, these components account for the previously reported halo-like emission. We also estimate the associated fluxes and spectral properties of the Flat Bubbles, the Inner Bubbles, and the Cocoons. This updated model for the \textit{Fermi} Bubbles provides new input toward understanding their origin.

\end{abstract}

\maketitle

\section{Introduction}
\label{sec:intro}

The gamma-ray sky observations by the \textit{Fermi}-Large Area Telescope (\textit{Fermi}-LAT) \cite{1999APh....11..277G, Fermi-LAT:2009ihh}, have attracted great interest since 2008, providing us with an unprecedented quality of data, allowing us to model the emission processes of GeV-scale gamma rays from the Milky Way and beyond. Furthermore, thousands of point and extended sources have been detected, most of which are distant galaxies \cite{Fermi-LAT:2010feq, Abdo:2010ge, Fermi-LAT:2011iqa, Fermi-LAT:2013oro, Fermi-LAT:2015bhf, Fermi-LAT:2019yla, Fermi-LAT:2019pir, Ballet:2020hze, Fermi-LAT:2022byn, Fermi-LAT:2022oww, Ballet:2023qzs}. Most interestingly, two previously unknown emission components have been discovered, both originating from the inner galaxy;  the Galactic Center Excess (GCE) \cite{Goodenough:2009gk, Vitale:2009hr, Hooper:2010mq, Abazajian:2010zy, Gordon:2013vta} and the \textit{Fermi} Bubbles \cite{Su:2010qj, Dobler:2009xz, Fermi-LAT:2014sfa}.

The evidence for the existence of GCE is robust~\cite{Calore:2014xka, Zhou:2014lva, Daylan:2014rsa, TheFermi-LAT:2015kwa, DiMauro:2021raz, Cholis:2021rpp, Hu:2026apn, DiMauro:2026fnp}. Recently, \textit{DAMPE}~\cite{DAMPE:2025ndf}, has independently identified the GCE with the spectrum and morphology as that derived in the analysis of Refs.~\cite{DiMauro:2021raz, Cholis:2021rpp}. 
There are three major interpretations for the GCE's origin. One is that it is a signal from particle dark matter annihilation \cite{Hooper:2011ti, Hooper:2013rwa, Gordon:2013vta, Daylan:2014rsa, Calore:2014xka, Calore:2014nla, Agrawal:2014oha, Berlin:2015wwa, TheFermi-LAT:2017vmf, Karwin:2016tsw, Leane:2019xiy, Zhu:2022tpr, McDermott:2022zmq, DiMauro:2023tho, Christy:2024gsl, Muru:2025vpz, Hu:2025thq, Hu:2026apn},  or at least predominantly from dark matter annihilation \cite{Cholis:2021rpp, Zhong:2024vyi}. Another is that it is the result of a population of millisecond pulsars \cite{Abazajian:2012pn, Petrovic:2014xra, Lee:2015fea, Bartels:2015aea, Bartels:2017vsx, Bartels:2018eyb, Buschmann:2020adf, Murgia:2020dzu, Gautam:2021wqn, Macias:2023qqc, Song:2024iup, Malyshev:2024obk}.  A third scenario is that it is the result of cosmic-ray outbursts from the center of the Milky Way \cite{Petrovic:2014uda, Carlson:2014cwa, Cholis:2015dea}. There have also been interpretations for the GCE that do not fall clearly within these three scenarios~\cite{OLeary:2015qpx, Dutta:2026rai}.

The spectral and morphological characteristics and the photon statistics of the GCE provide important information on the interpretation of its origin. The lack of a large number of detected millisecond pulsars from the inner galaxy and the smoothness of the GCE emission allow only for a small fraction of it to come from millisecond pulsars \cite{Zhong:2019ycb, Leane:2019xiy, Leane:2020pfc, List:2025qbx} (see also \cite{Ramirez:2024oiw}). 
Observations in radio, X-rays and gamma rays from other targets of the Milky Way allow us to model the expected luminosity distribution and spectrum of millisecond pulsars, and put tension on the hypothesis that a large population of unresolved millisecond pulsars can account for the GCE \cite{Cholis:2014lta, Haggard:2017lyq, Amerio:2024qor, Koechler:2026uvf}. 
Furthermore, the successive analysis results showing that the GCE morphology is nearly spherical and its spectrum pretty uniform within the various analysis sub-regions \cite{Daylan:2014rsa, Calore:2014xka, Cholis:2021rpp}, put some tension on the hypothesis that it is the result of cosmic-ray outburst activity from the supermassive black hole.

The dark matter interpretation of the GCE remains in agreement with both its morphology and spectrum. 
Limits from dwarf spheroidal galaxies \cite{Fermi-LAT:2016uux, DiMauro:2021qcf, McDaniel:2023bju}, 
are not strong enough to exclude the GCE as a dark matter signal \cite{Strigari:2018utn, Ando:2020yyk, DiMauro:2021qcf, McDaniel:2023bju, Circiello:2026inp}. Similarly, limits from the extragalactic gamma-ray background (e.g. \cite{Ajello:2015mfa, Cholis:2024hmd}) are similar to those from dwarf galaxies, and cannot exclude the dark matter interpretation. 
At the same time, two counterpart excess fluxes, one from Andromeda in gamma rays \cite{Karwin:2020tjw}, and the other in cosmic-ray antiprotons \cite{Cuoco:2016eej, Cui:2016ppb, Cholis:2019ejx} have been detected that are consistent with the dark matter interpretation of the GCE.

The \textit{Fermi} Bubbles in turn, are very likely associated with activity from the environment of the supermassive black hole at the center of our Galaxy taking place some million years ago \cite{Crocker:2010qn, Crocker:2010dg, Cheng:2011xd, Mertsch:2011es, Guo:2011eg, Guo:2011ip, Yang:2012fy, Carretti:2013sc, Lacki:2013zsa, Crocker:2014fla},
with clear associations at microwaves \cite{Finkbeiner:2003im, Dobler:2007wv, Dobler:2012ef, Planck:2012opn, 
 Planck:2015ica}, X-rays \cite{Snowden:1997ze, Bland-Hawthorn:2002rbf, Kataoka:2013tma, Tahara:2015mia}, and possibly also with cosmic-ray positrons \cite{Cholis:2022ajr}. Also, observations in X-rays have revealed even larger scale bubbles, the \textit{e-ROSITA} Bubbles \cite{Predehl:2020kyq}, that may be the effect of the forward shock \cite{Predehl:2020kyq, Yang:2022jck} (with the \textit{Fermi} Bubbles being the volume inside the contact discontinuity), or an earlier event of cosmic-ray activity from the environment of the supermassive black hole \cite{Predehl:2020kyq, Zhang:2025jzn}. Furthermore, cocoon structures have been identified within the same region of the galactic sky that the \textit{Fermi} Bubbles cover \cite{Su:2010qj, 2012ApJ...753...61S, Fermi-LAT:2014sfa, Balaji:2018rwz}, suggesting again that the supermassive black hole has likely had multiple outburst events over the past million years. 

While the detection of the \textit{Fermi} Bubbles is undebatable, their exact characteristics have remained an open question with only a small number of data analysis studies dedicated to quantifying their morphological and spectral properties \cite{Su:2010qj, 2012ApJ...753...61S, Fermi-LAT:2014sfa, Balaji:2018rwz}. Past analyses have suggested that the Bubbles have a hard spectrum of approximately $dN/dE \propto E^{-2}$ \cite{Su:2010qj, Fermi-LAT:2014sfa}, in the energy range of $E \approx 1-300$ GeV, with no conclusive results on their spectral properties beyond that energy range \cite{Fermi-LAT:2014sfa}. 
The Bubbles have non-trivial shapes but relatively clear edges. 
Their maximum extent in galactic longitude is between $-27^{\circ}\lesssim \ell \lesssim 16^{\circ}$ for the bubble in the southern hemisphere, and $-23^{\circ}\lesssim \ell \lesssim 18^{\circ}$ for the bubble in the northern hemisphere. 
They extend up to galactic latitudes of $|b| \simeq 50-55^{\circ}$, with a sharp end of their associated emission at those latitudes.
At galactic latitudes of $|b| \lesssim 10^{\circ}$, the dominant emission from the galactic disk is too bright to disentangle it from that of the Bubbles. Thus, we have yet to identify their morphology within the galactic center. 
Previous work on the Bubbles has claimed that, to first approximation, at latitudes $|b| \gtrsim 10^{\circ}$, they have a flat morphology within their defined boundaries \cite{Su:2010qj, Fermi-LAT:2014sfa}. 
However, at increasing latitudes the decreasing number of observed GeV-energy photons has made it more challenging to quantify how their brightness and spectrum evolve ---if at all--- with angular separation from the galactic center.

Recently, Ref.~\cite{Totani:2025fxx} has claimed the potential presence of a third emission component in the Stellar Halo region~\cite{2019ApJ...883..107C}, by analyzing the galactic sky within latitudes of $10^{\circ} \leq |b| \leq 60^{\circ}$ and longitudes of $|\ell| \leq 60^{\circ}$. 
This emission is claimed to have a spectrum that peaks around 20 GeV and extends up to at least 200 GeV, and a spherical morphology that resembles that of annihilating dark matter. If interpreted as a dark matter signal, this emission is bright enough in the high-latitude \textit{Fermi} data, that a natural question arises, why have not we observed it in earlier analyses, as e.g., Ref.~\cite{Chang:2018bpt}? Furthermore, if it is a signal of TeV-scale dark matter annihilation, how does it evade limits from dwarf galaxies, cosmic-ray antiprotons, as well as possible limits from direct dark matter detection and collider searches? 

In this work, we analyze the \textit{Fermi}-LAT observations from the same region of the galactic sky, which we refer to as the inner $60^{\circ}$ of the Galactic Center (GC) ($|\ell| \leqslant 60\degree$, $|b| \leqslant 60\degree$) region of interest (ROI), to validate that claim.
We first repeat the same analysis steps as in Ref.~\cite{Totani:2025fxx}, for the original ROI of that work. We find a similar result for an unaccounted-for emission component from the background/foreground diffuse emission assumptions to those used by Ref.~\cite{Totani:2025fxx}, with a similar but not identical morphology and spectrum. That unaccounted emission component we will refer to as halo-like emission (HLE) for the remainder of this work. 

We then expand the analysis in three manners. We break the inner $60^{\circ}$ of the GC into multiple smaller ROIs. This allows us to test the hypothesis that the HLE has the expected morphology of a signal mimicking that of dark matter annihilation.
Additionally, we use a wide range of galactic diffuse emission models from Refs.\cite{Fermi-LAT:2012edv} and~\cite{Cholis:2021rpp}, which allows us to test the robustness of the properties of the HLE. Finally, we test a range of modeling freedom in our template fitting procedure. This further allows us to see the effect that the modeling of the background/foreground emission from the \textit{Fermi} Bubbles, from Loop I \cite{Wolleben:2007pq, Fermi-LAT:2014sfa}, and from including or masking out the gamma-ray flux from point sources and extended sources has on the HLE resulting spectrum, morphology and significance of detection. 

This paper is organized as follows. In Section~\ref{sec:gammaraysky}, we discuss how we model the galactic diffuse emission with all the variations that we test to properly account for the underlying modeling uncertainties of that emission. 
We also discuss in that section the modeling of the emission from Loop I and our initial assumptions on the emission from the \textit{Fermi} Bubbles. 
Finally, only for the case that we aim to repeat the analysis of Ref.~\cite{Totani:2025fxx}, we include the gamma-ray emission from point and extended sources which we discuss towards the end of Section~\ref{sec:gammaraysky}.  
In Section~\ref{sec:analysis}, we describe the \textit{Fermi} data, the ROIs, the masking, and the fitting procedure that we use. We give our results on the HLE's properties in Section~\ref{sec:Results}. We find that the HLE's brightness \emph{does not} follow that of a dark matter annihilation profile when studying all the ROIs and background diffuse emission models. 
Instead, we find that its spectrum and brightness are much closer to those of the \emph{Fermi} Bubbles. Notably, the HLE has a non-flat morphology on the sky. We thus investigate whether the HLE captures the non-flat emission from the Bubbles, as one moves away from the galactic center.  
In Section~\ref{sec:NewFermiBubbles}, we test a set of alternative templates for that additional contribution to the \textit{Fermi} Bubbles.
This secondary contribution of the Bubbles in combination with the existence of the observed Cocoons in the south and less clearly in the north hemisphere, whose properties we also test, explains why it has been difficult to identify that as part of the gamma-ray emission from the Bubbles region. 
We provide an alternative description for the morphology of the \textit{Fermi} Bubbles including the Cocoons.
We give our conclusions in Section~\ref{sec:conclusions} and discuss further open questions on the emission from the inner galaxy and the galactic halo region.

\section{Modeling the gamma-ray sky}
\label{sec:gammaraysky}

The observed gamma-ray emission is the combination of multiple components: the galactic diffuse emission components that come from the interaction of high-energy cosmic rays with the interstellar medium (ISM), the emission from galactic and extragalactic point and extended sources, and the isotropic diffuse emission that is a combination of mostly extragalactic in origin emission with some minor contribution from the Milky Way at high latitudes. 
For the galactic diffuse emission and the isotropic emission we use templates, as we describe in sections~\ref{subsec:galacticdiffuse} and~\ref{subsec:isotropic}.  
In addition to those, we model separately through alternative choices of the templates used, the emission from Loop I (discussed in section~\ref{subsec:LoopI}), from the \textit{Fermi} Bubbles (see section~\ref{subsec:Bubbles}), and also, to make a direct comparison to the work of Ref.~\cite{Totani:2025fxx}, we make use of the positive and negative residual templates that were created by Ref.~\cite{Totani:2025fxx} (discussed in section~\ref{subsec:Residuals}).
We note that the emission from gamma-ray point and extended sources is handled in two alternative manners, (i) by using a spatial all point and extended gamma-ray sources template for $|b| \geq 10^{\circ}$, which we describe in section~\ref{subsec:PointSources}, and (ii) by masking all detected point and extended gamma-ray sources which we discuss in Section~\ref{sec:analysis}. Masking the point and extended gamma-ray sources is our standard analysis choice. 

\subsection{The Galactic diffuse emission}
\label{subsec:galacticdiffuse}

A significant fraction of the galactic diffuse emission comes from cosmic-ray protons and more massive nuclei that have hard inelastic collisions with the ISM gas. Such collisions produce a multitude of neutral mesons, mostly $\pi^{0}$ mesons, that subsequently decay giving gamma rays. That emission component directly tracks the ISM gas density and is typically and in this work referred to as the ``Pi0'' component. 
In addition to cosmic-ray protons and nuclei, cosmic-ray electrons interact with the ISM gas giving gamma rays via Bremsstrahlung. That mechanism also directly tracks the ISM gas density and we refer to it in this work as ``Bremss''. 
The combination of the galactic Pi0 and Bremss emission is responsible for most of the observed gamma rays in the energy range and sky regions that we study in this work.
The third galactic diffuse emission component in gamma rays is the result of cosmic-ray electrons up-scattering the cosmic microwave background, as well as photons in the infrared, visible, and UV, that compose the interstellar radiation field. Since this is inverse Compton scattering, we refer to it as the ``ICS'' component.

We use alternative templates to model these three galactic diffuse emission components. 
Those templates have been created in the works of Refs.~\cite{Cholis:2021rpp, Fermi-LAT:2012edv}, using the \texttt{GALPROP} publicly available code \cite{galprop, website}.
The morphology and spectral characteristics of each of these three emission components strongly depend on a sequence of underlying assumptions in the modeling of the Milky Way. 
For these reasons, we use the simulated diffuse emission templates from 100 different Milky Way models, in order to account for the astrophysical/modeling uncertainties regarding the properties of the Milky Way. 

Each of those Milky Way models makes alternative assumptions on (i) the diffusion/propagation volume of galactic cosmic rays, (ii) the diffusion coefficient that describes how fast cosmic rays diffuse in 3D space, (iii) the rigidity $R\equiv p/q$ (where $p$ is the momentum of a cosmic-ray particle and $q$ is its charge) dependence of that diffusion coefficient, as particles with increasing rigidity diffuse through the ISM faster. 
Furthermore, these models account for alternative choices on, (iv) the diffusive re-acceleration of cosmic rays (due to the presence of Alfv\'en waves) in the ISM, (v) the presence of large-scale convective winds originating from the galactic disk, that move cosmic rays away from the disk. 
Also, they include alternative modeling choices for, (vi) the morphology and amplitude of the Milky Way's magnetic field, and (vii) the distribution of energy of IR and visible photons in the Milky Way, known as the interstellar radiation field. 
The interstellar radiation field and the magnetic fields are crucial in modeling the propagation and energy losses of high-energy cosmic-ray electrons, which affect the Bremss and ICS emission components. 
The Milky Way models also test alternative choices on, (viii) the spatial distribution and type of cosmic-ray sources assumed to be located along the galactic disk, (ix) the injection spectra of various species of cosmic rays including electrons, Hydrogen and Helium from those sources, allowing for different injection spectra between species. 
Finally, these models test alternative assumptions on the distribution of (x) the atomic Hydrogen HI, ISM gas, (xi) the molecular Hydrogen H2, ISM gas, and (xii) the ionized Hydrogen HII, ISM gas, based on how emission-line observations in the millimeter and centimeter wavelengths are used to derive those gases' spatial distribution in the Milky Way. The distribution of ISM gases is crucial in modeling the Pi0 and Bremss components.

Those models have been created, relying on a range of observations, from cosmic-ray local measurements to observations in radio, millimeter, IR, the visible spectrum, the UV, and gamma-ray measurements across the sky. Each model makes predictions on the Pi0, Bremss and ICS template morphology and flux in gamma rays. For each model, both the predicted template fluxes and the template morphologies change with the energy of gamma rays, as the underlying distribution of the cosmic rays responsible for the gamma-ray emission depends on their (cosmic-ray) energy as well. For more details regarding those models, we refer the reader to the original works of Refs.~\cite{Cholis:2021rpp, Fermi-LAT:2012edv}.

Of the 100 models, 80 come from Ref.~\cite{Cholis:2021rpp}, with the galactic diffuse emission maps publicly available through \cite{2021TemplatesZenodo}. Those maps, model the galactic diffuse emission within a region of $|\ell| \leq 30^{\circ}$ and $|b| \leq 30^{\circ}$. 
The 20 models that come from Ref.~\cite{Fermi-LAT:2012edv}, model the entire $4\pi$ sky, with the galactic diffuse maps available in \cite{FermiTemplates}. Depending on the region of the galactic sky that we study, we use either all 100 models, or just the 20 models from Ref.~\cite{Fermi-LAT:2012edv}.

The work of Ref.~\cite{Totani:2025fxx}, used specifically one main template model from Ref.~\cite{Fermi-LAT:2012edv}, which we will refer to as ``SL\,Z6\,R30\,T150\,C2'' (from its original reference name), and subsequently tested its results against five more variations of that. All these six models are included in our analysis. 
We have found that these six models for the Milky Way used in \cite{Totani:2025fxx}, are quite similar to each other in their predictions regarding the galactic diffuse emission in the ROIs of study. Expanding to 100 models, allows us to perform a much more thorough test on the derived properties of the HLE. Furthermore, we find that among the additional Milky Way models, several give a better fit to the gamma-ray observations, compared to the fit using the SL\,Z6\,R30\,T150\,C2 or its five variations. 

\subsection{The isotropic diffuse emission}
\label{subsec:isotropic}

The isotropic gamma-ray background is mostly the result for the combined emission of unresolved extragalactic sources. In particular, BL Lacertae, flat spectrum radio quasars, star-forming galaxies, starburst galaxies and radio galaxies. Furthermore, at energies higher than 10 GeV, a significant fraction of the isotropic emission comes from ultra-high-energy cosmic rays interacting with low-energy photons in the intergalactic medium giving cascades of particles, including gamma rays. 
Finally, some unresolved galactic point sources at high latitudes may contribute a small fraction of the isotropic gamma-ray emission. See Refs.~\cite{Stecker:2019ybn, Blanco:2021icw, Korsmeier:2022cwp, Cholis:2024hmd}, for recent updates on the modeling of the isotropic background and its components.      

We use the isotropic gamma-ray background spectrum from Ref.~\cite{Fermi-LAT:2014ryh}.
In that work, the isotropic background spectrum is evaluated from 100 MeV to 820 GeV. Ref.~\cite{Fermi-LAT:2014ryh}, provides three alternative evaluations for that spectrum. This is done in order to account for modeling uncertainties on the galactic diffuse emission that needs to be subtracted (together with the resolved point and extended sources), from the \textit{Fermi}-LAT observations at latitudes of $|b| \geq 20^{\circ}$, to evaluate the isotropic gamma-ray flux. 
In our analysis, for the simulated isotropic background flux templates at different energies, we take the central values for the flux from the ``model A'' of Ref.~\cite{Fermi-LAT:2014ryh}. We also take the evaluated flux errors of ``model A''. 
Accounting for the errors of that emission provides a penalty in our fits (described in Section~\ref{subsec:fitting_procedure}), when the isotropic template's associated emission deviates significantly from the value of Ref.~\cite{Fermi-LAT:2014ryh}, at any given energy bin.
Given how small a contribution to the evaluated log-likelihoods these errors have on the combined fits, we have found that assuming instead  ``model B'' or ``model C'' for the isotropic background flux from Ref.~\cite{Fermi-LAT:2014ryh}, has an insignificant effect on our results. 

\subsection{The gamma-ray emission from Loop I}
\label{subsec:LoopI}

Loop I, sometimes referred to as the North Polar Spur, is a feature most prominent in the northern galactic hemisphere, observed first in radio waves \cite{1965Natur.208..741Q, 1971A&A....14..252B, 1979ApJ...229..533H, 1983BASI...11....1S}, and later on in gamma rays \cite{Su:2010qj, Fermi-LAT:2014sfa}. 
It is modeled as the combination of two close-by nearly spherical shells correlated with expanding
gas and dust shells and energized by a few million-year-old supernovae or stellar winds \cite{1971A&A....14..252B, 1979ApJ...229..533H, 1983BASI...11....1S, Wolleben:2007pq}. 

Ref.~\cite{Wolleben:2007pq}, using polarization data from the Dominion Radio Astrophysical Observatory (DRAO) survey at 1.4 GHz \cite{2006A&A...448..411W}, and from the Wilkinson Microwave Anisotropy Probe \textit{WMAP} at 23 GHz \cite{WMAP:2006rnx}, provided a ``geometric model'' for the morphology of Loop I. 
Following the updated assumptions of Ref.~\cite{Fermi-LAT:2014sfa} and in agreement with the uncertainties of Ref.~\cite{Wolleben:2007pq}, the two shells (shells ``1'' and ``2'') are taken to be centered on the galactic sky at $\ell_{1} = 341^{\circ}$, $b_{1} = 3^{\circ}$ and
$\ell_{2} = 332^{\circ}$, $b_{2} = 37^{\circ}$ respectively. 
The centers of these spherical shells are at distances from the Sun of $d_{1} = 78$ pc, and $d_{2} = 95$ pc. Shell ``1'' has an inner radius of $r_{1}^{\textrm{inner}} = 62$ pc and an outer radius of $r_{1}^{\textrm{outer}} = 81$ pc, while shell ``2'' is defined by $r_{2}^{\textrm{inner}} = 58$ pc and $r_{2}^{\textrm{outer}} = 82$ pc.

The radio observations represent the synchrotron emission from the associated energetic cosmic-ray electrons. The same electrons can give gamma rays through ICS with the local interstellar radiation field.     
Ref.~\cite{Fermi-LAT:2014sfa}, using the first 50 months of \textit{Fermi}-LAT observations, derived the gamma-ray spectrum from the combination of the two Loop I shells to be $dN/dE \propto E^{-2.4}$, treating them with a fixed relative normalization (i.e. as a single template). Instead, Ref.~\cite{Totani:2025fxx}, using the same geometric model, treated the two shells as separate templates, allowing for their relative brightness to vary. 

In our work, we follow both approaches in fitting to the \textit{Fermi} observations. 
In the radio observations, which measure the synchrotron emission, shell ``2'' is about three times brighter than shell ``1''. We will refer to these shells for the rest of this work as shell 1: ``Loop I Dim'' and shell 2: ``Loop I Bright'', to associate them with the dim and bright shells in radio, but also to retain the notion that even in the gamma-ray data, that ratio of 1:3 in brightness between the two shells should roughly be retained.  
We model the gamma-ray differential flux (before further normalization fitting) from these two shells as follows:
\begin{eqnarray}
\textrm{Loop I Dim} &:& \frac{dN}{dE} = 1.0 \times 10^{-8} \left(\frac{E}{7 \, \textrm{GeV}}\right)^{-2.4} \\
\textrm{and} \nonumber \\ 
\textrm{Loop I Bright} &:& \frac{dN}{dE} = 3.0 \times 10^{-8} \left(\frac{E}{7 \, \textrm{GeV}}\right)^{-2.4}.
\end{eqnarray}
These differential fluxes are given in units of GeV$^{-1}$ cm$^{-2}$ s$^{-1}$ sr$^{-1}$.
We note that in Ref.~\cite{Totani:2025fxx}, as a result of the fit to the \textit{Fermi} data, the ratio of brightness of the dim to the bright shell was found to be about 3:1 instead of 1:3 (shown in Fig. 14 of that work). In our opinion, this suggests some mismodeling of the observations that may originate from other templates. In our results, when we let the relative flux between the two shells vary, we rarely find the flux ratio between the two shells to be in such disagreement with what is expected from the radio data. 

\subsection{The emission from the Gamma-ray Bubbles}
\label{subsec:Bubbles}

The \textit{Fermi} Bubbles have been known to be an emission component on both the northern and southern galactic sky, with clear edges since the work of Ref.~\cite{Su:2010qj}, with the earlier work of \cite{Dobler:2009xz}, identifying the additional flux component but not finding its edges.  
Furthermore, there are clear indications for a cocoon in the southern sky \cite{Su:2010qj, 2012ApJ...753...61S, Fermi-LAT:2014sfa}, with a hint for a cocoon in the northern hemisphere as well \cite{2012ApJ...753...61S, Balaji:2018rwz}. 

As a first step, we take the Bubbles with their morphology from Ref.~\cite{Su:2010qj} (their Table 1), which claims clear edges for these structures. For their spectrum between 0.1 and 600 GeV, we take the data from Table 2 of Ref.~\cite{Fermi-LAT:2014sfa}, which used a significantly higher number of photon statistics than~\cite{Su:2010qj}, while finding a very similar morphology for the Bubbles. 
Furthermore, we take the claimed errors of the Bubbles flux from Ref.~\cite{Fermi-LAT:2014sfa}, which, like the isotropic emission, can add a penalty factor in the calculated log-likelihood of the combined fit to the \textit{Fermi} data, when the flux component associated with the Bubbles deviates from the claimed results of~\cite{Fermi-LAT:2014sfa}, (see Section~\ref{subsec:fitting_procedure}).   

Ref.~\cite{Totani:2025fxx}, claimed also a Bubbles template but that template, contains also emission from other regions beyond the Bubbles and we will describe it in the following Section~\ref{subsec:Residuals}. 

We note that what is described in this subsection, is the \textit{initial} model for the \textit{Fermi} Bubbles' morphology and spectrum. In Section~\ref{sec:NewFermiBubbles}, we provide an updated model on the Bubbles, as we do conclude that the HLE is in fact part of the Bubbles emission not accounted for in the earlier results. 

\subsection{The emission associated with the positive and negative residual templates of Ref.~\cite{Totani:2025fxx}}
\label{subsec:Residuals}

Ref.~\cite{Totani:2025fxx}, in its analysis used two templates directly evaluated from the \textit{Fermi}-LAT data, at energies of a few GeV, referred to as the positive and negative residual templates (denoted in this work as ``Pos Res" and ``Neg Res" respectively). These templates are created to provide a better fit to the data and to account for ``discrepancies'' between the \texttt{GALPROP} galactic diffuse emission model and the gamma-ray data (for more details we refer the reader to Ref.~\cite{Totani:2025fxx}).
We use these positive and negative residual templates, evaluated from the fit to the $|\ell| \leq 60^{\circ}$, $|b| \leq 60^{\circ}$ sky region, at 4.3 GeV~\cite{TotaniPrivateComm}. \footnote{These residual templates give predictions for latitudes even lower than $|b| \leq 10^{\circ}$.} While the positive residual template was created to account for mostly the \textit{Fermi} Bubbles, it does contain many positive flux pixels in directions on the galactic sky that are well outside the Bubbles region. 

These residual maps were evaluated using the SL\,Z6\,R30\,T150\,C2 galactic diffuse emission model and cannot be directly used with any of the other 99 diffuse emission models, as any positive or negative residual map would have to be re-evaluated once the galactic diffuse emission prediction is also changed. 
For that reason, we only use the positive and negative residual templates in combination with the SL\,Z6\,R30\,T150\,C2 diffuse emission model. 
That is done only in order for us to check the compatibility of our results with those of Ref.~\cite{Totani:2025fxx}, when repeating the analysis of that work. 

We do not create any positive or negative residual templates using a methodology similar to that of Ref.~\cite{Totani:2025fxx}, for any of the other galactic diffuse emission models. Instead, we seek to derive our main results using the entire collection of galactic diffuse emission models and use that as our criterion of robustness of our  conclusions on the physical properties of the HLE. 

\subsection{The emission from the point and extended gamma-ray sources}
\label{subsec:PointSources}

In most of our analysis, we mask the point and extended gamma-ray sources of the 4FGLDR4 catalog~\cite{Ballet:2023qzs}, as described in Section~\ref{subsec:mask}. However, when comparing to the analysis of Ref.~\cite{Totani:2025fxx}, we create a template that contains the flux of all the point and extended gamma-ray sources.
To create that template, we used the FITS files provided by the \textit{Fermi} Collaboration \cite{4FGLDR4url}, which give for every point and extended gamma-ray source its best-fit spectral properties.
The majority of these sources can be described by a simple power-law spectrum, while most of the remaining sources are described by a log-parabolic spectrum. A small fraction of sources are best described by a power-law with a super-exponential cutoff.  
We use all three of these classes of spectra in creating the point and extended template, placing each of these sources on their respective best-fit location on the sky. 
In the fit (like with every other template) we convolve this template with the respective point spread function (PSF), that is energy dependent \cite{FermiPSFurl}.  

We note that, formally, the flux of each of these sources is evaluated by the \textit{Fermi} Collaboration using a different background model to that of SL\,Z6\,R30\,T150\,C2, and that in fact an analysis of the gamma-ray sky with the point and extended sources included should be using relevant \textit{Fermi} Collaboration tools \cite{FermiTOOLSurl} instead. 

We do not attempt to claim that having this energy-dependent template for the point and extended sources is the better way. However, we use that point and extended sources template only to check our ability to repeat the analysis of Ref.~\cite{Totani:2025fxx}. For the rest of our main results, we simply mask these sources.

\subsection{The emission from HLE}
For the HLE template, we model it as the integration along the line of sight of the prompt emission from annihilating dark matter, with the dark matter density following a generalized Navarro-Frenk-White  profile~\cite{Navarro:1996gj} (gNFW),
\beq
\label{eq:NFW}
\rho(r) = \rho_s \left(\frac{r}{r_s}\right)^{-\gamma} \left(1+\frac{r}{r_s}\right)^{-3+\gamma},
\eeq
where $\rho_s$, $r_s$, and $\gamma$ are the scale density, scale radius, and cuspiness, respectively. We refer to the $\gamma=1$ case simply as the NFW profile. We take $r_s=20$ kpc and the normalization of $\rho_s$ such that it gives a local (at $r=8.5$ kpc) dark matter density of 0.4 GeV cm$^{-3}$ \cite{Catena:2009mf, Salucci:2010qr, Bovy:2012tw}.

\section{Analyzing the \textit{Fermi}-LAT sky in alternative Regions of Interest}
\label{sec:analysis}

\subsection{Data}
\label{subsec:data}

We analyze \textit{Fermi}-LAT Pass~8 data~\cite{2013arXiv1303.3514A}, collected between August 4, 2008 (mission week 9) and April 29, 2025 (week 882). The event selection and exposure-map generation are performed with the \textit{Fermi} \texttt{ScienceTools P8v2.4.0}. To achieve the best angular resolution, we retain only the CLEAN, FRONT-converted events. 
We apply standard quality cuts of \(z_{\text{max}} = 100^\circ\), \(\texttt{DATA\_QUAL} = 1\), \(\texttt{LAT\_CONFIG} = 1\), and \(|\texttt{ROCK\_ANGLE}| < 52^\circ\). The data are binned into sixteen energy bins spanning $0.275~\text{GeV} \leqslant E_{\gamma} \leqslant 928.0~\text{GeV}$ (shown in~\tabref{mask}).

\begin{table}[h]
     \centering
     \topcaption{The energy binning and the point-source mask radii $r_\text{s}$ ($r_\text{l}$) applied to sources with $\text{TS} < 49$ ($\text{TS} \geqslant 49$) for 4FGLDR4 mask.}
     \label{tab:mask}
     \begin{tabular}{c c c c}
     \hline
        Energy~Bin & $E_{\min}-E_{\max}$ [GeV] & $r_\text{s}$[\degree] & $r_\text{l}$[\degree] \\
     \hline \hline
      0 & 0.275 -- 0.357 & 1.125 & 3.75 \\
      1 & 0.357 -- 0.464 & 0.975 & 3.25 \\
      2 & 0.464 -- 0.603 & 0.788 & 2.63 \\
      3 & 0.603 -- 0.784 & 0.600 & 2.00 \\
      4 & 0.784 -- 1.02 & 0.450 & 1.50 \\
      5 & 1.02 -- 1.32 & 0.375 & 1.25 \\
      6 & 1.32 -- 1.72 & 0.300 & 1.00 \\
      7 & 1.72 -- 2.24 & 0.225 & 0.750 \\
      8 & 2.24 -- 2.91 & 0.188 & 0.625 \\
      9 & 2.91 -- 3.78 & 0.162 & 0.540 \\
      10 & 3.78 -- 4.91 & 0.125 & 0.417 \\
      11 & 4.91 -- 10.8 & 0.100 & 0.333 \\
      12 & 10.8 -- 23.7 & 0.060 & 0.200 \\
      13 & 23.7 -- 51.9 & 0.053 & 0.175 \\
      14 & 51.9 -- 114.0 & 0.053 & 0.175 \\
      15 & 114.0 -- 928.0 & 0.053 & 0.175 \\
     \hline
\end{tabular}
\end{table}

\subsection{Region of Interest}
\label{subsec:roi}

\begin{figure*}[t]
    \centering
    \includegraphics[width=0.96\linewidth]{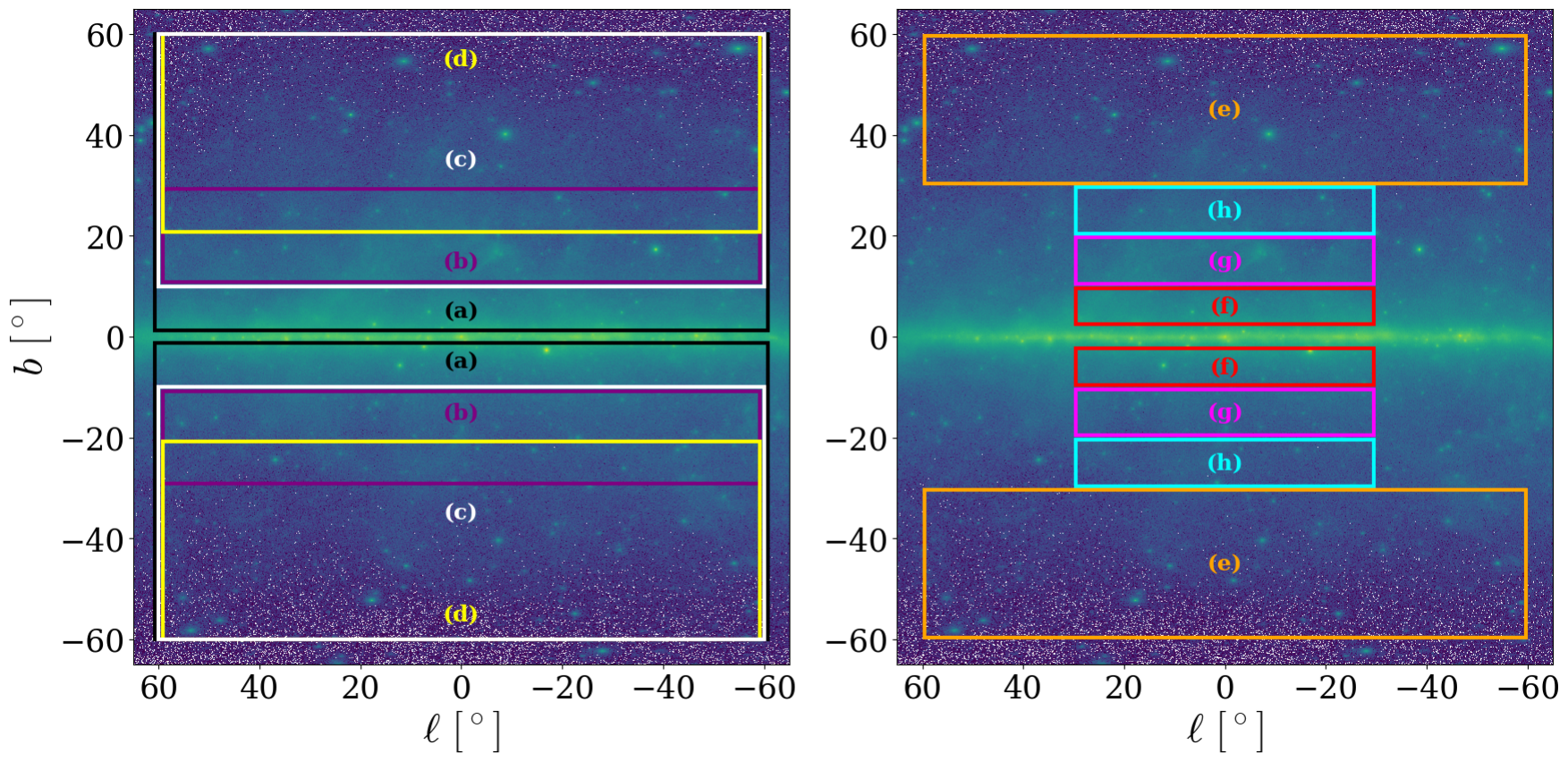}
    \caption{(\emph{left}) The ROIs $\textbf{a}$, $\textbf{b}$, $\textbf{c}$, and $\textbf{d}$ and (\emph{right}) $\textbf{e}$, $\textbf{f}$, $\textbf{g}$, and $\textbf{h}$, overlaid on the \emph{Fermi}-LAT sky (see also~\tabref{roi}, for their description). 
    }
    \label{fig:roi}
\end{figure*}

We focus on a region that is $120\degree \times 120^{\circ}$, centered at the Galactic center, and binned in Cartesian pixels of size $0.1\degree \times 0.1\degree$. 
Within this ROI, we define a set of sub-ROIs, listed in~\tabref{roi} and shown in~\figref{roi}.

\subsection{Mask}
\label{subsec:mask}
\begin{figure}[h]
    \centering
\includegraphics[width=\linewidth]{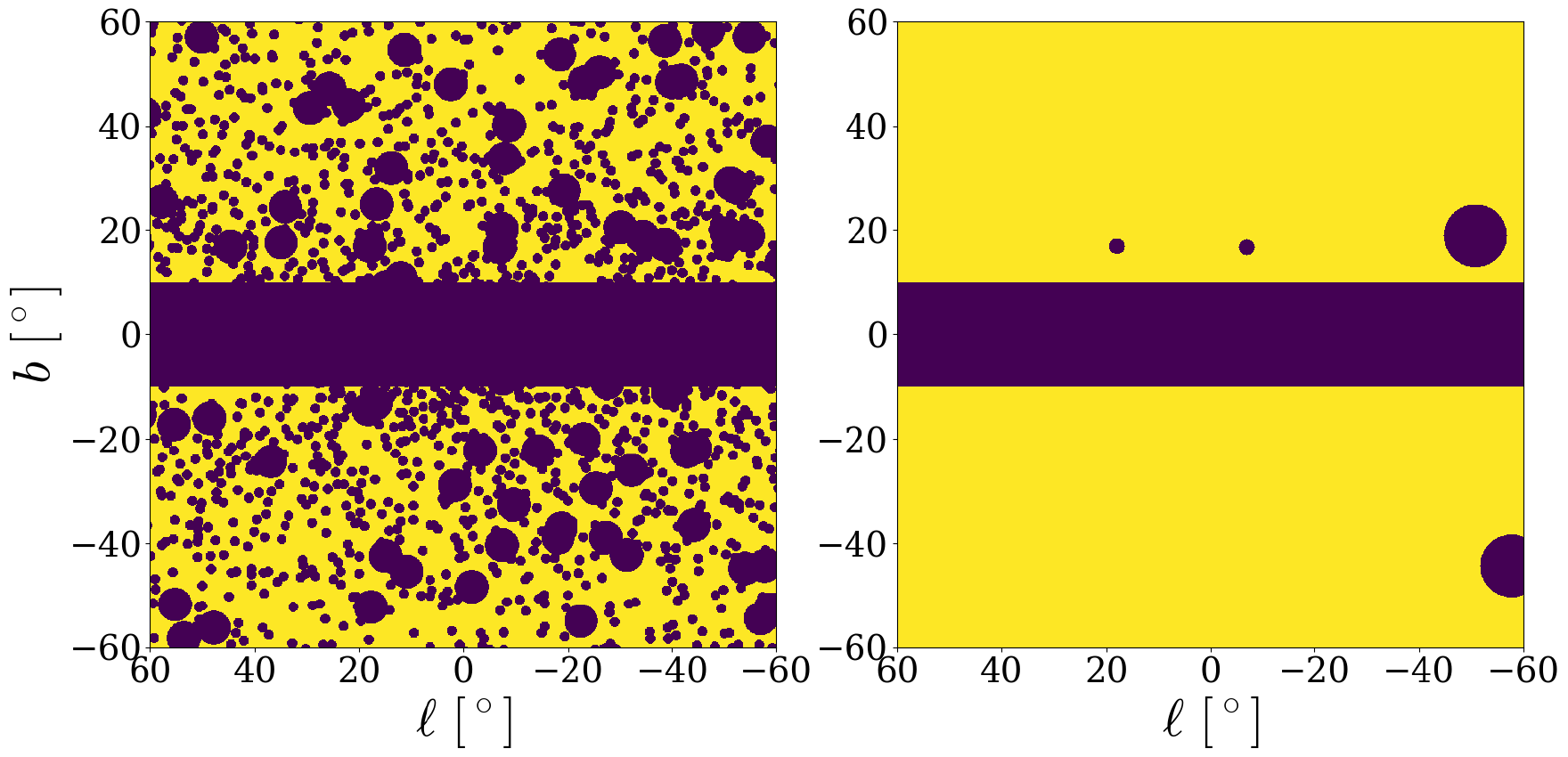}
    \caption{(\emph{Left}) The 4FGLDR4 mask with radii given in Table~\ref{tab:mask}, combined with a disk mask covering $|b|<10^\circ$. (\emph{Right}) The Totani  mask used in Ref.~\cite{Totani:2025fxx}, which combines the same disk mask with only a small number of masked \textit{Fermi} sources. Both masks are plotted for energy bin $0.357\leq E \leq 0.464$ GeV. 
    }
    \label{fig:mask_comparison}
\end{figure}

Within each ROI, we apply an energy-dependent mask to all point and extended sources of the 4FGLDR4 catalog~\cite{Ballet:2023qzs}. The masking radius depends on the source significance, with separate radii for sources ($r_\text{s}$ or $r_\text{l}$) of low or high test statistic ($\text{TS} < 49$ or $\text{TS} \geqslant 49$), as listed in~\tabref{mask}. We refer to it as ``4FGLDR4 mask". In addition, we consider the mask adopted in Ref.~\cite{Totani:2025fxx}, which we refer to as the ``Totani mask'' in what follows. Fig.~\ref{fig:mask_comparison}, compares the two masks.

\begin{table}[h]
    \centering
    \topcaption{Summary of the ROIs analyzed in this work.}
    \label{tab:roi}
    \begin{tabular}{cc}
    \hline
        Name & Ranges  \\
         \hline \hline
        $\textbf{a}$ & $|\ell| \leqslant 60\degree$, $2\degree \leqslant |b| \leqslant 60\degree$  \\
        $\textbf{b}$ & $|\ell| \leqslant 60\degree$, $10\degree \leqslant |b| \leqslant 30\degree$  \\
        $\textbf{c}$ & $|\ell| \leqslant 60\degree$, $10\degree \leqslant |b| \leqslant 60\degree$  \\
        $\textbf{d}$ & $|\ell| \leqslant 60\degree$, $20\degree \leqslant |b| \leqslant 60\degree$  \\
        $\textbf{e}$ & $|\ell| \leqslant 60\degree$, $30\degree \leqslant |b| \leqslant 60\degree$  \\
        \textbf{f} & $|\ell| \leqslant 30\degree$, $2\degree \leqslant |b| \leqslant 10\degree$  \\
        \textbf{g} & $|\ell| \leqslant 30\degree$, $10\degree \leqslant |b| \leqslant 20\degree$  \\
        $\textbf{h}$ & $|\ell| \leqslant 30\degree$, $20\degree \leqslant |b| \leqslant 30\degree$  \\
        \hline
    \end{tabular}
\end{table}

\subsection{Fitting Procedure}
\label{subsec:fitting_procedure}

For each energy bin $j$, we model the total diffuse flux in every unmasked pixel $p$ of a given ROI as a weighted superposition of diffuse emission templates,
\beq
\Phi_{j,p}^{\rm TOT} =\sum_{T} c_{j}^{T} \Phi_{j,p}^{T},
\eeq
where the index $T$ runs over the five to nine templates included in a given fit: Pi0+Bremss, ICS, Isotropic, Loop I (or the Loop I Dim and Loop I Bright templates), and the \emph{Fermi} Bubbles template from~\cite{Su:2010qj} or, alternatively, the Pos Res and Neg Res templates from~\cite{Totani:2025fxx}, (or the new Bubbles and Cocoons templates developed later in this work) and finally just for the nine-parameter fit the combined 4FGLDR4 sources.
When the fit includes the HLE, $T$ also runs over the HLE template. The coefficients $c_{j}^{T}$ are the template normalizations to be fitted. We always set $c_{j}^\text{Pi0} = c_{j}^{\text{Bremss}}$, since the two templates trace similar spatial distributions (albeit with some differences).  The model count map is then obtained by multiplying the total flux map by the exposure map.

The total log-likelihood $\ln (\mathcal L)$ is evaluated over the ROI's pixels with a mask ${\cal M}_{j,p}$, by comparing the model counts (${\cal C}_{j,p}$) with the observed \emph{Fermi}-LAT counts (${\cal D}_{j,p}$). The resulting fit statistic is $-2 \ln (\mathcal L) = -2 \sum_j \ln (\mathcal L_j)$ with,
\begin{align}
\label{eq:log-likelihood_function}
-2\ln ({\cal L}_j) ={}& \Bigg\{ 2\sum_{p}\Bigg[{\cal M}_{j,p}{\cal C}_{j,p} + \ln\left[\left({\cal M}_{j,p}{\cal D}_{j,p}\right)!\right] \nonumber\\
& - {\cal M}_{j,p}{\cal D}_{j,p} \ln\left({\cal M}_{j,p}{\cal C}_{j,p}\right) \Bigg]  \Bigg\} \nonumber\\
& + \chi^{2}_{{\rm Bubbles},j} + \chi^{2}_{{\rm Iso},j} + \chi^{2}_{{\rm bright},j} + \chi^{2}_{{\rm dim},j}
\end{align}
The penalty terms $\chi^{2}_{{\rm Bubbles},j}$, $\chi^{2}_{{\rm Iso},j}$, $\chi^{2}_{{\rm bright},j}$, and $\chi^{2}_{{\rm dim},j}$ constrain the normalizations of the \emph{Fermi} Bubbles, Isotropic, Loop I Bright, and Loop I Dim components, respectively, to match their measured high-latitude spectra; see Refs.~\cite{Fermi-LAT:2014sfa, Fermi-LAT:2014ryh}. Ref.~\cite{Fermi-LAT:2014sfa} provides a constraint only on the overall normalization of Loop I, without separately constraining the Loop I Bright and Loop I Dim components. We therefore impose the same normalization constraint on both Loop I Bright and Loop I Dim as that derived for the total Loop I emission.

To sample the parameter space, we first explore the posterior distributions of the normalization coefficients $c_{j}^{T}$ with the nested-sampling package \texttt{DYNESTY}~\cite{2020MNRAS.493.3132S}, and then refine the inference with a No-U-Turn Sampler/Hamiltonian Monte Carlo algorithm implemented in \texttt{NumPyro}~\cite{2019arXiv191211554P}. Both steps are executed within the \texttt{gcepy} pipeline~\cite{McDermott:2022zmq, Cholis:2021rpp}. Finally, we process the resulting chains with \texttt{ChainConsumer}~\cite{2016JOSS....1...45H}, to extract the central values and $1\sigma$ uncertainties of the $c_{j}^{T}$ posteriors (we typically present the $2\sigma$ uncertainties of the $c_{j}^{T}$).

Table~\ref{tab:fit_schemes}, summarizes the fitting schemes, i.e., the combinations of ROI, mask and templates used in Section~\ref{sec:Results}. We name them according to the number of independent normalizations of the fit as ``six-parameter", ``seven-parameter", ``eight-parameter", or ``nine-parameter" fit. For fits in ROIs \textbf{a} through \textbf{e}, we adopt the 20 diffuse emission model combinations of templates from Ref.~\cite{Fermi-LAT:2012edv}, including model SL\,Z6\,R30\,T150\,C2 templates. 
For fits in ROIs \textbf{f}, \textbf{g}, and \textbf{h}, we use the same 20 galactic diffuse emission models together with the templates predicted by the 80 emission models from \cite{Cholis:2021rpp}. For the HLE, we consider three scenarios (i) no presence, (ii) the dark matter density follows NFW (gNFW with $\gamma=1.0$), and (iii) the dark matter density follows gNFW with $\gamma = 1.25$.

\begin{table*}[t]
    \centering
    \renewcommand{\arraystretch}{1.3}
    \topcaption{Summary of fitting schemes.}
    \label{tab:fitting}
    \newcommand{\name}[1]{\parbox[t]{0.1\textwidth}{\raggedright #1\strut}}%
    \newcommand{\roi}[1]{\parbox[t]{0.1\textwidth}{\raggedright #1\strut}}%
    \newcommand{\mask}[1]{\parbox[t]{0.1\textwidth}{\raggedright #1\strut}}%
    \newcommand{\tmpl}[1]{\parbox[t]{0.5\textwidth}{\raggedright #1\strut}}%
    \newcommand{\gde}[1]{\parbox[t]{0.2\textwidth}{\raggedright #1\strut}}%
    \begin{tabular}{l l l l l}
    \hline
     Name & ROI & Mask & Templates & GDE templates\\
    \hline \hline 
    \name{6-param} &
\roi{\textbf{a},\textbf{b},\textbf{c},\textbf{d},\textbf{e}} & \mask{4FGLDR4} & \tmpl{Pi0+Bremss, ICS, Isotropic, Bubbles, Loop I, HLE} & SL\,Z6\,R30\,T150\,C2\\
            \hline
\name{7-param} & \roi{\textbf{a},\textbf{b},\textbf{c},\textbf{d},\textbf{e}, \\\textbf{f},\textbf{g},\textbf{h}} & \mask{4FGLDR4} & \tmpl{Pi0+Bremss, ICS, Isotropic, Bubbles, Loop I Bright, Loop I Dim, HLE} & \gde{Refs.~\cite{Fermi-LAT:2012edv} and~\cite{Cholis:2021rpp} \\ (up to 100 models)}\\
            \hline
\name{8-param} & \roi{\textbf{a},\textbf{b},\textbf{c},\textbf{d},\textbf{e}} & \mask{4FGLDR4} & \tmpl{Pi0+Bremss, ICS, Isotropic, Pos Res, Neg Res, Loop I Bright, Loop I Dim, HLE} & SL\,Z6\,R30\,T150\,C2\\
            \hline
           \name{9-param}  & \roi{\textbf{c}} & \mask{Totani} & \tmpl{Pi0+Bremss, ICS, Isotropic, Pos Res, Neg Res, Loop I Bright, Loop I Dim, 4FGLDR4 sources, HLE} & SL\,Z6\,R30\,T150\,C2\\
    \hline
    \end{tabular}
     \label{tab:fit_schemes}
\end{table*}

\section{Results}
\label{sec:Results}

\subsection{Testing the results of Ref.~\cite{Totani:2025fxx}}

First, we present the results following the methodology of Ref.~\cite{Totani:2025fxx}, i.e. our nine-parameter fit using the ``Totani mask'', described also in Table~\ref{tab:fit_schemes}. 

\begin{figure}[ht]
    \centering
    \includegraphics[width=0.98\linewidth]{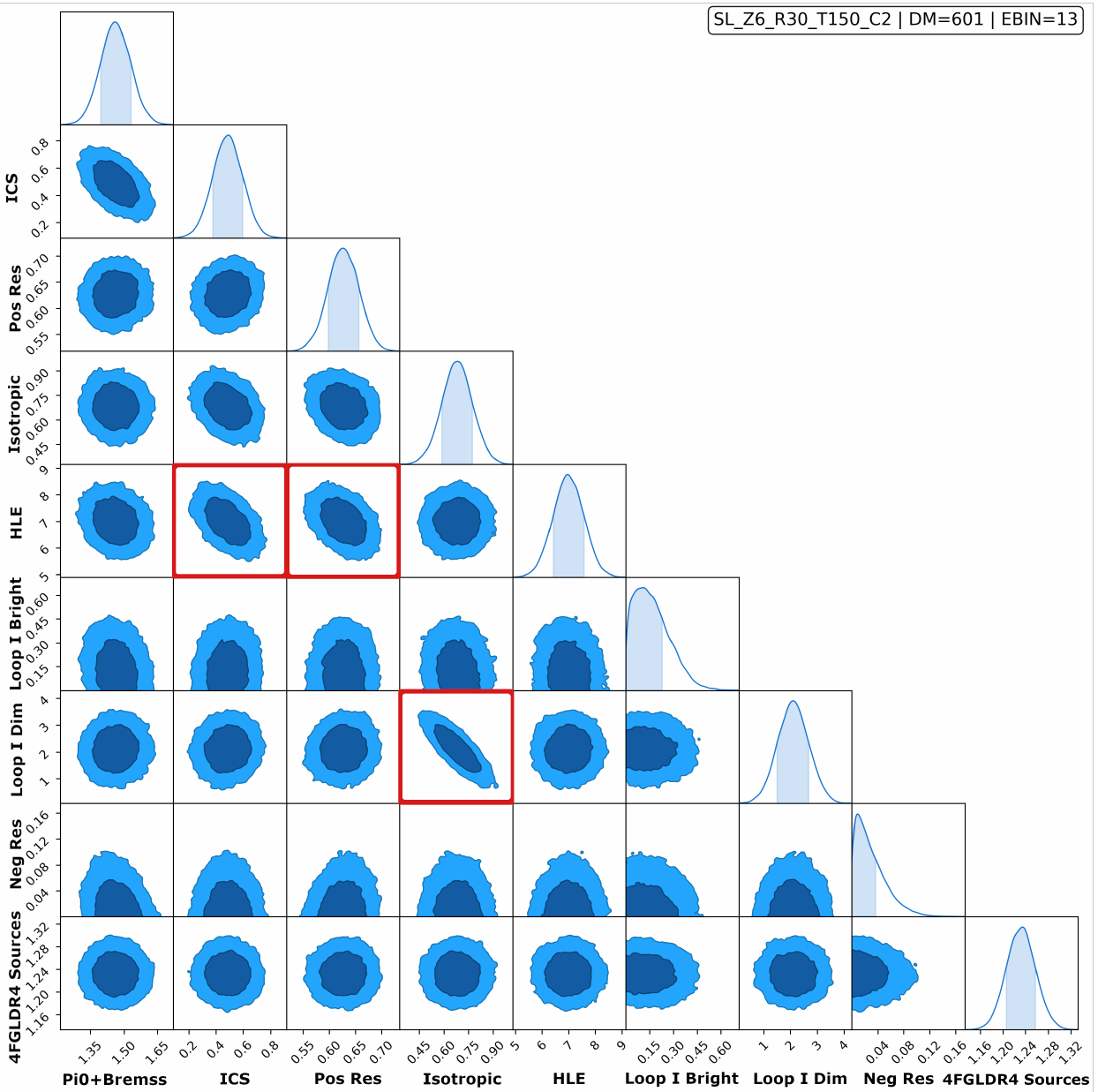}
    \caption{The nine-parameter fit to the ROI \textbf{c} using Totani mask. We show the results on the template normalizations 1- (darker blue) and 2-$\sigma$ (lighter blue) ranges at the energy $23.7\leq E\leq 51.9$ GeV. We find some level of degeneracy between the normalization parameters as e.g. between those for the ICS and the HLE templates, or for the Isotropic and the Loop I Dim templates (highlighted with red boxes).}
    \label{fig:CornerPlotTotaniMethod}
\end{figure}

In Fig.~\ref{fig:CornerPlotTotaniMethod}, we show the corner plot of the ranges for the nine normalization parameters, evaluated for the energy bin 13, $23.7\leq E\leq 51.9$ GeV. We show the 1- and 2-$\sigma$ ranges for each of these nine normalization parameters, for the ROI \textbf{c}, with the Totani mask (as in Ref.~\cite{Totani:2025fxx}). 
We use the same galactic diffuse emission model, SL\,Z6\,R30\,T150\,C2, as Ref.~\cite{Totani:2025fxx}.

Our fitting procedure can easily find the best-fit combination for the nine normalizations of the templates in each of the sixteen energy bins.
However, we do notice certain degeneracies between some of these normalization parameters. 
In particular, there is a significant level of degeneracy between the normalization for the Isotropic template component and that associated to the ``Loop I Dim'' shell template. 
Similarly, there is a strong level of degeneracy between the ICS and the HLE templates, and some degeneracy between the HLE and the positive residuals (``Pos Res'') templates.
We find these degeneracies to be the case not just in the energy bin shown, but across all energies (especially the degeneracy between the Isotropic and the Loop I Dim templates). The degeneracy between the HLE and the positive residuals template appears only above 2 GeV. 
As we describe in further detail in presenting the fit results from the various smaller ROIs, the degeneracy between some of these templates may explain the true nature of the HLE. 

\begin{figure}[ht]
    \centering
    \includegraphics[width=0.98\linewidth]{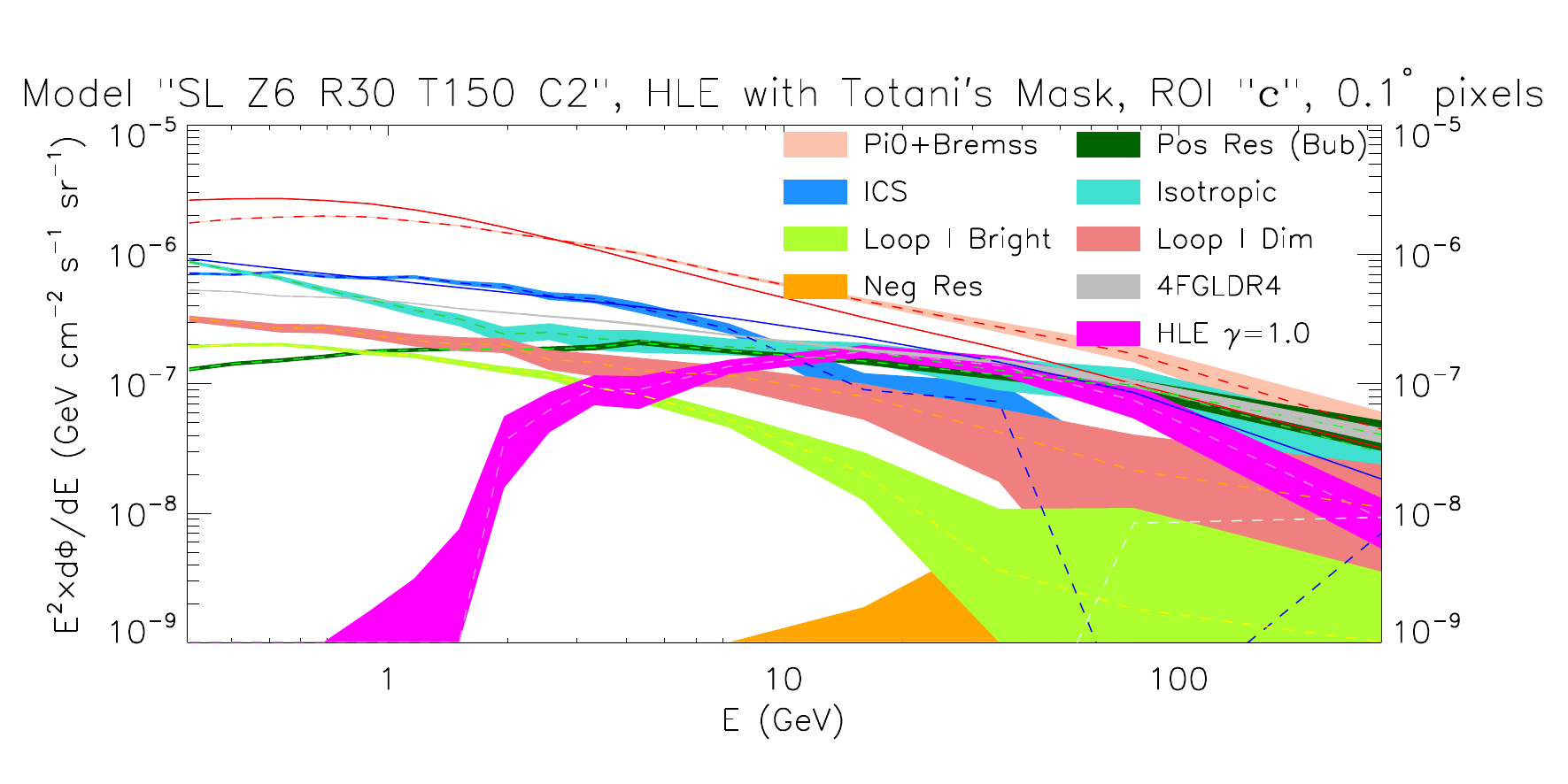}\\
    \vspace{-0.3cm}
    \includegraphics[width=0.98\linewidth]{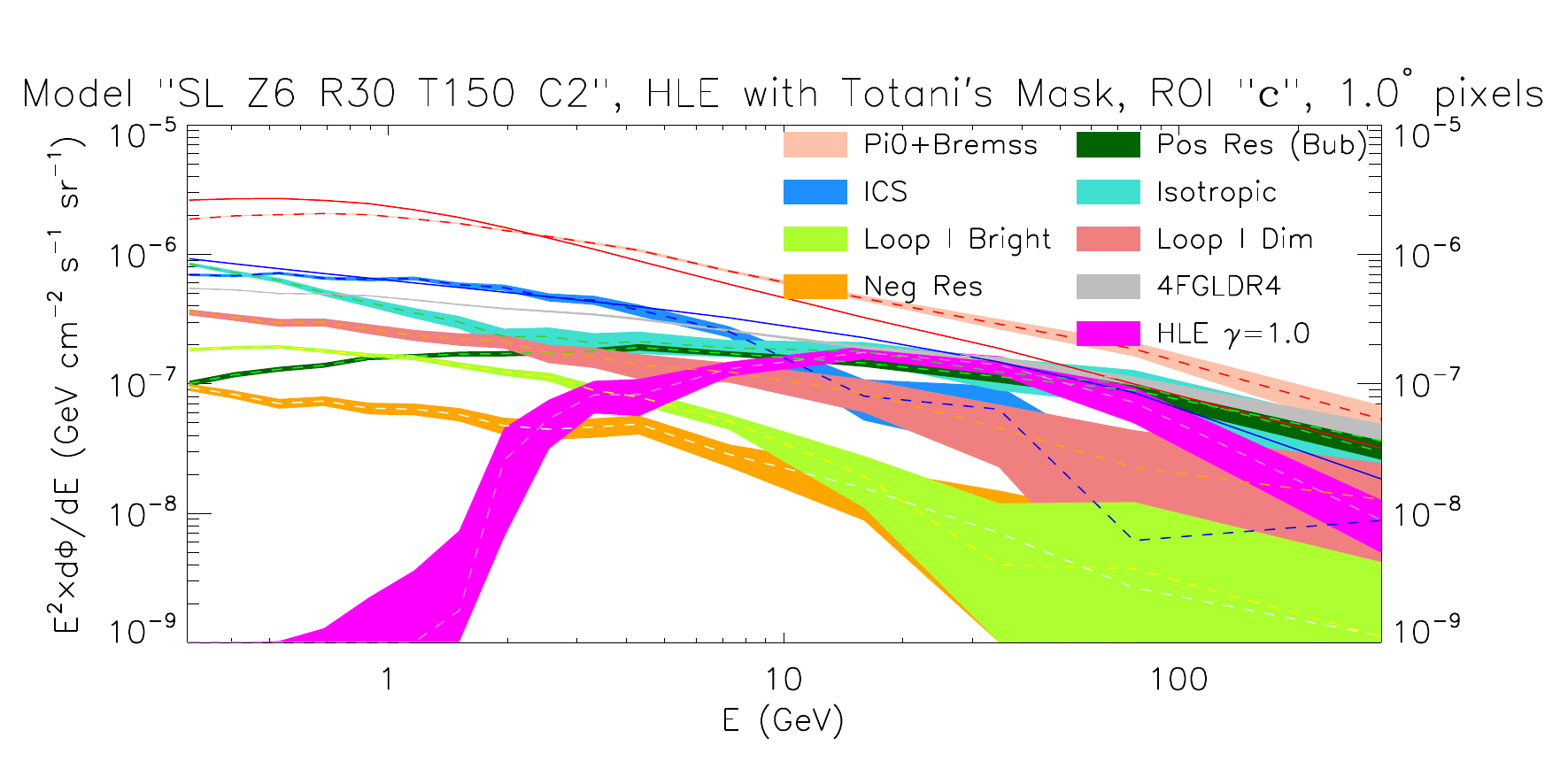}\\
    \vspace{-0.3cm}
    \includegraphics[width=0.98\linewidth]{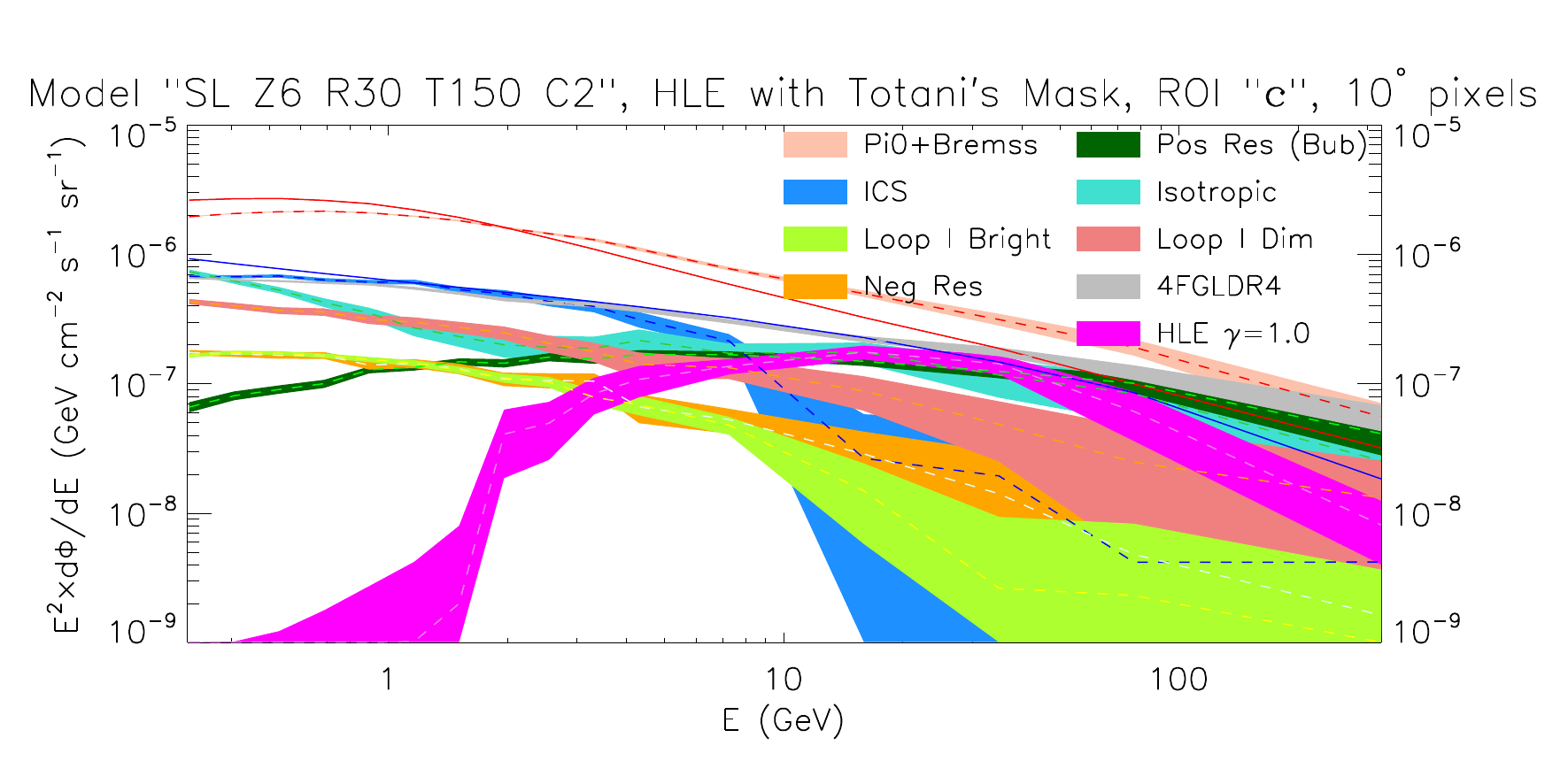}
    \vspace{-0.3cm}
    \caption{Using the reference galactic diffuse emission model and the negative and positive residual templates of Ref.~\cite{Totani:2025fxx}, with the 4FGLDR4 point sources as an additional template. We use the same ROI (``\textbf{c}'') and the Totani mask as Ref.~\cite{Totani:2025fxx}. (\emph{Top}) using the $0.1^{\circ}\times0.1^{\circ}$ pixels. (\emph{Middle}) grouping the $0.1^{\circ}\times0.1^{\circ}$ pixels to have a size of $1^{\circ}\times1^{\circ}$, following the procedure in Ref.~\cite{Totani:2025fxx}.
    (\emph{Bottom}) grouping $0.1^{\circ}\times0.1^{\circ}$ pixels to have size of $10^{\circ}\times10^{\circ}$, which is the same pixel size as the one used in the fit of~\cite{Totani:2025fxx}. We get a roughly similar flux associated with the HLE template as claimed by Ref.~\cite{Totani:2025fxx}. We notice that as we group the pixels, the HLE template is increased as the normalizations of several templates change across all energy bins. Furthermore, the negative residuals template's normalization gets increased as we increase the size of the pixels.}
    \label{fig:HLE_TotaniMask}
\end{figure}

In the fit of Fig.~\ref{fig:CornerPlotTotaniMethod}, we used our standard $0.1^{\circ}\times0.1^{\circ}$ pixelization for the ROI \textbf{c}, which differs to the prescription of Ref.~\cite{Totani:2025fxx}, which grouped all pixels before doing the template fit to much larger pixels of $10^{\circ}\times10^{\circ}$ in size. In Fig.~\ref{fig:HLE_TotaniMask}, we show the associated gamma-ray spectra from all nine components, for all energy bins, $0.275\leq E \leq 928$ GeV. We show fit results with a progressive level of pixel grouping, from our standard pixels of $0.1^{\circ}\times0.1^{\circ}$, i.e. no grouping (in the top panel), to $1.0^{\circ}\times1.0^{\circ}$ (middle panel), to $10^{\circ}\times10^{\circ}$ (bottom panel). We used again the SL\,Z6\,R30\,T150\,C2 galactic diffuse emission model to get the Pi0, Bremss, and ICS template predictions, which we used as input in our fit. We also adopted the Totani mask through all that grouping analysis. 

In all panels of Fig.~\ref{fig:HLE_TotaniMask}, Pi0+Bremss and the ICS original flux predictions are shown as solid red and dark blue lines. The best-fit flux spectra are shown through the equivalent color (red and blue) dashed lines; with the light red and blue bands around them being the 2-$\sigma$-fit normalization ranges.
For all the other templates, we just show the best-fit fluxes (dashed lines) and 2-$\sigma$-fit ranges as there is no model-specific prediction.\footnote{As we described in Section~\ref{sec:gammaraysky} and in Eq.~\ref{eq:log-likelihood_function}, prior assumptions on the isotropic gamma-ray spectrum, the spectrum of the \textit{Fermi} Bubbles and of Loop I, are taken into account in the fit.} 
The comparison between the original \texttt{GALPROP} model prediction from Ref.~\cite{Fermi-LAT:2012edv}, and the subsequent fit result is important.
While the Pi0+Bremss prediction spectrum before the fit and the Pi0+Bremss spectrum after the fit disagree only at the level of $\mathcal O(10\%)$, for the ICS template the fitted flux at energies above 10 GeV significantly deviates ---by as much as an order of magnitude--- from its original prediction, i.e., from the original physical assumptions on the properties of the Milky Way that were inputted in calculating it. This is an example of the degeneracies that exist in these energies between the different templates that are being fitted.
In the case of the used ICS template, it is most degenerate to the HLE template. 

Furthermore, we notice that in all panels of Fig.~\ref{fig:HLE_TotaniMask}, the template denoted as ``Loop I Bright'', i.e., the shell of Loop I that is brighter in the radio observations, is actually dimmer than the ``Loop I Dim'' in the gamma-ray data. 
To first order, the same energy electrons emitting via synchrotron radiation in radio waves emit via ICS in gamma rays. 
Thus, to first order, the expectation is that the gamma-ray flux modeled in the fit by the ``Loop I Bright'' template should be higher than the gamma-ray flux modeled by the ``Loop I Dim'' template.
In the radio data, the ``Loop I Bright'' shell has a factor of three higher associated radio flux than the ``Loop I Dim''. 
However, in the gamma-ray fit, the relative flux of Loop I Bright: Loop I Dim instead of being 3:1, is closer to 1:2. 
While the two shells may have different underlying spectra of the cosmic-ray electrons between them, that difference in electron spectra should have also been observed in the gamma-ray fit, i.e. the light green band and the brown band should suggest significantly different electron spectra by having different power laws, which they do not.
The fact that as we increase the grouping of the pixels the tension between the radio observations expectation and the gamma-ray fit result becomes stronger (the ``Bright'' shell becomes even more dim) suggests an artifact of the fit.

As an additional note, the negative flux that is associated with the negative residuals (Neg Res) template increases in amplitude, i.e. there is more negative flux as we group more pixels (see the orange band in the three panels of Fig.~\ref{fig:HLE_TotaniMask}).
This is a natural consequence of our fitting procedure that does not allow for the total number of predicted photons in any single pixel to be negative, which sets a restriction on the overall normalization of the negative component. After grouping, the total number of pixels is reduced, and this restriction is subsequently relaxed. 
The negative residuals map is meant to account for ``mismodeling'' of the galactic diffuse emission as is also the positive residual (Pos Res) template. Interestingly, as we group more pixels, the positive flux associated with the Pos Res template does not increase as well to balance the greater negative flux; it decreases instead. The normalizations of other templates change to account for the changing associated positive and negative fluxes.
The fact that the normalizations of the positive and negative templates change significantly with the grouping of the pixels and the fact that the negative residuals template can become a significant component once we reach the $10^{\circ}\times10^{\circ}$ grouping of pixels, raise questions on the robustness in the use of negative/positive residual templates evaluated at a fixed energy (see discussion on Section~\ref{subsec:Residuals}). 
Even for a fixed galactic diffuse emission model and for a fixed ROI and mask, the fit-normalizations change with alternative pixelizations, even though the actual positive and negative templates' high-resolution modeled morphologies do not. 

\begin{figure*}[t]
    \centering
    \includegraphics[width=0.49\linewidth]{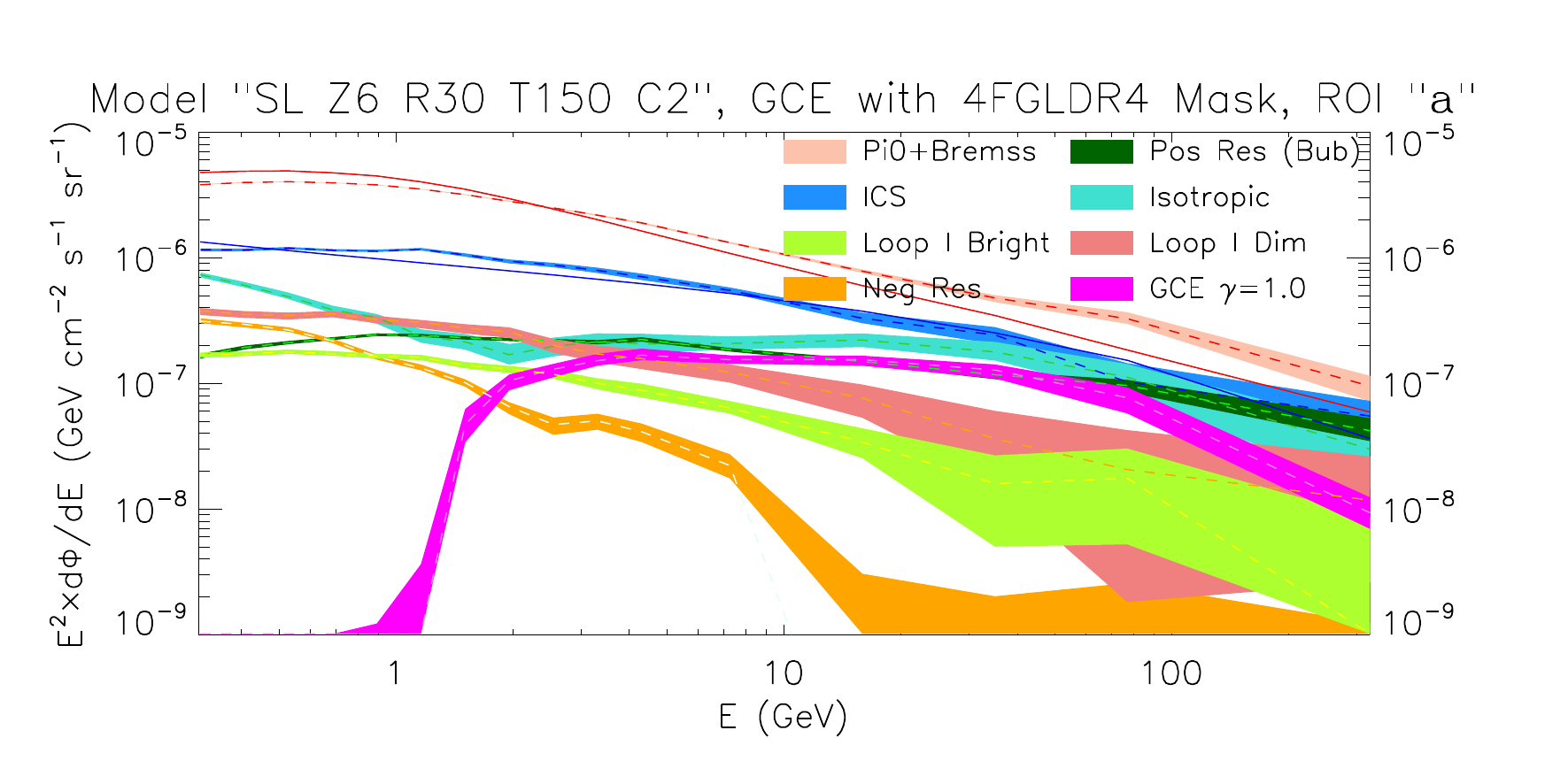}~    \includegraphics[width=0.49\linewidth]{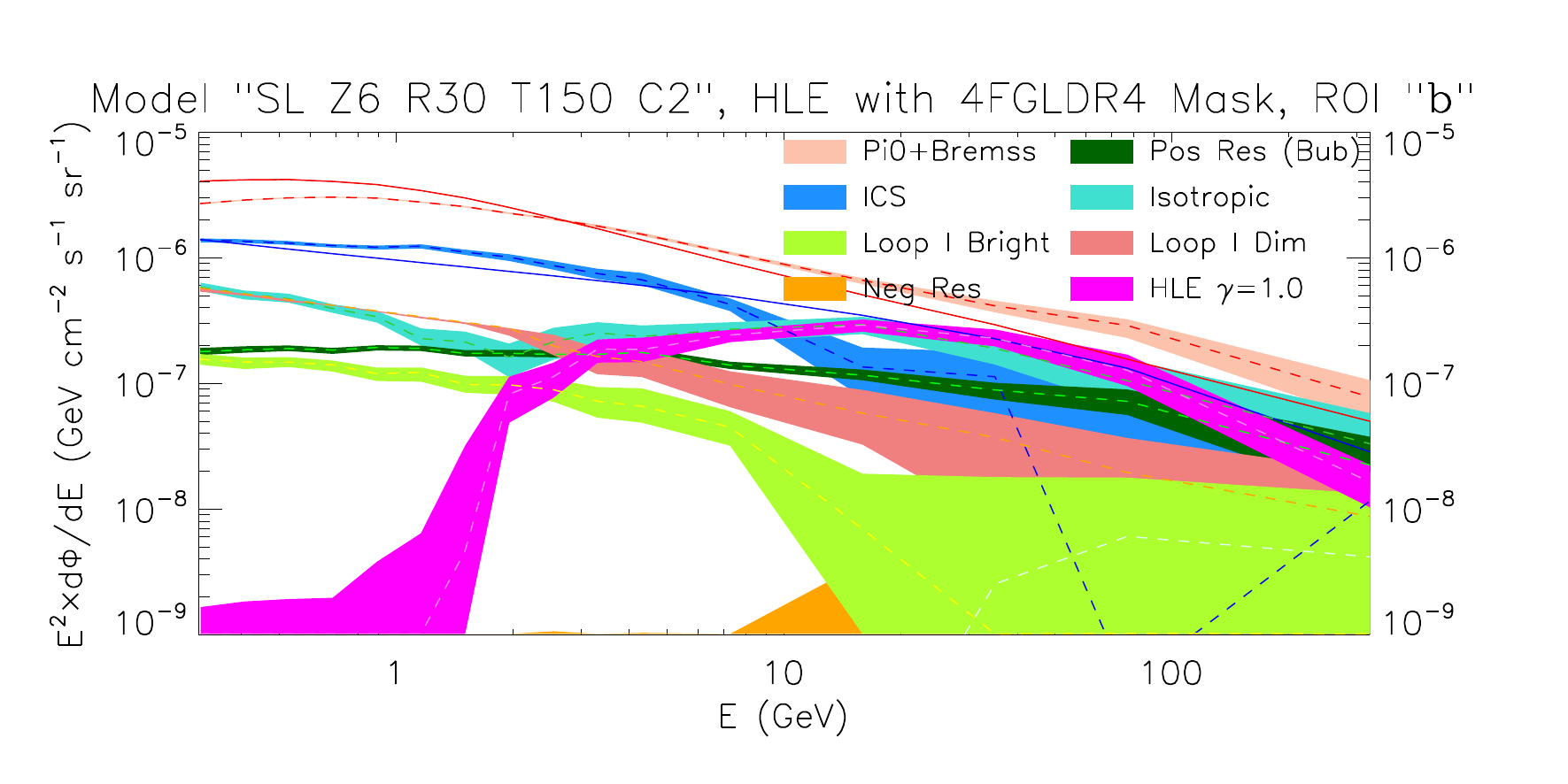}\\
    \vspace{-0.3cm}
    \includegraphics[width=0.49\linewidth]{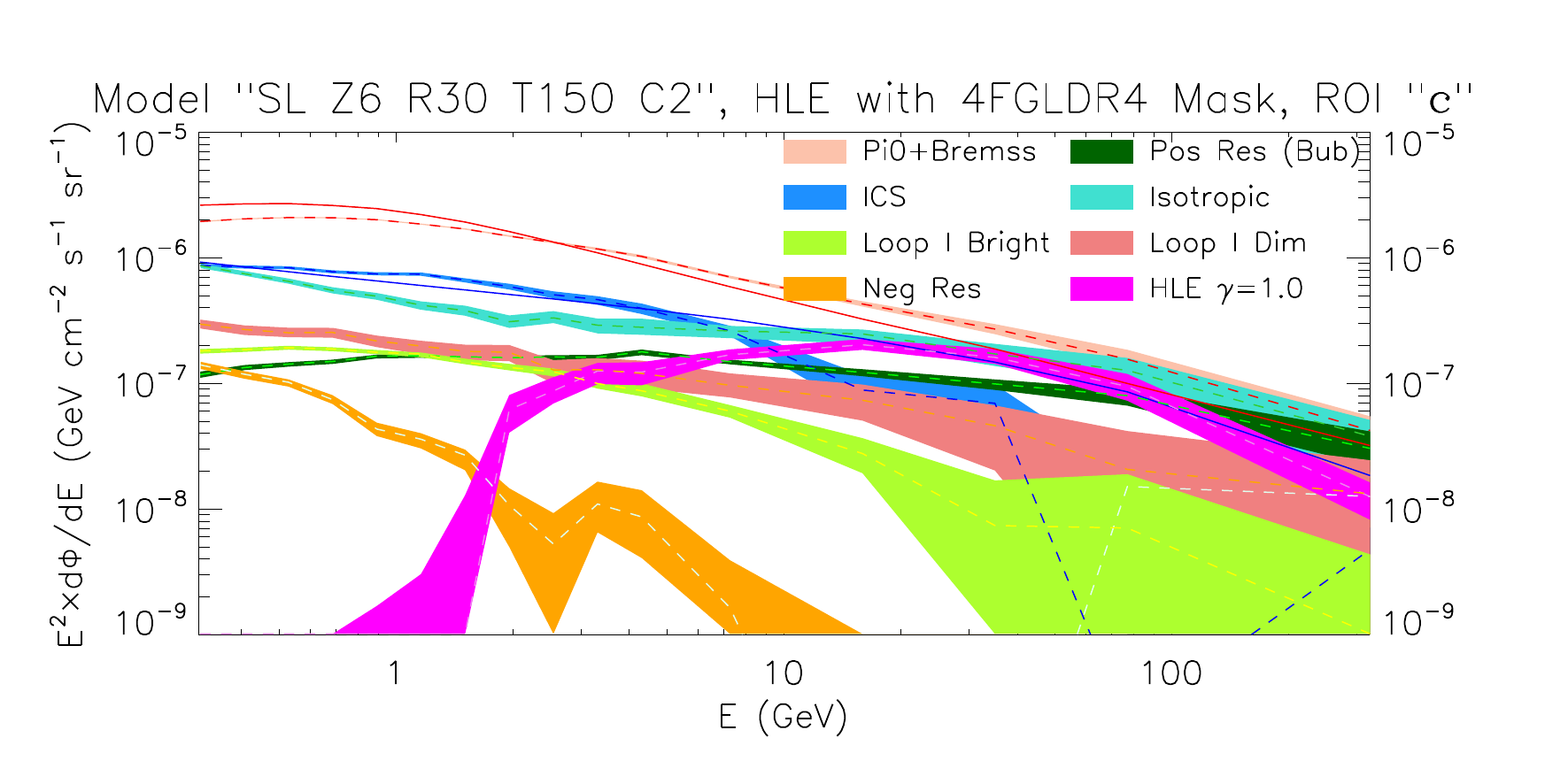}~
    \includegraphics[width=0.49\linewidth]{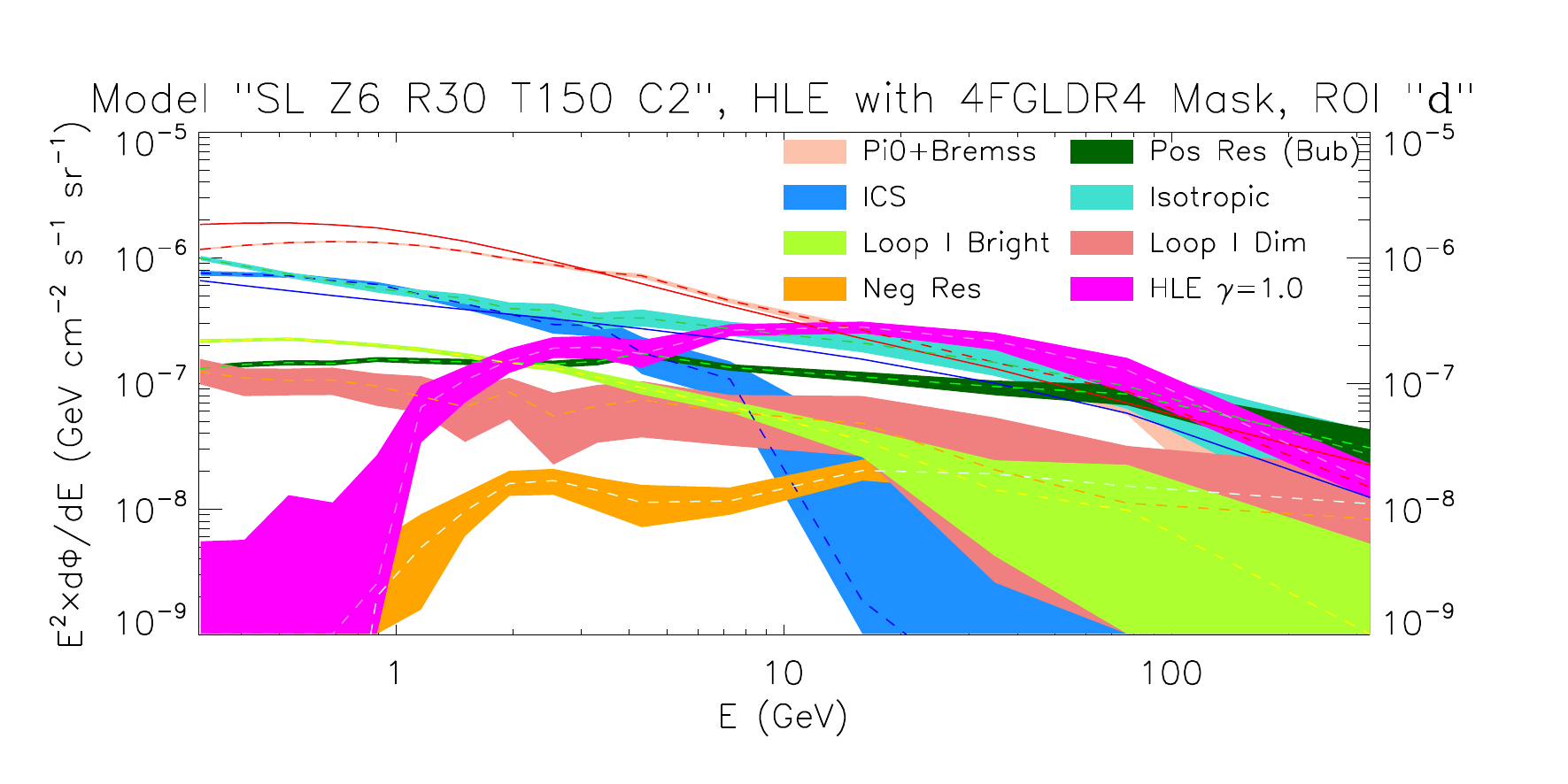}\\
    \vspace{-0.3cm}
    \includegraphics[width=0.49\linewidth]{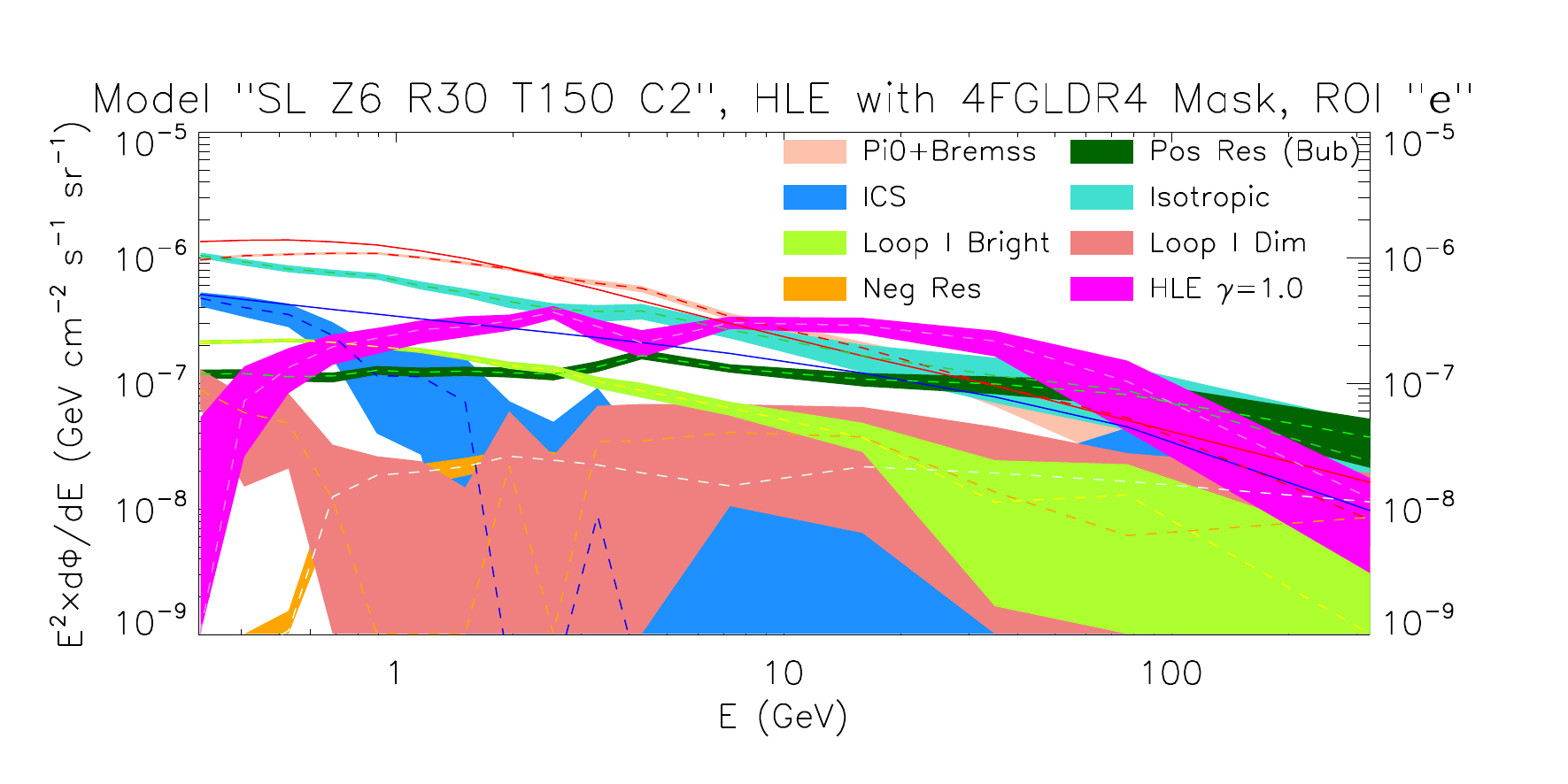}
    \vspace{-0.3cm}
    \caption{The galactic diffuse emission spectra for the $120^{\circ} \times 120^{\circ}$ ROI, with the 4FGLDR4 \textit{Fermi} point and extended sources masked in all cases in the fit. We show five alternative galactic latitude ranges.
    We include the same galactic diffuse emission model and Loop I modeling as Ref.~\cite{Totani:2025fxx}, and we use the negative and positive residual templates of Ref.~\cite{Totani:2025fxx}.
    We show: 
    i) in the top left panel ROI ``\textbf{a}'', ii) in the top right panel ROI ``\textbf{b}'', iii) in the middle left panel ROI ``\textbf{c}'', iv) in the middle right panel ROI ``\textbf{d}'' and v) in the bottom panel ROI ``\textbf{e}''.
    Changing ROIs from \textbf{a} to \textbf{e}, we show regions where the fit probes the emission from increasingly higher latitudes.
    By adding the negative residual template in the fit we find the spectrum of gamma rays associated with the dark-matter annihilation-like template to expand to higher energies in all studied ROIs. However, their amplitude is inconsistent with what would be expected for a dark matter annihilation signal between most ROIs (see text for details).}
    \label{fig:HLE_120x120_WITH_ResMaps}
\end{figure*}

In order to test the robustness of the HLE results, instead of having the point and extended sources template, we choose to mask these 4FGLDR4 sources (shown in the left panel of Fig.~\ref{fig:mask_comparison}), and repeat the analysis. 
Masking the point sources is in fact a more robust approach to using a single template for their combined flux.
Furthermore, we test not just ROI ``\textbf{c}'', i.e. $10^{\circ} \leq |b| \leq 60^{\circ}$, $|\ell| \leq 60^{\circ}$, but five choices of ROIs (``\textbf{a}'' to ``\textbf{e}'' of Table~\ref{tab:roi}). 
At low latitudes, we know that there is the GCE, which will be picked in the fit.
For this reason when we include the $2^{\circ} \leq |b|\leq 60^{\circ}$ latitudes (ROI ``\textbf{a}''), we refer to the dark-matter-like diffuse emission template component as the GCE; while when we mask the $|b|\leq 10^{\circ}$, or fit even higher latitudes regions, we refer to dark-matter-like diffuse emission template as the HLE.

\begin{figure*}[ht]
    \centering
    \includegraphics[width=0.49\linewidth]{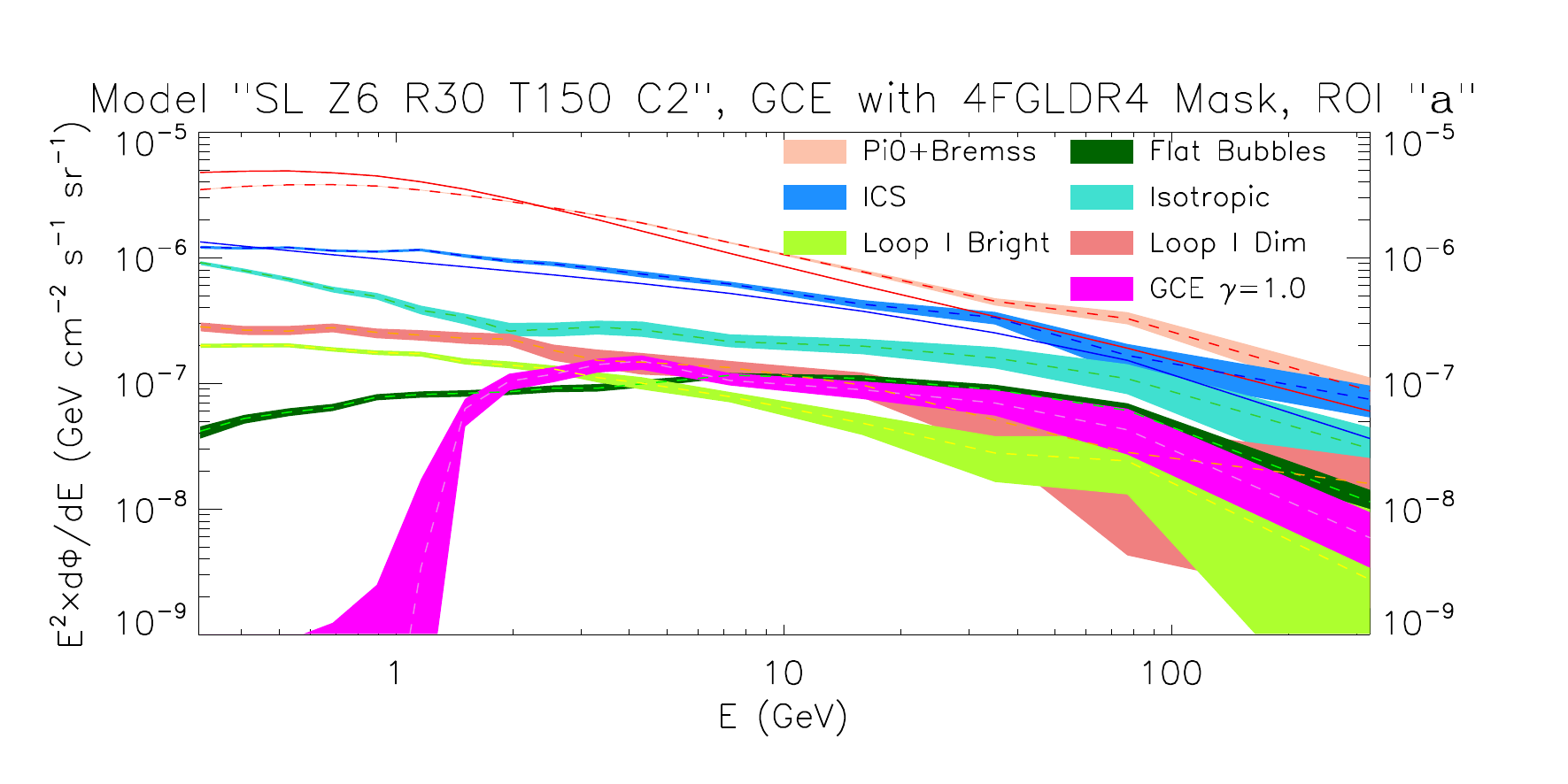}~    \includegraphics[width=0.49\linewidth]{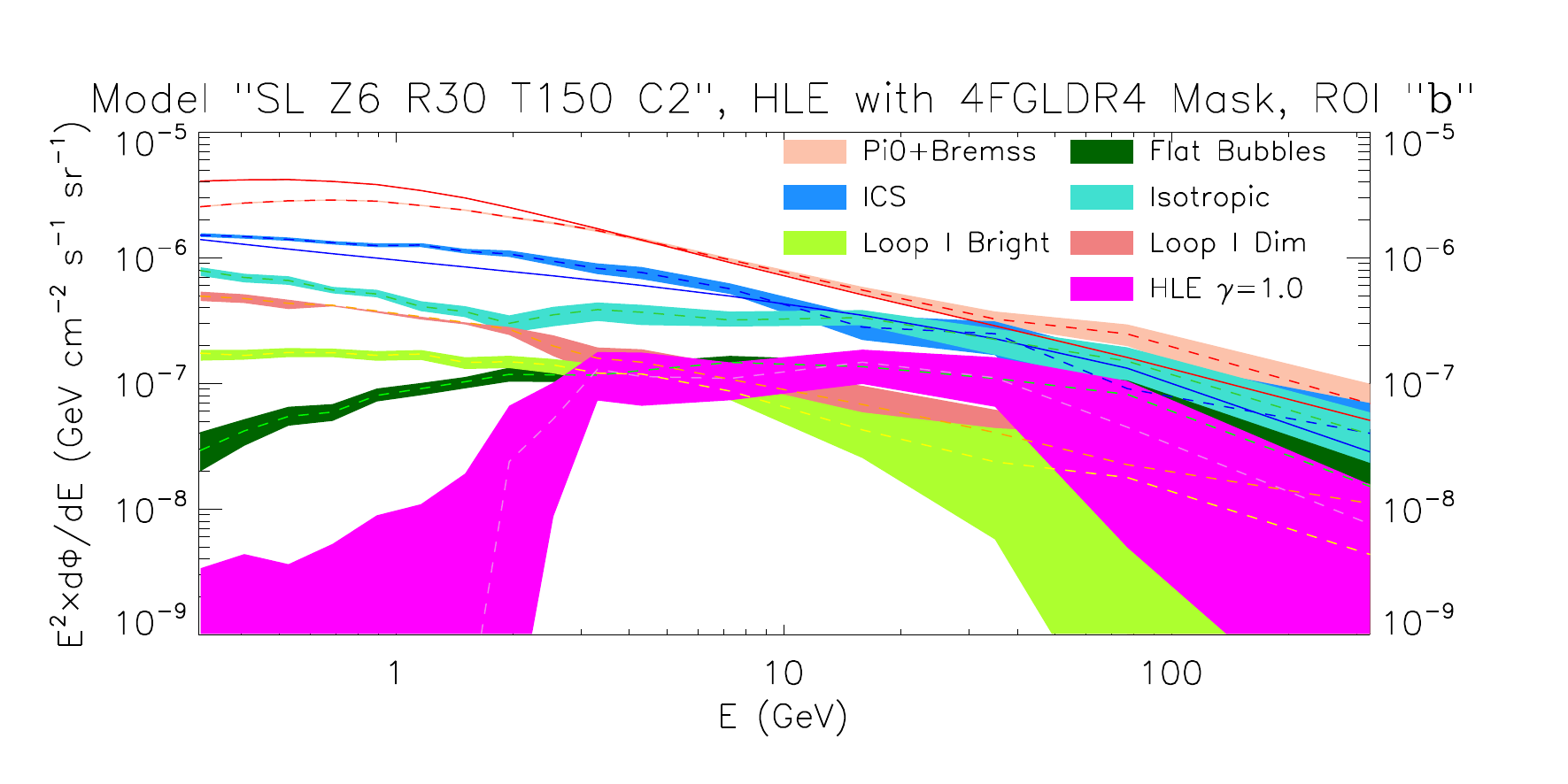}\\
    \vspace{-0.3cm}
    \includegraphics[width=0.49\linewidth]{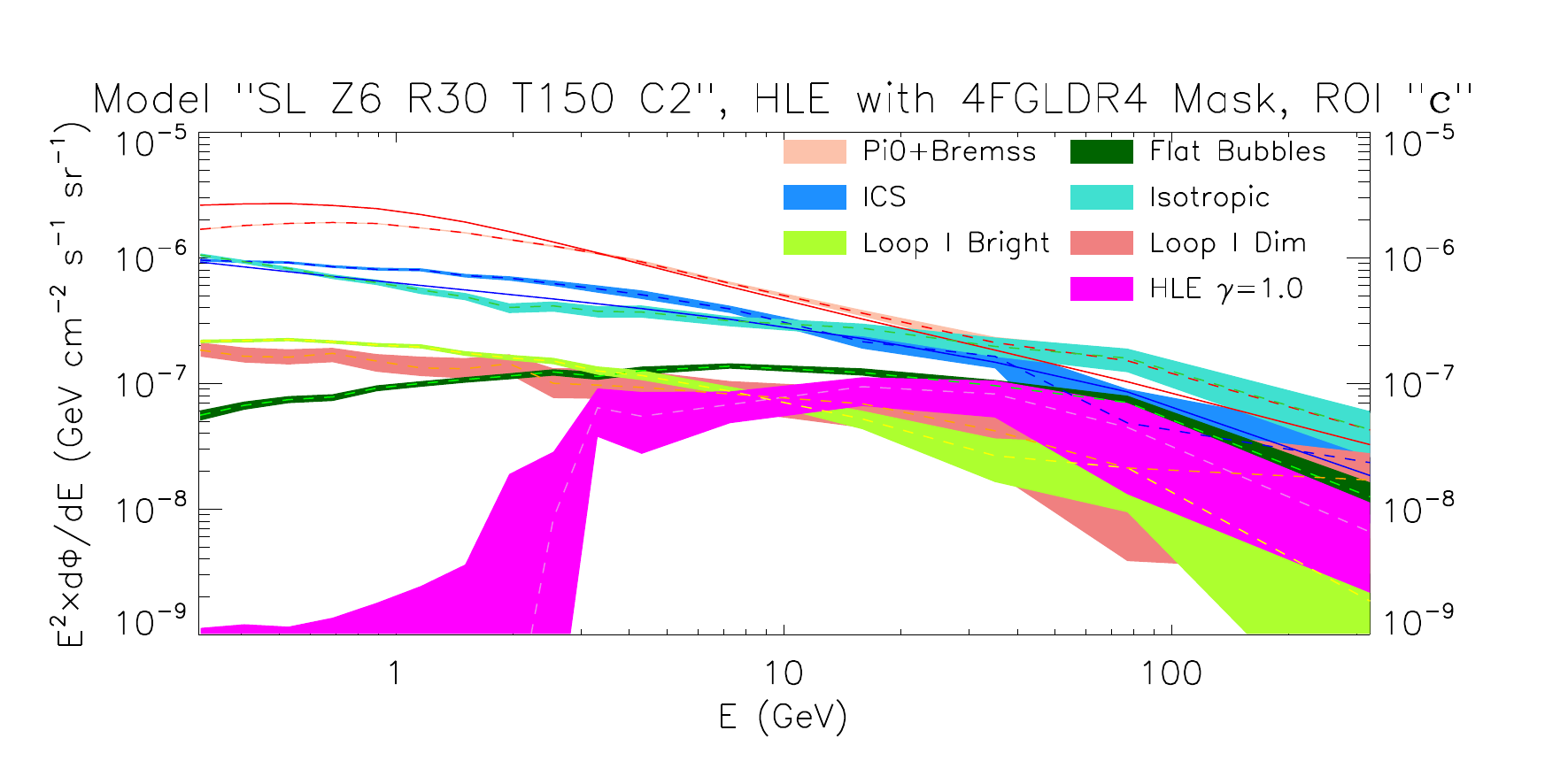}~
    \includegraphics[width=0.49\linewidth]{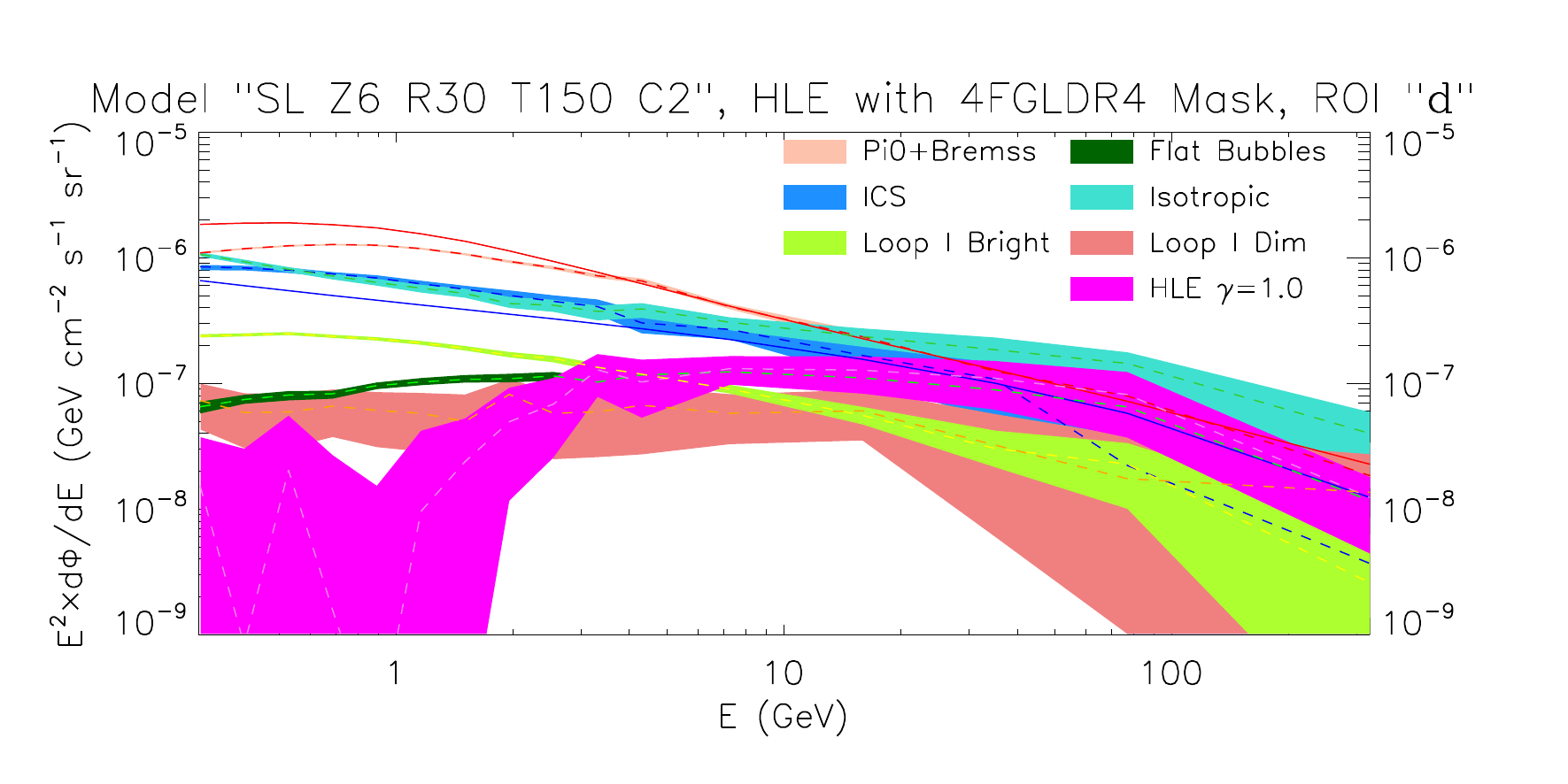}\\
    \vspace{-0.3cm}
    \includegraphics[width=0.49\linewidth]{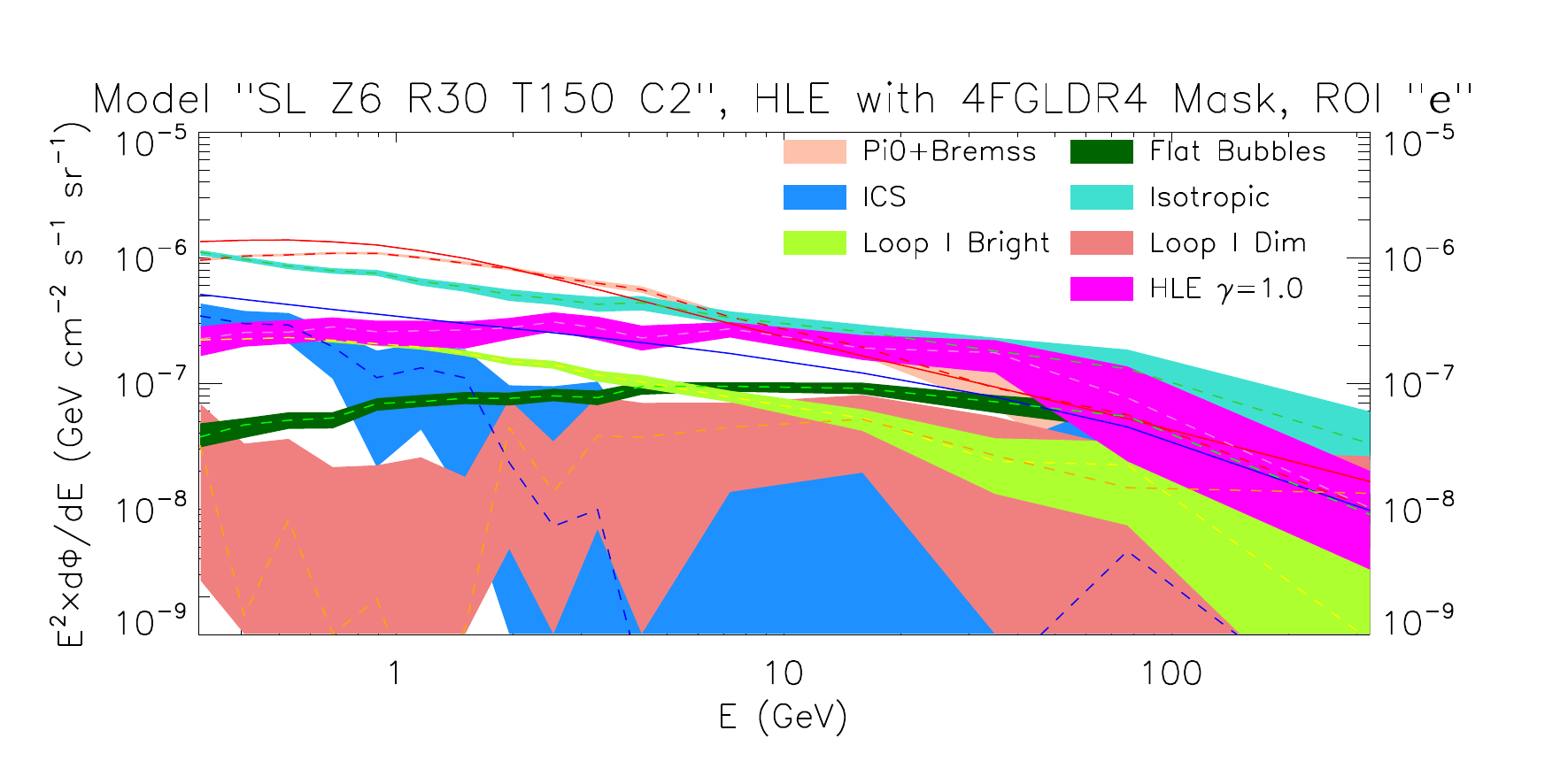}
    \vspace{-0.3cm}
    \caption{As in Fig.~\ref{fig:HLE_120x120_WITH_ResMaps}, the galactic diffuse emission spectra for the $120^{\circ} \times 120^{\circ}$ ROI with the 4FGLDR4 \textit{Fermi} point and extended sources and the five choices of galactic latitude ranges/ROIs (in the same order).
    We use the galactic diffuse emission and Loop I modeling as Ref.~\cite{Totani:2025fxx}. However, instead of the negative and positive residual templates of Ref.~\cite{Totani:2025fxx}, we use the Flat Bubbles from~\cite{Su:2010qj}. 
    In the top row and middle left panels the excess of gamma rays has an amplitude that changes between these ROIs roughly as expected for a dark-matter annihilation signal.
    However, the HLE template's flux as we increasingly exclude in our ROIs higher latitudes close to the galactic disk, does not decrease as expected for a dark matter annihilation signal; but instead absorbs part of the Bubbles and the ICS.}
    \label{fig:HLE_120x120_WO_ResMaps}
\end{figure*}

In Fig.~\ref{fig:HLE_120x120_WITH_ResMaps}, we show the associated gamma-ray
spectra from all eight diffuse emission components, for all energy bins, $0.275\leq E \leq 928$ GeV. In that fit, we use  the SL\,Z6\,R30\,T150\,C2 galactic diffuse emission model for  the Pi0+Bremss, and ICS templates prediction, and perform the fit only in the highest resolution $0.1^{\circ}\times0.1^{\circ}$ pixels. 
In the fit, we mask the 4FGLDR4 point and extended sources within each ROI, defined in Table~\ref{tab:roi}, by its longitude and latitude limits. 
However, in the various panels of Fig.~\ref{fig:HLE_120x120_WITH_ResMaps} (and also Figs.~\ref{fig:HLE_120x120_WO_ResMaps},~\ref{fig:SLZ6R30T150C2_spectra_6parameters},~\ref{fig:HLE_spectrum_FermiModels_ROIe}, and~\ref{fig:HLE_spectrum_AllModels_AltROIs}), for simplicity, 
we include the predicted flux from the modeled galactic diffuse emission and from the other diffuse templates (Loop I, Bubbles) originating from the masked regions that are centered at the point and extended sources and that are within each ROI.

We find that while the HLE is always present in the fit, its normalization is dramatically different across the various ROIs compared to what should be expected for a signal of dark matter annihilation that is the line-of-sight integral of $\rho(r)^{2}$, where $\rho(r)$ is given by~\eqref{NFW}. In fact, instead of the HLE decreasing in its amplitude as we exclude the lower latitudes, the dark matter annihilation template flux claimed to explain the HLE increases from ROI ``\textbf{a}'' to ROI ``\textbf{e}''. We find that, when comparing the HLE flux between the  ROIs that contain the lower latitudes (ROIs \textbf{a}, \textbf{b} and \textbf{c}), to those that exclude them (ROIs \textbf{d} and \textbf{e}), their relative amplitude is nowhere close to compatible with the relative amplitude for a dark matter annihilation signal. For comparison, a dark matter annihilation signal from a NFW profile with an inner slope of $\gamma=1.0$, in units of averaged flux per steradian at the ROI ``\textbf{a}'' should be as high as in ROI ``\textbf{b}'',\footnote{ROI ``\textbf{a}'' contains compared to ROI ``\textbf{b}'' both lower latitudes where the dark matter signal is more prominent, but also higher latitudes where dark matter would contribute very little. As a result to within a few \% their flux in per steradians is the same.} 1.5 times higher than ROI ``\textbf{c}'', 2.0 times higher than in ROI ``\textbf{d}'' and 2.6 times higher compared to ROI ``\textbf{e}''. 
This is in strong contradiction with the results from the fits. For instance, in ROI ``\textbf{e}'', the HLE flux between 2 and 10 GeV is roughly seven times higher than it should be, if it were a dark matter annihilation signal and we are to trust the results from ROIs \textbf{a} or \textbf{b}, where such a signal would be easier to detect. 
The discrepancy on the HLE flux between ROI \textbf{e} and ROIs \textbf{a} and \textbf{b} is even greater at lower energies. 
We also tested in our fits the case for a gNFW profile with $\gamma=1.25$. The tension between the results for ROIs \textbf{a} and \textbf{b} and ROIs \textbf{d} and \textbf{e}, is even more stark.   
We believe these discrepancies, happen in part because of the significant level of degeneracy between the ICS, the HLE and the Isotropic templates.

In addition, if one compares the results for ROI ``\textbf{d}'' that are for latitudes of $20^{\circ} \leq |b| \leq 60^{\circ}$, to those for ROI ``\textbf{e}'' that are for $30^{\circ} \leq |b| \leq 60^{\circ}$, one finds the ICS fitted flux to be much reduced compared to the \texttt{GALPROP} original prediction for that region. That discrepancy between the original ICS prediction and the fitted results is not at all present in the ROI ``\textbf{a}'', while in ROIs ``\textbf{b}'' and ``\textbf{c}'' the discrepancy appears at the higher energies. 
This reduced ICS flux (from the original prediction) shows up in the fit at the same ROIs where the HLE becomes brighter as does the isotropic flux. The isotropic flux is one component that we expect to be the same in the units presented across all ROIs.  
We also notice that while the positive residuals template has a roughly constant predicted flux between the five ROIs, the negative residuals template spectrum changes dramatically between ROIs, even when the fit results from these ROIs should be similar, e.g., between ROI ``\textbf{b}'' and ROI ``\textbf{c}''. Again, that suggests a lack of robustness introduced by having this negative residuals template. 

\subsection{Studying the \textit{Fermi} sky using positive-flux templates}

For the remainder of this section, we avoid the use of the negative and positive residual templates of Ref.~\cite{Totani:2025fxx}, and instead use the more conventional Flat Bubbles of \cite{Su:2010qj}, as the way to model the Bubbles' emission.\footnote{We revisit the modeling on the Bubbles in Section~\ref{sec:NewFermiBubbles}.}
We focus on understanding the properties of the HLE using the library of 100 alternative galactic diffuse emission models, eight alternative ROIs and also comparing our results to the alternative fitting procedures. We do that in order to see what conclusions we can robustly derive from all these tests. 
The questions we ask for the remaining analysis of this section are: (i) is the HLE a robust signal, (ii) if it is robust, is it a dark matter signal or, (iii) is its presence informing us of the need for a change/addition to the regular Bubbles template.  

\begin{figure*}[t]
    \centering
    \includegraphics[width=0.49\linewidth]{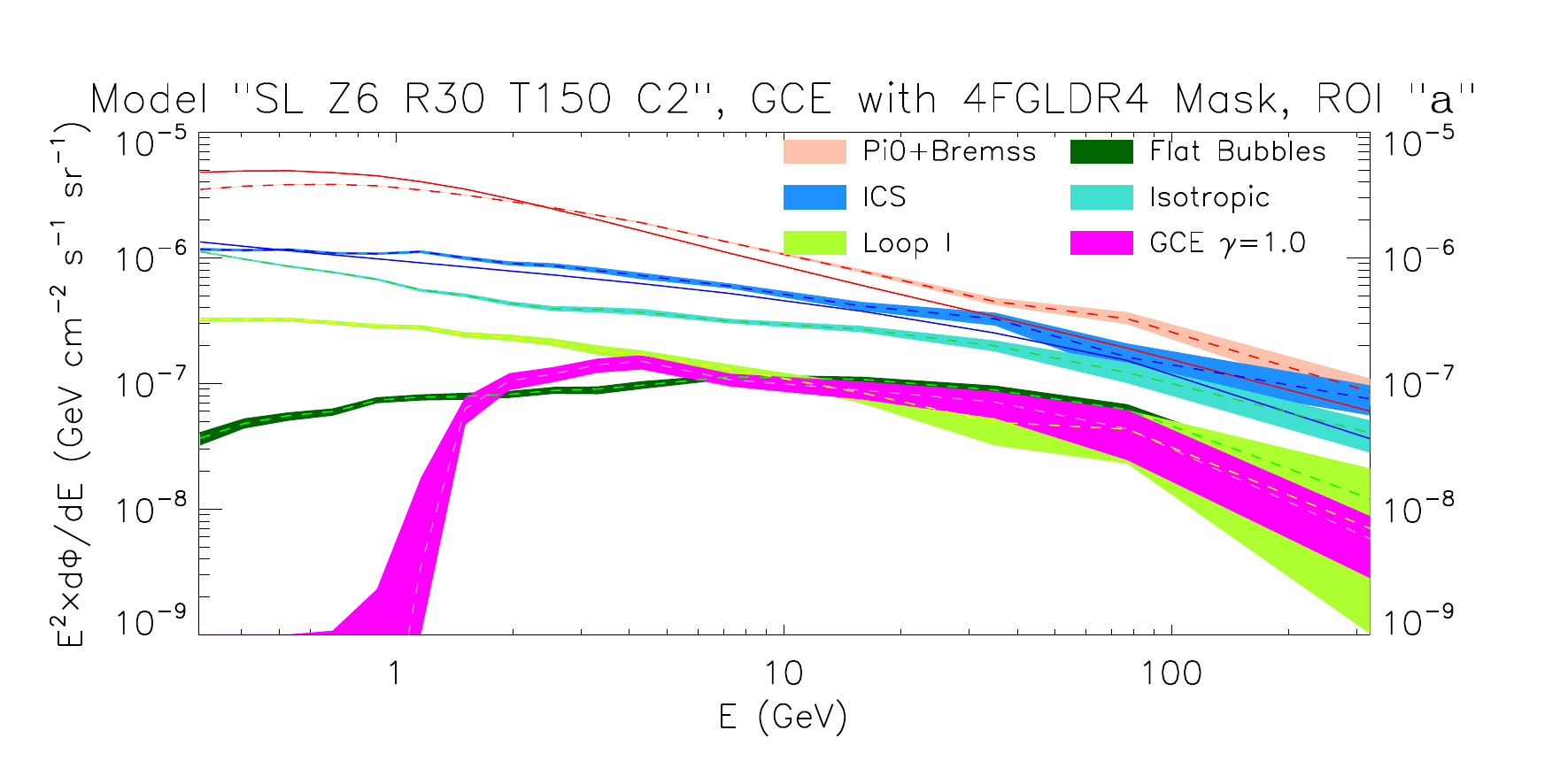}~
    \includegraphics[width=0.49\linewidth]{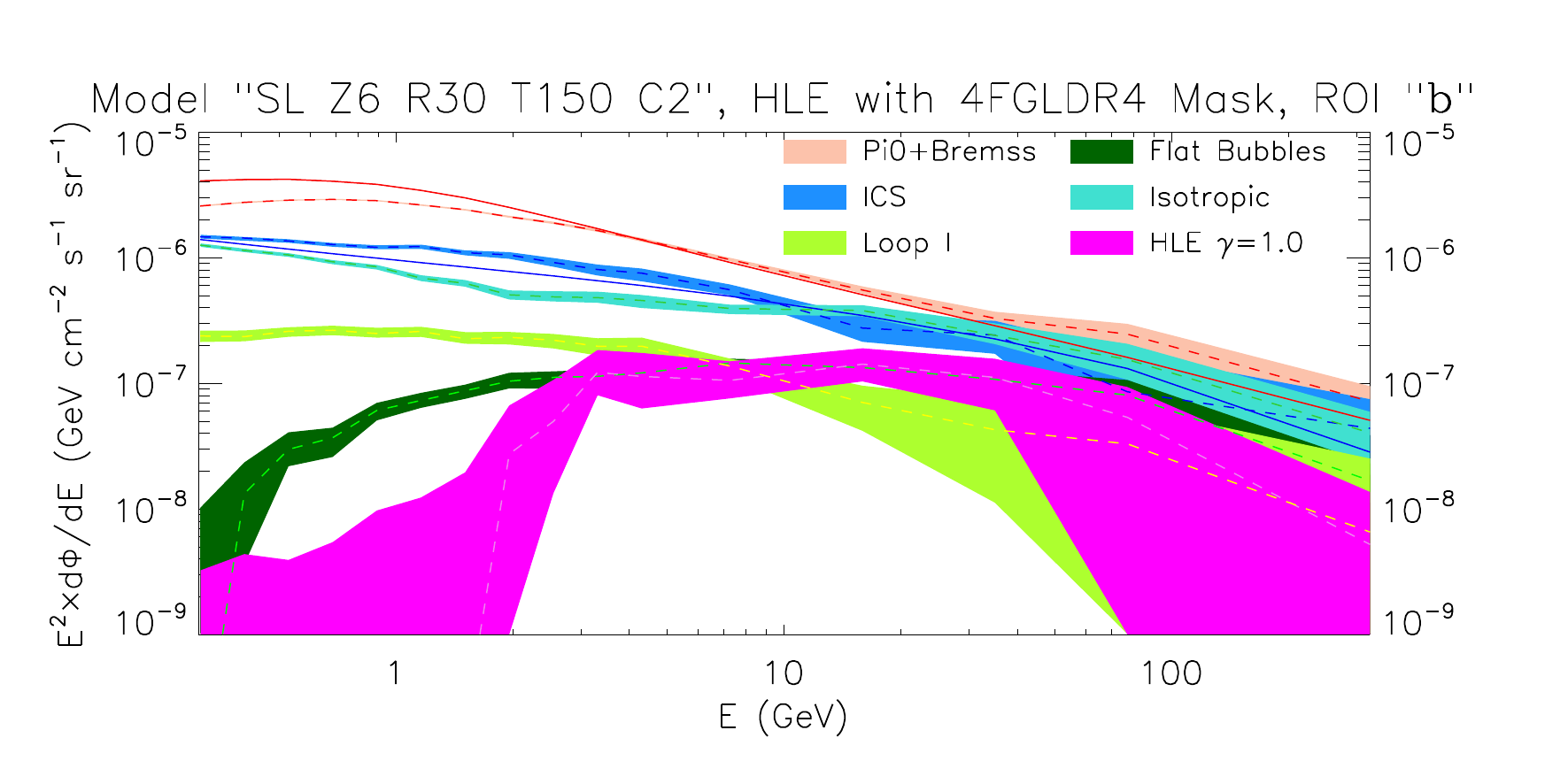}\\
    \vspace{-0.3cm}
    \includegraphics[width=0.49\linewidth]{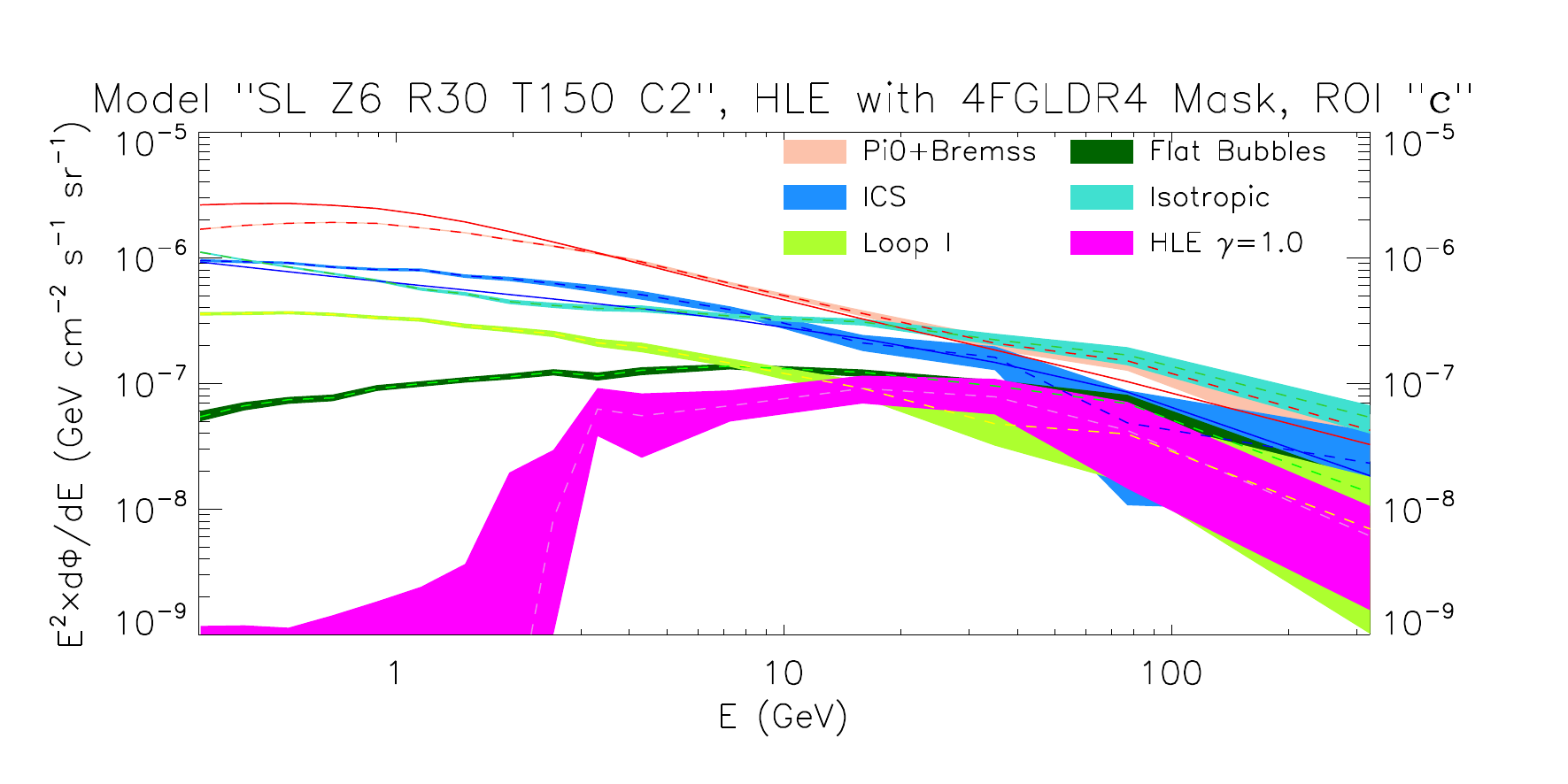}~
    \includegraphics[width=0.49\linewidth]{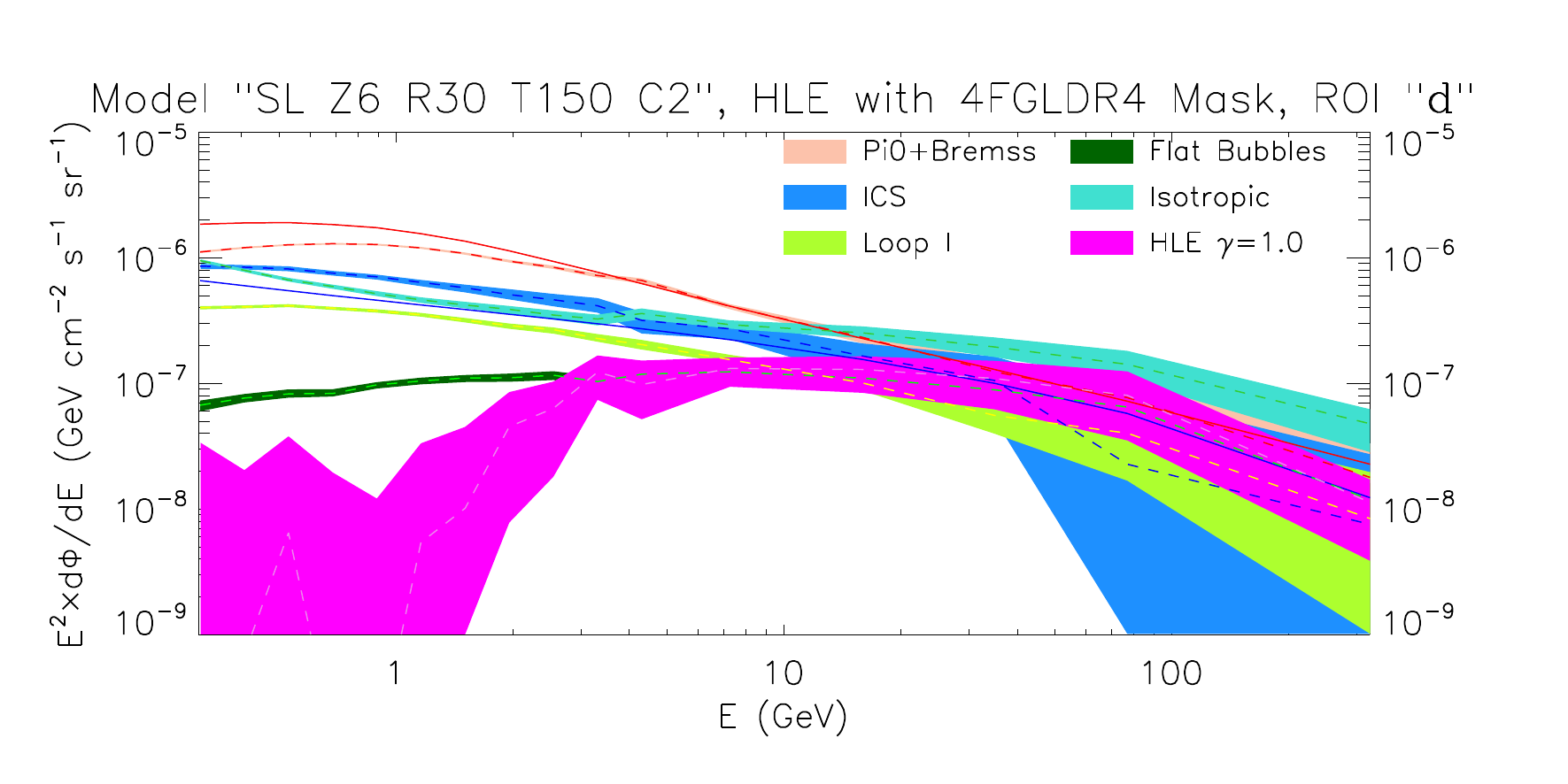}\\
    \vspace{-0.3cm}
    \includegraphics[width=0.49\linewidth]{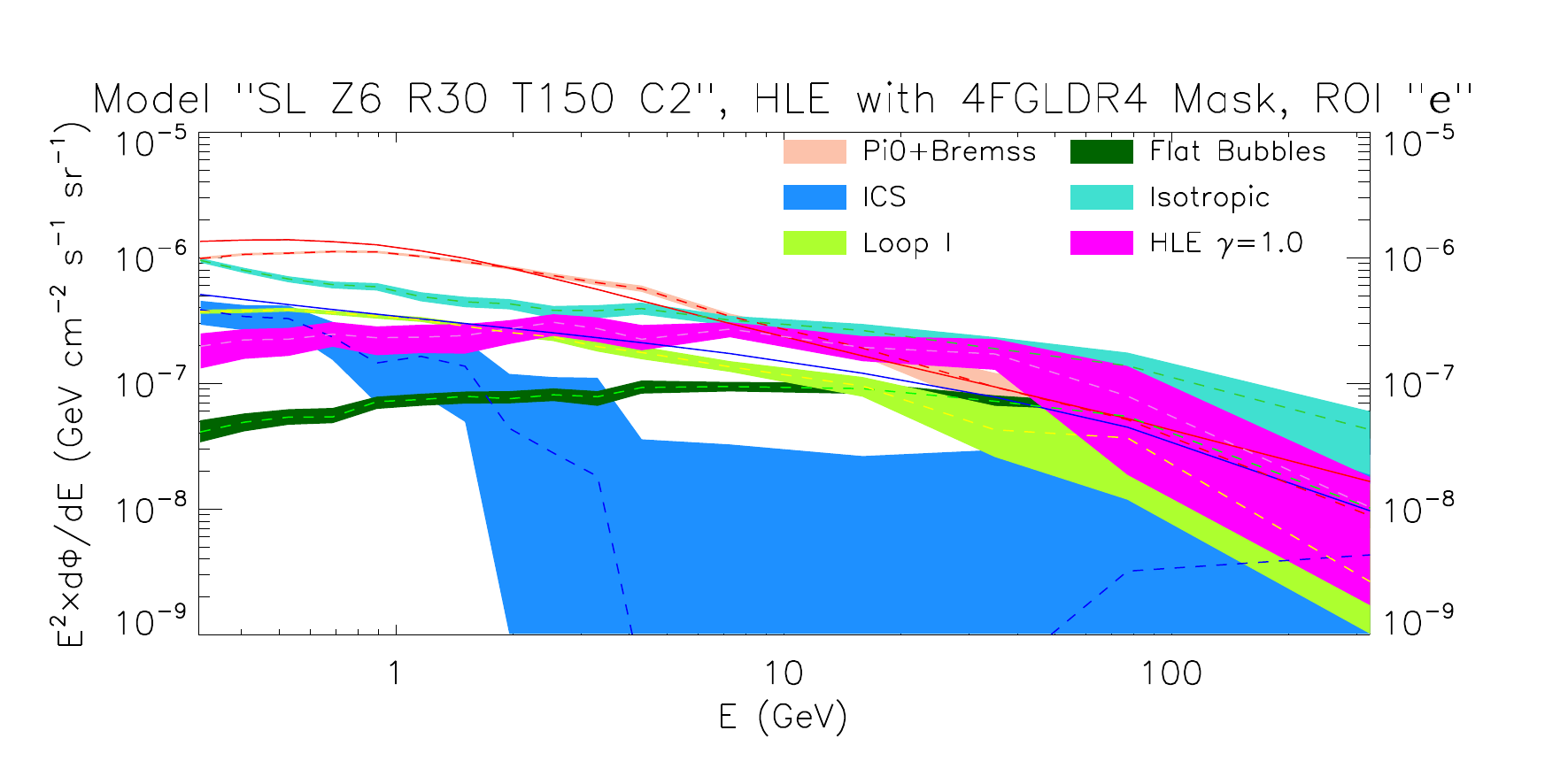}
    \vspace{-0.3cm}
    \caption{Using the reference galactic diffuse emission model of Ref.~\cite{Totani:2025fxx}, but with a combined \textit{single} template for Loop I, and the \textit{Fermi} Flat Bubbles of Ref.~\cite{Su:2010qj}.
    We show the best-fit results for the five alternative ROIs within the $120^{\circ} \times 120^{\circ}$ region, in the same order as in Figs.~\ref{fig:HLE_120x120_WITH_ResMaps} and~\ref{fig:HLE_120x120_WO_ResMaps}. We do not include the residual templates of Ref.~\cite{Totani:2025fxx}. In all panels we mask the 4FGLDR4 \textit{Fermi} point and extended sources. The HLE template's associated flux as we exclude lower latitudes in our ROIs (i.e. $|b|\leq 20^{\circ}$ and $|b|\leq 30^{\circ}$) does not decrease as one would expect for a dark matter annihilation signal; in fact at the higher-latitude regions its flux absorbs part of the Bubbles and the ICS (see text for details).}
    \label{fig:SLZ6R30T150C2_spectra_6parameters}
\end{figure*}

In Fig.~\ref{fig:HLE_120x120_WO_ResMaps}, we show as was done in Fig.~\ref{fig:HLE_120x120_WITH_ResMaps}, the gamma-ray spectra
from all seven diffuse emission components (see also Table.~\ref{tab:fit_schemes}) for $0.275\leq E \leq 928$ GeV. 
For the Pi0+Bremss and ICS templates prediction we use the SL\,Z6\,R30\,T150\,C2 model, masking the 4FGLDR4 sources and fitting the $0.1^{\circ}\times0.1^{\circ}$ pixelized sky in the ROIs \textbf{a}--\textbf{e}. 

In these cases, we find that the dark matter annihilation-like template's flux normalization (denoted as GCE or HLE depending on the ROI), is roughly compatible between ROIs \textbf{a}, \textbf{b} and \textbf{c} with what would be expected for such a signal. 
However, still in the ROIs \textbf{d} and \textbf{e}, it becomes significantly brighter, by a factor of 2 and 5 respectively, contradicting the hypothesis that the HLE is a dark matter annihilation signal. 
We also notice that with the exception of ROI \textbf{e}, the ICS \texttt{GALPROP} flux prediction before the fit and the ICS flux after the fit are in good agreement, while also the Pi0+Bremss flux before and after the fit are in good agreement in all ROIs. The flux absorbed by the Isotropic template in the fit is also roughly of the same normalization and spectrum among all panels as is expected, an improvement to the results of Fig.~\ref{fig:HLE_120x120_WITH_ResMaps}. 
Also, the relative ratio of the Loop I Bright: Loop I Dim is closer to 1:1 in ROIs \textbf{a}-\textbf{c}.

In ROIs \textbf{d} and \textbf{e}, the degeneracies between the HLE and the galactic ICS and possibly the \textit{Fermi} Bubbles and the Loop I templates become more evident.  
\emph{These results suggest that above latitudes of $|b| \geq 20^{\circ}$, the HLE is absorbing/fitting a new component that is not a real dark matter annihilation signal, but some other emission connected to the \textit{Fermi} Bubbles, the galactic ICS and Loop I.} 

\begin{figure}[t]
    \centering
    \includegraphics[width=0.98\linewidth]{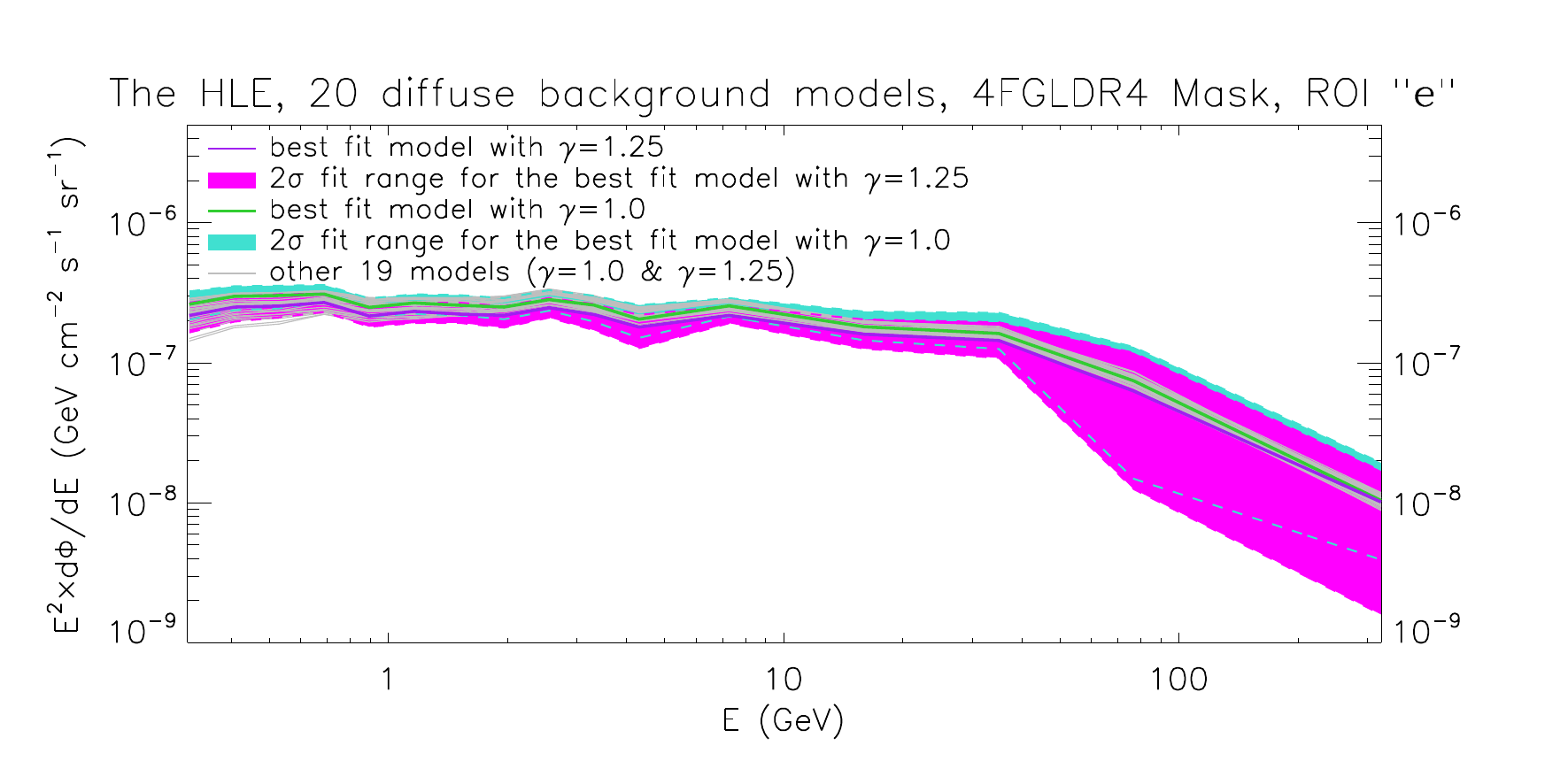}
    \vspace{-0.3cm}
    \caption{The HLE spectrum in the ROI ``\textbf{e}'', masking the 4FGLDR4 \textit{Fermi} point and extended sources. We use the 20 galactic diffuse emission models from Ref.~\cite{Fermi-LAT:2012edv}. We show the HLE spectrum from the galactic diffuse emission model that gives the best fit for the $\gamma = 1.25$ (purple line) and the $\gamma = 1.0$ (green line) dark matter annihilation template. For that diffuse emission model in the magenta (turquoise) band we also give the 2-$\sigma$ fit range for the HLE spectrum. The HLE spectral results for the remaining 19 models used from Ref.~\cite{Fermi-LAT:2012edv}, are given by the gray lines.}
    \label{fig:HLE_spectrum_FermiModels_ROIe}
\end{figure}

\begin{figure}[t]
    \centering
    \includegraphics[width=0.98\linewidth]{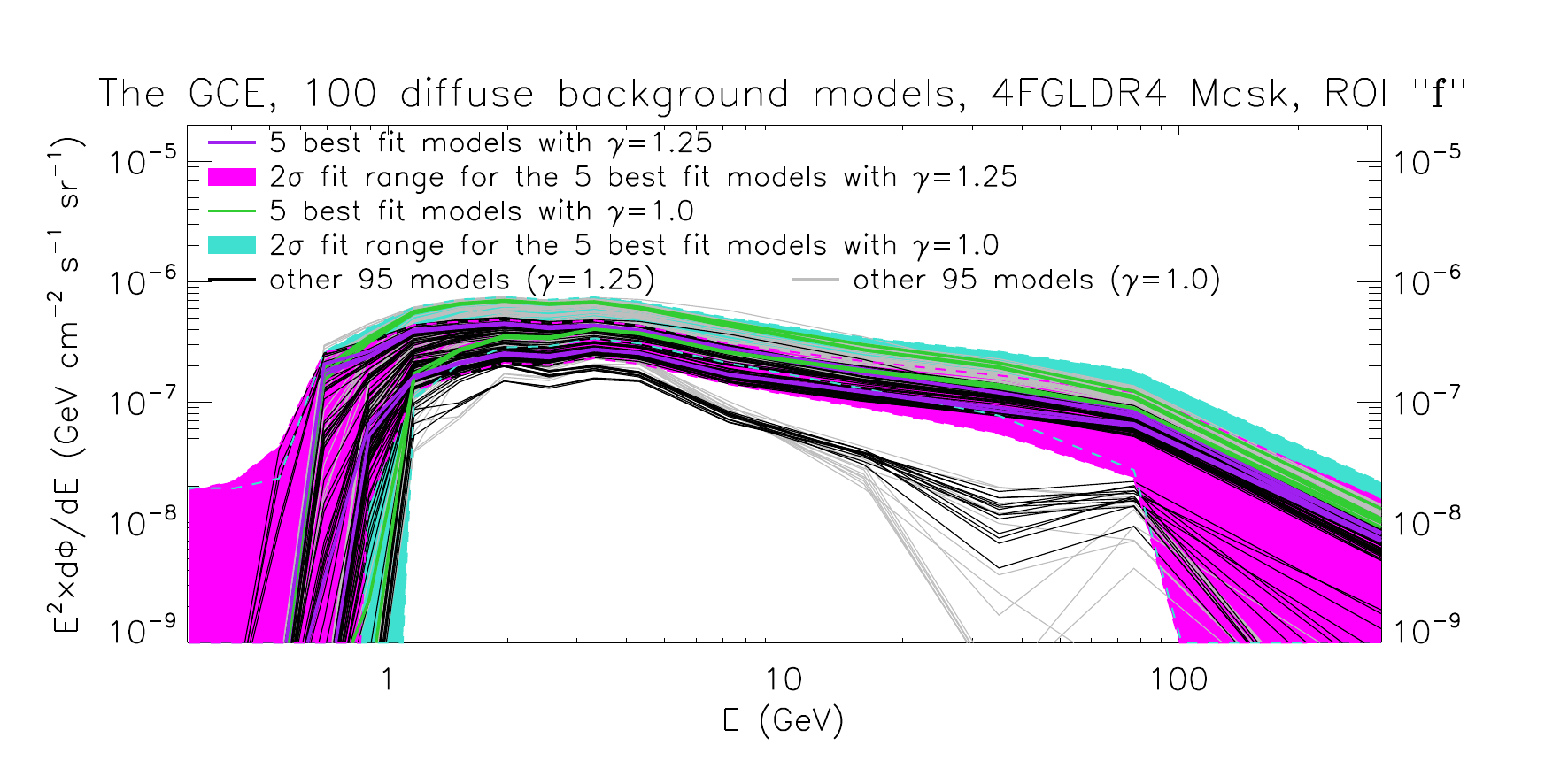}\\
    \vspace{-0.3cm}
    \includegraphics[width=0.98\linewidth]{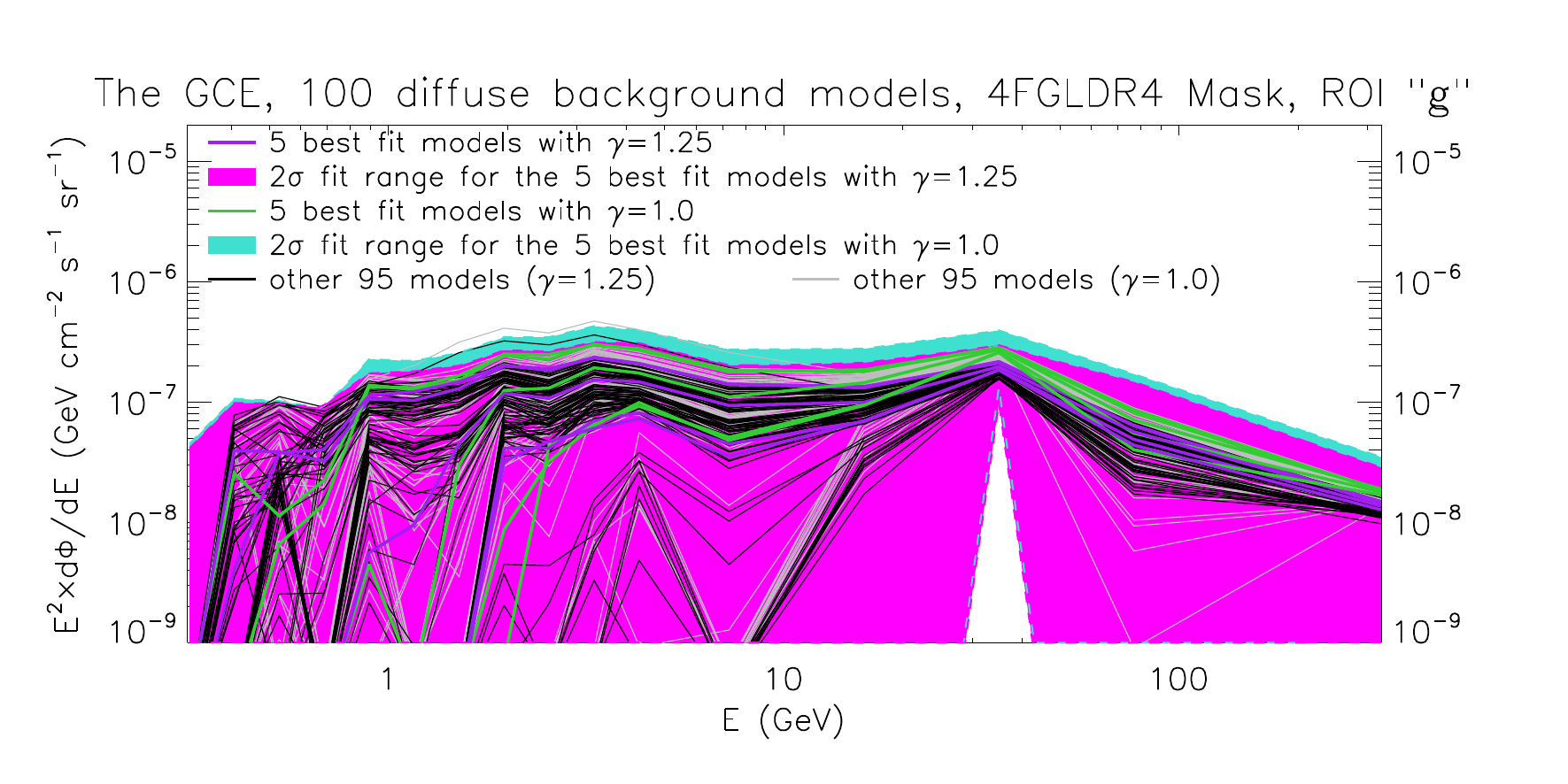}\\
    \vspace{-0.3cm}
    \includegraphics[width=0.98\linewidth]{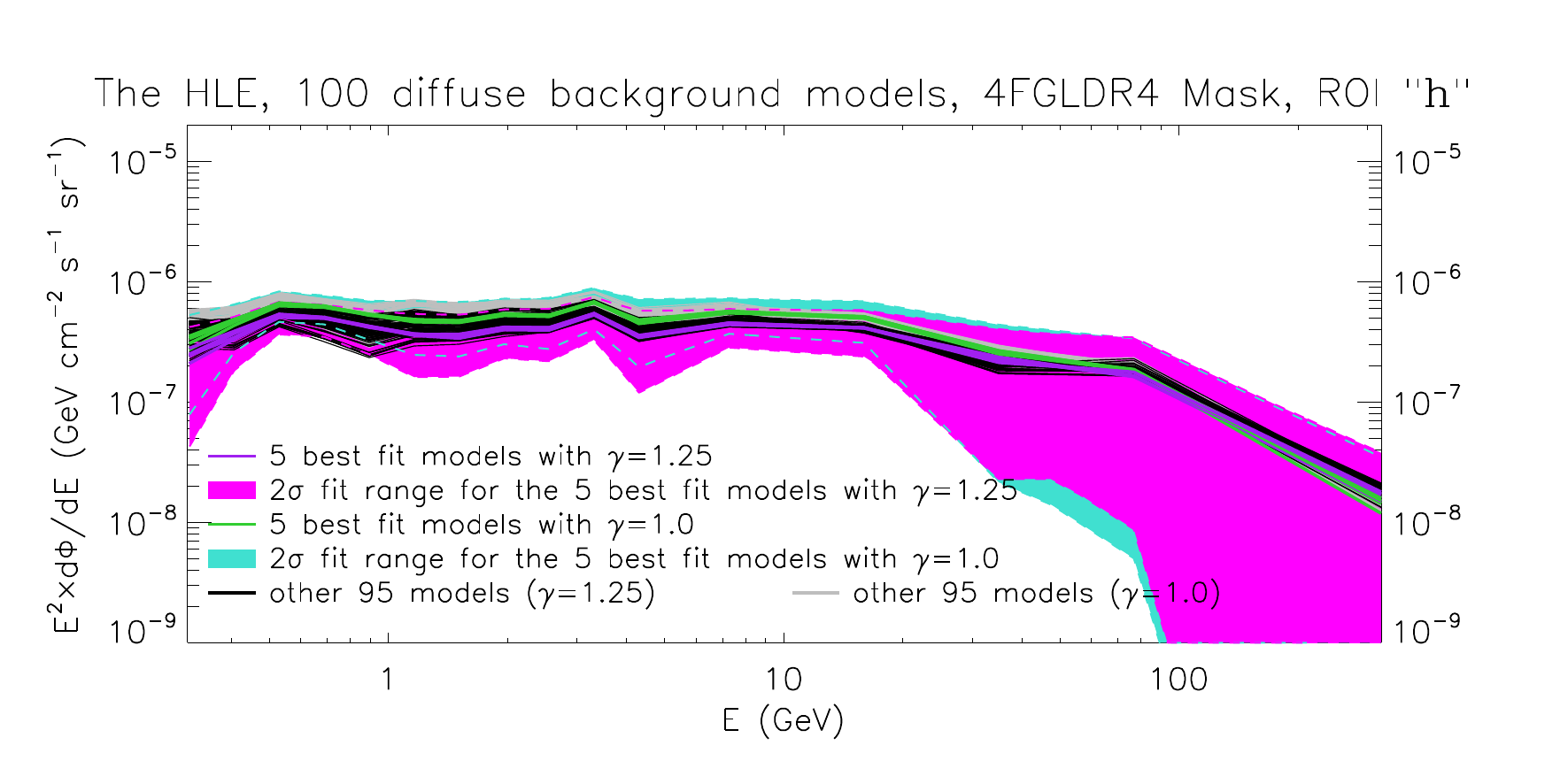}
    \vspace{-0.3cm}
    \caption{In the top panel, we give the GCE spectral results after fitting from the Inner Galaxy, i.e. the $60^{\circ} \times 60^{\circ}$ window, only the $2^{\circ} \leq |b| \leq 10^{\circ}$ latitudes (ROI ``\textbf{f}''). In the middle panel, we give the same results when fitting only the $10^{\circ} \leq |b| \leq 20^{\circ}$ latitudes of the Inner Galaxy (ROI ``\textbf{g}''). We consider this to be the GCE (and not the HLE). 
    In the bottom panel we give the equivalent results for the HLE spectrum fitting only the $20^{\circ} \leq |b| \leq 30^{\circ}$ latitudes (ROI ``\textbf{h}''). 
    We also mask the 4FGLDR4 \textit{Fermi} sources.
    In all panels, we use the 80 galactic diffuse emission models from Ref.~\cite{Cholis:2021rpp} and the 20 diffuse models from Ref.~\cite{Fermi-LAT:2012edv}. We fit the two Loop I templates separately. We show results for both the $\gamma = 1.25$ (black lines) and the $\gamma = 1.0$ (gray lines) dark matter annihilation templates. The GCE/HLE spectra from the five galactic diffuse emission models that give the best-fit results for the $\gamma = 1.25$ ($\gamma = 1.0$) templates are given by purple (green) lines. The combined 2-$\sigma$ fit ranges for the GCE/HLE templates for these five best-fit galactic diffuse emission models are given by the magenta (turquoise) bands (see text for further details).}
    \label{fig:HLE_spectrum_AllModels_AltROIs}
\end{figure}

To further test that hypothesis in Fig.~\ref{fig:SLZ6R30T150C2_spectra_6parameters}, we first simplify the modeling of the emission from Loop I by combining the two templates into one, and in Figs.~\ref{fig:HLE_spectrum_FermiModels_ROIe} and~\ref{fig:HLE_spectrum_AllModels_AltROIs}, we include the alternative galactic diffuse emission models to avoid biasing our conclusions on the HLE by simply using one galactic diffuse emission model.

In Fig.~\ref{fig:SLZ6R30T150C2_spectra_6parameters}, we combine the two Loop I templates into one, taking the averaged gamma-ray flux within ROI \textbf{a}, to be 3:1 for the Loop I Bright: Loop I Dim shells. Other than that, we repeat the same analysis in the same ROIs as presented in Figs.~\ref{fig:HLE_120x120_WITH_ResMaps}
and~\ref{fig:HLE_120x120_WO_ResMaps}, using again the SL\,Z6\,R30\,T150\,C2 galactic diffuse emission model. We thus have six diffuse emission components (see Table.~\ref{tab:fit_schemes}). Accordingly, we replace the penalty terms $\chi^{2}_{{\rm bright},j}$ and $\chi^{2}_{{\rm dim},j}$ with a single penalty term $\chi^{2}_{{\rm Loop I},j}$, adopted from Ref.~\cite{Fermi-LAT:2014sfa}.

Our results are similar to the seven-parameter fit. The HLE's normalization and spectrum between ROIs \textbf{a} to \textbf{c} follow roughly what is expected for a dark matter annihilation signal. However, in ROIs \textbf{d} and \textbf{e} the HLE template absorbs an additional emission giving it a flux up to a factor of 5 higher than expected for a dark matter annihilation signal with $\gamma = 1.0$. 
The flux associated with the Isotropic template is self-consistent in all ROIs, and the ICS template's fitted spectrum is consistent with the \texttt{GALPROP} prediction in ROIs \textbf{a} to \textbf{d}, being suppressed due to degeneracies with the other templates only in ROI \textbf{e}.

We conclude that the exact modeling of the Loop I, i.e., being treated as two separate templates one for each shell, or a combined template has only a small effect on the results. For this reason, in the results of Figs.~\ref{fig:HLE_spectrum_FermiModels_ROIe} and~\ref{fig:HLE_spectrum_AllModels_AltROIs}, where we test alternative galactic diffuse emission models we allow for two Loop I templates in the fit.

In performing fits with the six to nine alternative free normalizations (which we refer to as different fitting schemes), we also evaluate the log-likelihood for each alternative assumption. Table~\ref{tab:results}, presents the results for the six- to eight-parameter fits, with the 4FGLDR4 point and extended gamma-ray sources masked. We remind the reader that for these fits the maps have a pixel resolution of $0.1^{\circ} \times 0.1^{\circ}$. The total log-likelihood is obtained by combining all 16 energy bins spanning the energy range from 0.275 to 928 GeV.
We emphasize that changing the ROI can substantially alter the number of pixels included in the fit. Consequently, the absolute log-likelihood values cannot be directly compared across different ROIs. Instead, within each ROI, we compare the log-likelihoods among different fitting schemes and study how these relative preferences vary with the ROI. This allows us to draw conclusions about the underlying properties of the HLE/GCE.

For each ROI, we evaluate the differences in log-likelihood (in $-2\Delta \! \ln (\mathcal L)$), based on the hypothesis that there is no additional HLE/GCE template.
This allows us to have a quantitative measure of the significance for the HLE component. 
Furthermore, in each case we test both the NFW standard inner slope of $\gamma=1.0$ and the cuspier profile of $\gamma=1.25$. 
We set as a reference point for each ROI, the log-likelihood that we get for our simplest six-parameter fit without a dark matter emission component. This gives a quantitative measure on the statistical importance of adding modeling complexity in the fit. We caution that a statistical improvement does not cancel the fact that when adding positive/negative residual maps one introduces by construction a huge improvement on the gamma-ray map at a given energy range; which as we have explained earlier produces other artifacts. 

In all cases, adding the HLE/GCE template improves the fit by a pretty significant $2\Delta \! \ln (\mathcal L)$.
We find in ROI \textbf{a}, that there is a significant preference for the $\gamma=1.25$ gNFW profile for the six- and seven-parameter fitting scheme, while for the eight-parameter fit we get similar likelihood results for $\gamma=1.0$ and $\gamma=1.25$. This is in agreement with the existing literature results on the GCE's morphology. In ROI \textbf{a}, it is the GCE that is being probed by the dark matter annihilation template. 
However, for ROIs \textbf{b} to \textbf{e}, there is a slight preference for the $\gamma=1.0$ inner slope for the NFW profile, becoming typically most prominent in ROI \textbf{e}. 
Also, in all ROIs the added complexity in the fitting schemes improves the fit for either the hypothesis of no additional dark matter annihilation-like template and for the case that such a template assumes a gNFW dark matter profile with $\gamma=1.0$ or $\gamma=1.25$. 

\begin{table*}[t]
\caption{Comparison of models for the analysis of the GCE/HLE. In all cases we mask the 4FGLDR4 point and extended sources. All these results are based only on the SL\,Z6\,R30\,T150\,C2 galactic diffuse emission model. }
\begin{center}
\begin{tabular}{c c l}
\hline
ROI \& \# of Templ. & GCE/HLE Templ. & $-2\Delta \! \ln (\mathcal L)$ (w.r.t. 6 d.o.f.) \\ \hline \hline
\textbf{a} with 6 d.o.f. & No Excess & 0 \\
\textbf{a} with 6 d.o.f. & $\gamma = 1.0$ & $-1642$\\
\textbf{a} with 6 d.o.f. & $\gamma = 1.25$ & $-2274$\\
\textbf{a} with 7 d.o.f. & No Excess & $-1794$ \\
\textbf{a} with 7 d.o.f. & $\gamma = 1.0$ & $-3434$\\
\textbf{a} with 7 d.o.f. & $\gamma = 1.25$ & $-4028$\\
\textbf{a} with 8 d.o.f. & No Excess & $-5.466\times 10^{4}$ \\
\textbf{a} with 8 d.o.f. & $\gamma = 1.0$ & $-7.043\times 10^{4}$\\
\textbf{a} with 8 d.o.f. & $\gamma = 1.25$ & $-7.035\times 10^{4}$\\
\hline
\textbf{b} with 6 d.o.f. & No Excess & 0 \\
\textbf{b} with 6 d.o.f. & $\gamma = 1.0$ & $-161$\\
\textbf{b} with 6 d.o.f. & $\gamma = 1.25$ & $-136$\\
\textbf{b} with 7 d.o.f. & No Excess & $-2677$ \\
\textbf{b} with 7 d.o.f. & $\gamma = 1.0$ & $-2839$\\
\textbf{b} with 7 d.o.f. & $\gamma = 1.25$ & $-2816$ \\
\textbf{b} with 8 d.o.f. & No Excess & $-1.531\times10^{4}$ \\
\textbf{b} with 8 d.o.f. & $\gamma = 1.0$ & $-1.644\times10^{4}$\\
\textbf{b} with 8 d.o.f. & $\gamma = 1.25$ & $-1.629\times10^{4}$ \\
\hline 
\textbf{c} with 6 d.o.f. & No Excess & 0 \\
\textbf{c} with 6 d.o.f. & $\gamma = 1.0$ & $-214$\\
\textbf{c} with 6 d.o.f. & $\gamma = 1.25$ & $-162$\\
\textbf{c} with 7 d.o.f. & No Excess & $-48$ \\
\textbf{c} with 7 d.o.f. & $\gamma = 1.0$ & $-264$\\
\textbf{c} with 7 d.o.f. & $\gamma = 1.25$ & $-212$\\
\textbf{c} with 8 d.o.f. & No Excess & $-1.749\times10^{4}$ \\
\textbf{c} with 8 d.o.f. & $\gamma = 1.0$ & $-2.301\times 10^{4}$\\
\textbf{c} with 8 d.o.f. & $\gamma = 1.25$ & $-2.278\times 10^{4}$\\
\hline
\textbf{d} with 6 d.o.f. & No Excess & 0 \\
\textbf{d} with 6 d.o.f. & $\gamma = 1.0$ & $-210$\\
\textbf{d} with 6 d.o.f. & $\gamma = 1.25$ & $-174$ \\
\textbf{d} with 7 d.o.f. & No Excess & $-369$ \\
\textbf{d} with 7 d.o.f. & $\gamma = 1.0$ & $-591$\\
\textbf{d} with 7 d.o.f. & $\gamma = 1.25$ & $-549$ \\
\textbf{d} with 8 d.o.f. & No Excess & $-8661$ \\
\textbf{d} with 8 d.o.f. & $\gamma = 1.0$ & $-9926$\\
\textbf{d} with 8 d.o.f. & $\gamma = 1.25$ & $-9796$\\
\hline
\textbf{e} with 6 d.o.f. & No Excess & 0 \\
\textbf{e} with 6 d.o.f. & $\gamma = 1.0$ & $-951$\\
\textbf{e} with 6 d.o.f. & $\gamma = 1.25$ & $-888$\\
\textbf{e} with 7 d.o.f. & No Excess & $-535$ \\
\textbf{e} with 7 d.o.f. & $\gamma = 1.0$ & $-1648$\\
\textbf{e} with 7 d.o.f. & $\gamma = 1.25$ & $-1546$\\
\textbf{e} with 8 d.o.f. & No Excess & $-5004$ \\
\textbf{e} with 8 d.o.f. & $\gamma = 1.0$ & $-6285$\\
\textbf{e} with 8 d.o.f. & $\gamma = 1.25$ & $-6223$\\
\hline
\end{tabular}
\end{center}
\label{tab:results}
\end{table*}%

In performing the nine-parameter fit, we use the Totani mask instead of the 4FGLDR4 mask. This means the 4FGLDR4 point sources and some of the extended gamma-ray sources are not masked (see~\figref{mask_comparison}). 
Thus, we cannot directly compare how much of an improvement in the log-likelihood there is by adding the ninth fitting degree of freedom per energy bin, associated with the normalization of the 4FGLDR4 sources.
However, we do find as with the results from the other fitting schemes in the ROI \textbf{c}, that there is a $2 \Delta \! \ln \mathcal L =107$ in favor of the $\gamma = 1.0$ vs. the $\gamma =1.25$ gNFW profile in the nine-parameter fit, when using the Totani mask in the $0.1^{\circ} \times 0.1^{\circ}$ pixelized maps. 

In Fig.~\ref{fig:HLE_spectrum_FermiModels_ROIe}, using our seven-parameter fitting scheme, we present the HLE spectrum from ROI \textbf{e}, using the 20 galactic diffuse emission models from \cite{Fermi-LAT:2012edv}. We provide information on the physical parameters of these 20 models in Appendix~\ref{app:FermiModels}. 
We test both the case for dark matter annihilation templates for the HLE from a gNFW profile with an inner slope of $\gamma=1.0$ and $\gamma=1.25$. We find within a factor of less than 2 the same best-fit spectrum and flux normalization for the HLE across $0.275\leq E \leq 928$ GeV, from all 20 diffuse emission models. We also show in the magenta and turquoise bands the 2-$\sigma$ fit range for the HLE spectrum coming from the galactic diffuse emission model that gives the best-fit result in ROI \textbf{e}. That model is not the SL\,Z6\,R30\,T150\,C2 used in \cite{Totani:2025fxx}, but instead model SL\,Z6\,R20\,T100000\,C5 described in Appendix~\ref{app:FermiModels}, not previously tested in the context of the HLE. We find that even for that galactic diffuse emission model, as well as for all 20 tested models there is a degeneracy between the ICS and the HLE components in ROI \textbf{e}.  

The models from Ref.~\cite{Cholis:2021rpp}, test a much more extended range of modeling assumptions about the conditions of the Inner Galaxy, in order to test any claims for a dark matter-like signal in that region with respect to the background/astrophysical modeling uncertainties.
They can also be used to further test the GCE/HLE claim, with the limitation however, that they cover only the $|\ell| \leq 30^{\circ}$ and $|b| \leq 30^{\circ}$ window.  Thus, we focus on analyzing ROIs ``\textbf{f}'', ``\textbf{g}'' and ``\textbf{h}'' (see Table~\ref{tab:roi}). 
In Fig.~\ref{fig:HLE_spectrum_AllModels_AltROIs}, we show the results for the GCE spectrum from the fit of the regions \textbf{f} and \textbf{g} and the HLE spectrum from the fit of ROI \textbf{h}, compared to the entire combination of 100 models from Refs.~\cite{Cholis:2021rpp} and~\cite{Fermi-LAT:2012edv}. 
We test both the case for a gNFW $\gamma = 1.25$ and $\gamma = 1.0$.
A value of $\gamma = 1.25$ describes better the GCE morphology that is picked up in ROIs \textbf{f} and \textbf{g}, while for ROI \textbf{h} the HLE fit still shows a slight preference for $\gamma = 1.0$. 
In Fig.~\ref{fig:HLE_spectrum_AllModels_AltROIs}, we also show in purple and green lines and magenta and turquoise bands, the best fit and 2-$\sigma$ fit ranges for the GCE/HLE fluxes, from the five galactic diffuse emission models that provide the best log-likelihood fits. These best-fit galactic emission models for each of the three ROIs come from both Refs.~\cite{Cholis:2021rpp} and~\cite{Fermi-LAT:2012edv}. 

The fact that the GCE and the HLE are two distinct emission components, is also shown by the difference in their spectra and predominantly their relative normalizations. Much like our previously discussed results for Figs.~\ref{fig:HLE_120x120_WITH_ResMaps},~\ref{fig:HLE_120x120_WO_ResMaps}, and~\ref{fig:SLZ6R30T150C2_spectra_6parameters}, if in ROIs \textbf{f} to \textbf{h} there was one dark matter annihilation-like signal, then for an inner slope of $\gamma=1.0$ ($\gamma=1.25$), in units of averaged flux per steradian, this flux at ROI \textbf{f} should be 1.6 (2.0) times higher than in ROI \textbf{g}, and 2.3 (3.2) times higher compared to ROI \textbf{h}.
The results in Fig.~\ref{fig:HLE_spectrum_AllModels_AltROIs}, from ROIs \textbf{f} and \textbf{g}, show that the relative flux normalization for the GCE scales as is expected for a dark matter annihilation-like signal. However, the per steradian averaged flux from ROI \textbf{h}, is as high as that from ROI \textbf{f} (and for several models even higher), making the normalization of the ROI \textbf{h} a factor of $\simeq 3$ higher than it should if we are to trust the results from ROIs \textbf{f} and \textbf{g}.    

\begin{figure}[ht]
    \includegraphics[width=0.98\linewidth]{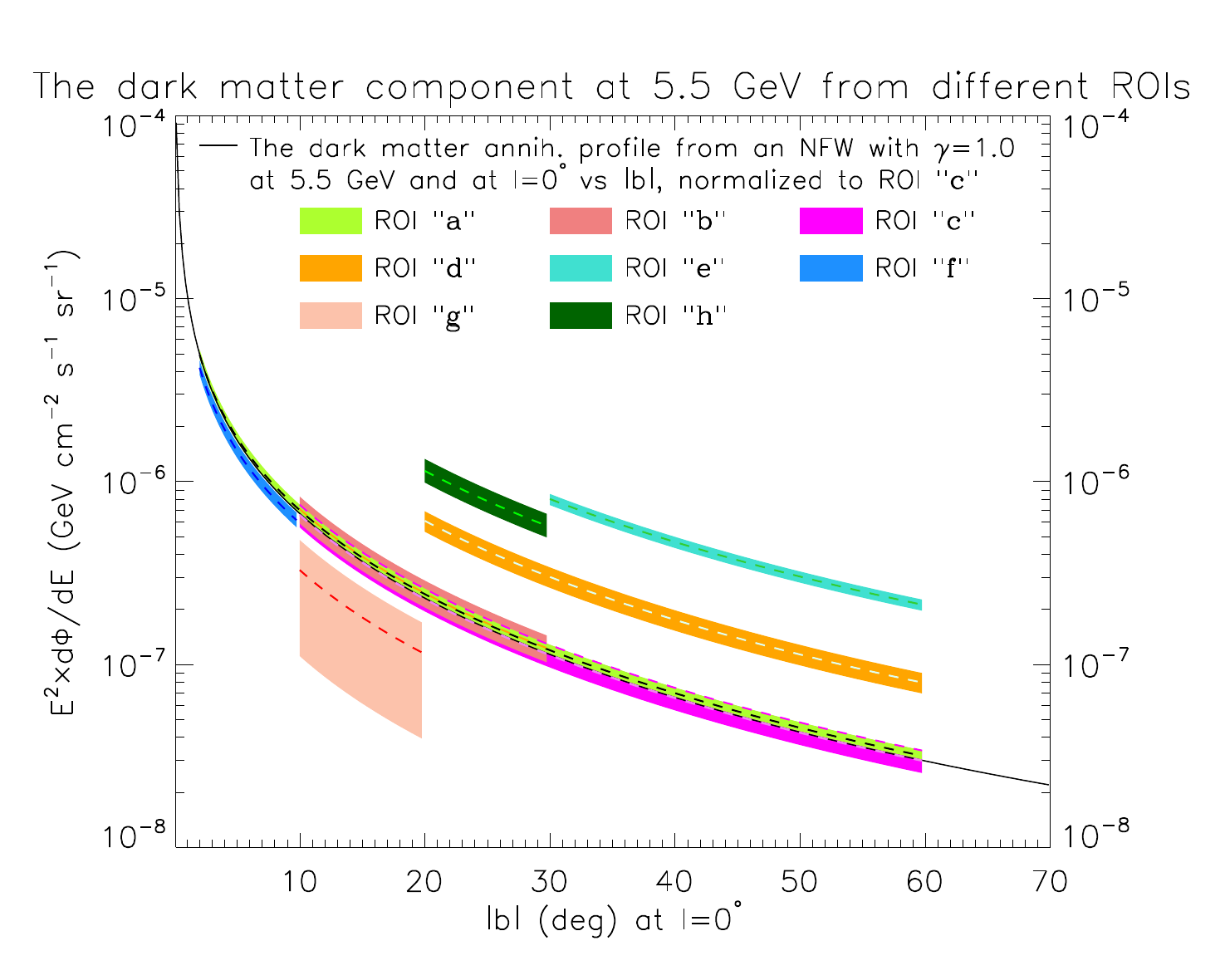}
    \vspace{-0.3cm}
    \caption{The associated flux from the dark matter annihilation template component at 5.5 GeV as function of Galactic latitude $|b|$, evaluated at $l=0^{\circ}$. The black solid line gives the expected dark matter annihilation profile from a NFW profile, as suggested for the HLE. That line is normalized to the fit from ROI \textbf{c}. Different color bands represent the results from the alternative ROIs, with the width representing the 1-$\sigma$ fit range. That width is suggestive of the number of photons associated in the fit to the dark matter annihilation template component, i.e. the associated statistical errors. In most ROIs those bands are narrow. There is a significant discrepancy in the fit-amplitude of the dark matter annihilation template between the ROIs containing latitudes only at $|b| \leq 20^{\circ}$ and the ROIs containing latitudes only at $|b| \geq 20^{\circ}$.}
    \label{fig:DMprofile_for_ROIs}
\end{figure}
In Fig.~\ref{fig:DMprofile_for_ROIs}, we show all our results on the fitted amplitude of the dark matter annihilation template from ROIs \textbf{a}-\textbf{h}. We show the results from our seven-parameter fits using background model SL\,Z6\,R30\,T150\,C2. We plot the associated dark matter annihilation per steradian flux at 5.5 GeV, an energy at which the HLE is always significant. The black line gives the expectation for the dark matter flux at $\ell=0^{\circ}$ vs $|b|$, from a NFW profile, normalized to the best-fit template normalization for the HLE (the dark matter annihilation template) in ROI \textbf{c}.
The results from the alternative ROIs are shown in different bands. Each band's width along the $|b|$-axis represents the latitudes covered in that ROI (see also Fig.~\ref{fig:roi} and Table~\ref{tab:roi}). The width of each band along the $E^{2}d\Phi/dE$-axis gives the 1-$\sigma$ fit range of the dark matter template. In most cases these widths are narrow, indicative of the large enough number of photons associated with the HLE template in the fits. As discussed earlier there is a discrepancy in the normalization of the  dark matter annihilation template between the ROIs containing latitudes only at $|b| \leq 20^{\circ}$ (ROIs \textbf{f} and \textbf{g}) and the ROIs containing latitudes only at $|b| \geq 20^{\circ}$ (ROIs \textbf{d}, \textbf{e} and \textbf{h}). 
We find very similar results across alternative galactic diffuse emission models. 
By comparison, for the GCE it has been shown in multiple references that its amplitude is consistent with a dark matter annihilation profile from a gNFW profile with $\gamma \simeq 1.25$ across alternative ROIs, many of which much smaller in size and with fewer years of \textit{Fermi}-LAT observations (see e.g. Refs.~\cite{Daylan:2014rsa, Calore:2014xka, Calore:2014nla, TheFermi-LAT:2017vmf, Cholis:2021rpp}). 

We also want to note that if there were two dark matter annihilation-like signals present, then the brighter HLE signal should dominate (outshine the GCE) in the ROIs \textbf{f} and especially \textbf{g} (as the GCE signal drops more steeply with increasing latitude). For whatever reason, the HLE is not scaling in these inner regions of the Milky Way in the same manner that a dark matter annihilation-like signal would. While it is always present in the results, and thus represents some emission not modeled by the other templates, it does not have the morphology of a dark matter annihilation-like template when comparing the $|b| \leq 20^{\circ}$ latitudes to the latitudes of $|b| \geq 20^{\circ}$. For this reason, and also because using the library of the 100 galactic diffuse emission models we systematically find a degeneracy between the HLE, the ICS and the \textit{Fermi} Bubbles, we explore in the following Section~\ref{sec:NewFermiBubbles}, the need for a new template to describe the emission coming from the \textit{Fermi} Bubbles.

\section{A new geometric model for the \textit{Fermi} Bubbles and the Cocoons}
\label{sec:NewFermiBubbles}

In this section we discuss how we construct a new model for the \textit{Fermi} Bubbles, including the Southern and Northern Cocoons that as seen from our location on the Milky Way, span a region of the Galactic sky that is inside the main Bubbles. 

We test three aspects of the \textit{Fermi} Bubbles and Cocoons. In short, 
\begin{enumerate}
    \item {First we test the boundaries of the \textit{Fermi} Flat Bubbles. After trying alternative choices, we find the boundaries of the Flat Bubbles to be larger from those of Ref.~\cite{Su:2010qj}, by up to a few degrees in some directions, but otherwise in good agreement with this earlier result. Such a flat brightness structure with clear edges, may originate from the reverse shock generated by the fast-moving wind that created the \emph{Fermi} Bubbles~\cite{Fujita:2022wny}.}
    \item{We test the properties of the Cocoons, modeling them as a coaxial pair of ellipsoids, each with a semi-major axis $a_{\textrm{cocoon}}$, semi-minor axis $b_{\textrm{cocoon}}$, their centers separated from the galactic center by $z_{0, \, \textrm{cocoon}}$. Furthermore, each Cocoon has a constant gamma-ray emissivity inside its volume and zero outside. We test templates where the two Cocoons have different emissivities. We find that the Southern Cocoon has a significantly larger emissivity within its volume than the Northern one. Finally, we impose two rotations described by a yaw $\phi_{\textrm{cocoon}}$ and a pitch $\theta_{\textrm{cocoon}}$ angle.} 
    \item We replace the HLE template with a template that describes the inner morphology of the main Bubbles, which we denote as the ``Inner Bubbles'' (IB) template.  
     As with the HLE template, it is added on top of the Flat Bubbles template rather than replacing it. Unlike the Flat Bubbles of Ref.~\cite{Su:2010qj}, which are used for most of the results of Section~\ref{sec:Results}, the Inner Bubbles are not flat on the sky. Instead, as for the Cocoons, we assume a constant gamma-ray emissivity throughout their volume, so that the projected intensity is proportional to the line-of-sight path length through each ellipsoid; unlike the Cocoons, however, the North and South Inner Bubbles are assigned the same emissivity. To first order, we model the Inner Bubbles as a coaxial pair of ellipsoids, each with a semi-axis $a_\textrm{IB}$ along the $Z$-axis and a semi-axis $b_\textrm{IB}$ along both the $X$- and $Y$-axes, so that each ellipsoid is a spheroid whose symmetry axis is the $Z$-axis. No rotation is applied to the pair. The two centers lie on the $Z$-axis, displaced symmetrically from the Galactic Center to $(0,\,0,\,\pm z_{0,\,\textrm{IB}})$. Finally, the Inner Bubbles template is set to zero outside the optimal Bubble boundaries (edges) found in the first test.

\end{enumerate}

\subsection{The boundaries of the \textit{Fermi} Bubbles}
\label{subsec:BubblesBoundaries}

\begin{figure}[t]
    \centering
    \includegraphics[width=0.98\linewidth]{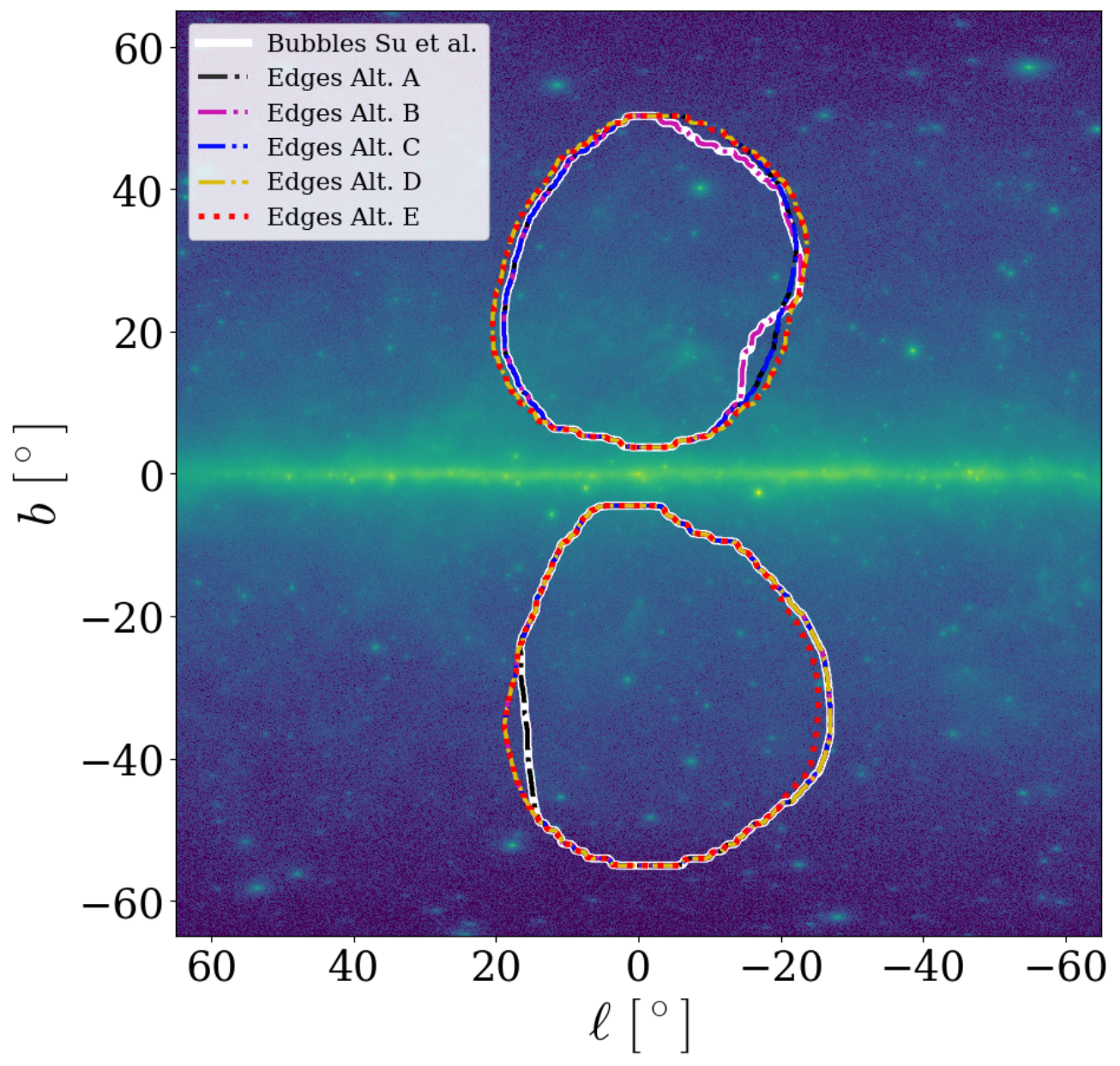}
    \vspace{-0.3cm}
    \caption{The boundaries/edges of the \textit{Fermi} Flat Bubbles, i.e. flat in morphology template from Ref.~\cite{Su:2010qj} (Su \textit{et al.} in solid white), together with five alternative edges that we test. We find alternative option ``E'' (in red dotted line), to give the best fit to the gamma-ray sky within ROI \textbf{c}, of all six options. This result is valid both when testing against the data, in addition to the regular template components, a combination of the flat-only Bubbles template with the HLE template component, and when instead we test the combination of Flat Bubbles $\&$ Inner Bubbles $\&$ Cocoons templates (see text for more details).}
    \label{fig:alt_flat_bubbles}
\end{figure}

In performing our first test,  on the boundaries/edges of the \textit{Fermi} Bubbles, we initially assume they are flat in brightness, i.e. we test the boundaries of the Flat Bubbles.
However, after performing our other tests, on the properties of the Cocoons and the Inner Bubbles, we confirm the validity of the new boundaries even for the case of this more elaborate model on the Bubbles' morphology.
In Fig.~\ref{fig:alt_flat_bubbles}, we show the boundaries of the \textit{Fermi} Bubbles from Ref.~\cite{Su:2010qj}, alongside five alternative options that we test. 

\begin{table*}[t]
\caption{Comparison of the alternative options for the boundaries of the \textit{Fermi} Bubbles. In all cases, we mask the 4FGLDR4 point and extended sources and perform the fit in ROI \textbf{c} that excludes $|b| \leq 10^{\circ}$. The reported $-2\Delta \! \ln (\mathcal L)$ results come for fits using the SL\,Z6\,R20\,T100000\,C5 galactic diffuse emission model that provides the overall best fit to the gamma-ray sky in this ROI. Negative values indicate an improvement to the fit. By ``Template combination'' we mean the template components that we add to the existing combination of Pi0+Bremss, ICS, Isotropic, Loop I Bright and Loop I Dim components. The first six lines come from a seven-parameter fit and the second six lines from an eight-parameter fit. Differences in log-likelihood are evaluated from the choice of Ref.~\cite{Su:2010qj}, for each type of fit.}
\begin{center}
\begin{tabular}{c c l}
\hline
Boundaries/Edges & Template combination & $-2\Delta \! \ln (\mathcal L)$ \\ \hline \hline
Ref.~\cite{Su:2010qj} & Flat Bubbles $\&$ HLE w. $\gamma = 1.0$ & $0$ \\
Alt. Edges A & Flat Bubbles $\&$ HLE w. $\gamma = 1.0$ & $+182$ \\
Alt. Edges B & Flat Bubbles $\&$ HLE w. $\gamma = 1.0$ & $-66$ \\
Alt. Edges C & Flat Bubbles $\&$ HLE w. $\gamma = 1.0$ & $+114$ \\
Alt. Edges D & Flat Bubbles $\&$ HLE w. $\gamma = 1.0$ & $+44$ \\
Alt. Edges E & Flat Bubbles $\&$ HLE w. $\gamma = 1.0$ & $-444$ \\
\hline
Ref.~\cite{Su:2010qj} & Flat Bubbles $\&$ Inner Bubbles $\&$ Cocoons & $0$ \\
Alt. Edges A & Flat Bubbles $\&$ Inner Bubbles $\&$ Cocoons &  $-230$ \\
Alt. Edges B & Flat Bubbles $\&$ Inner Bubbles $\&$ Cocoons &  $-40$ \\
Alt. Edges C & Flat Bubbles $\&$ Inner Bubbles $\&$ Cocoons &  $-268$ \\
Alt. Edges D & Flat Bubbles $\&$ Inner Bubbles $\&$ Cocoons &  $-568$ \\
Alt. Edges E & Flat Bubbles $\&$ Inner Bubbles $\&$ Cocoons & $-572$ \\
\hline
\end{tabular}
\end{center}
\label{tab:Bubbles_Edges}
\end{table*}

We test the five alternative choices against the results of Ref.~\cite{Su:2010qj}, for the boundaries of the Bubbles, first using a seven-parameter fit (as in Fig.~\ref{fig:HLE_120x120_WO_ResMaps}), just for ROI \textbf{c}. Our seven templates are described in Table~\ref{tab:fit_schemes}.
In these tests, as the first round of Bubbles modeling, the Bubbles are assumed flat always, and it is the HLE that absorbs any non-flatness from the region they cover in this ROI. We note, however, that the HLE component extends in all cases beyond the Bubbles' edges. 

Our ROI \textbf{c} includes gamma rays from up to $|b| = 60^{\circ}$, but excludes the $|b| \leq 10^{\circ}$ low latitudes.  Tackling the properties of the Bubbles at even lower latitudes has always been challenging. Any fit including latitudes close to the disk is dominated by the much more prominent galactic diffuse emission along the disk that is unassociated to the emission of the Bubbles themselves. Moreover, the number of point sources increases significantly at $|b| \leq 10^{\circ}$. 
As can be seen in Fig.~\ref{fig:alt_flat_bubbles} (and in later figures), the Bubbles predict an emission even at lower latitudes. At the end of this work, in Section~\ref{sec:conclusions}, we revisit the GCE present at the lower latitudes using the updated model for the \textit{Fermi} Bubbles. 

In Table~\ref{tab:Bubbles_Edges}, we show the log-likelihood difference (in $-2\Delta \! \ln (\mathcal L)$) between each of the alternative Bubbles edges and the edges proposed in Ref.~\cite{Su:2010qj}. 
In the first six lines we show the results from the seven-parameter fit for the SL\,Z6\,R20\,T100000\,C5 galactic diffuse emission model that provides the best fit to the gamma-ray data within ROI \textbf{c} among the 20 models of \cite{Fermi-LAT:2012edv}.
We also test the alternative edges using models SL\,Z6\,R30\,T150\,C2
(used in Figs.~\ref{fig:CornerPlotTotaniMethod}-\ref{fig:SLZ6R30T150C2_spectra_6parameters}), SL\,Z10\,R30\,T150\,C2, SO\,Z4\,R30\,T150\,C5 and SY\,Z6\,R30\,T100000\,C5, to avoid biasing our results on the Bubble's properties by using one specific galactic diffuse emission model.
In the seven-parameter fit, only alternative edges ``B'' and ``E'' for the Flat Bubbles give a better fit than the original Su \textit{et al.}, a result valid for all five galactic diffuse emission models. 
Instead, alternative edges ``A'', ``C'' and ``D'' perform worse than Su \textit{et al.} edges for most diffuse emission models. 
Alternative edges E improves the fit compared to Su \textit{et al.}  by a $2\Delta \! \ln (\mathcal L) : 230-520$. 

Compared to Ref.~\cite{Su:2010qj}, the alternative edges E increase the size of the Northern Bubble on both the positive and the negative longitudes sides. Instead, the Southern Bubble is larger on the positive longitudes (the same side as the Southern Cocoon is), but a little smaller on the side of the negative longitudes. In Appendix~\ref{app:FermiBubblesEdgesE}, we give in the form of a table the digitized boundaries of the Bubbles E alternative. 

In the bottom six lines of Table~\ref{tab:Bubbles_Edges}, we show the results from an eight-parameter fit, where in addition to the five template components for the Pi0+Bremss, ICS, Isotropic, Loop I Bright, and Loop I Dim, we have the Flat Bubbles, the Inner Bubbles, and the Cocoons (described in the following sections).
Using the same galactic diffuse emission model as in the top six lines of that table, i.e. SL\,Z6\,R20\,T100000\,C5, we confirm that the alternative edges E option is still preferred to the original of Ref.~\cite{Su:2010qj}, and to the other four options, for the final configuration for the Bubbles and the Cocoons best-fit morphologies. The diffuse emission model SL\,Z6\,R20\,T100000\,C5, even with the updated version of the \textit{Fermi} Bubbles \& Cocoons still gives the overall best-fit result to the \textit{Fermi} data. We note that when using the updated morphology for the Bubbles and the Cocoons, all alternative versions for the Bubbles' edges perform better than those of Ref.~\cite{Su:2010qj}. 
Of these alternative edges, option ``B'' gives the most similar log-likelihood values to the original Bubbles for all five models.  Alternative edges ``A'' and ``C'' perform better than edges ``B'' and the options ``D'' and ``E'' are the ones that, for all cases, perform the best. 
Alternative edges E improves the fit compared to Su \textit{et al.}  by a $2\Delta \! \ln (\mathcal L) : 500-720$, while alternative edge B only by $40-180$. 

\begin{table*}[t]
    \centering
    \renewcommand{\arraystretch}{1.3}
    \topcaption{Summary of fitting schemes with the additions of the Cocoons.}
    \label{tab:fitting}
    \newcommand{\name}[1]{\parbox[t]{0.1\textwidth}{\raggedright #1\strut}}%
    \newcommand{\roi}[1]{\parbox[t]{0.1\textwidth}{\raggedright #1\strut}}%
    \newcommand{\mask}[1]{\parbox[t]{0.1\textwidth}{\raggedright #1\strut}}%
    \newcommand{\tmpl}[1]{\parbox[t]{0.5\textwidth}{\raggedright #1\strut}}%
    \newcommand{\gde}[1]{\parbox[t]{0.2\textwidth}{\raggedright #1\strut}}%
    \begin{tabular}{l l l l l}
    \hline
     Name & ROI & Mask & Templates & GDE templates\\
    \hline \hline 
\name{8-param} & \roi{\textbf{c}} & \mask{4FGLDR4} & \tmpl{Pi0+Bremss, ICS, Isotropic, Flat Bubbles, HLE, Loop I Bright, Loop I Dim, Cocoons} & \parbox[t]{0.25\textwidth}{\raggedright 
From Ref.~\cite{Fermi-LAT:2012edv}:\\
SL\,Z6\,R30\,T150\,C2, \\
SL\,Z6\,R20\,T100000\,C5, \\ 
SL\,Z10\,R30\,T150\,C2, \\ 
SO\,Z4\,R30\,T150\,C5, \\ 
SY\,Z6\,R30\,T100000\,C5}\\
\\
           \hline
    
\name{8-param} & \roi{\textbf{c}} & \mask{4FGLDR4} & \tmpl{Pi0+Bremss, ICS, Isotropic, Flat Bubbles, Inner Bubbles, Loop I Bright, Loop I Dim, Cocoons} & \parbox[t]{0.25\textwidth}{\raggedright Ref.~\cite{Fermi-LAT:2012edv} \\ 
(20 models)} \\ 
\\
            \hline
            
    \end{tabular}
     \label{tab:fit_schemes_2}
\end{table*}

\subsection{The properties of the Southern and Northern Cocoons}
\label{subsec:Cocoons}

\begin{figure}[t]
    \centering
    \includegraphics[width=0.98\linewidth]{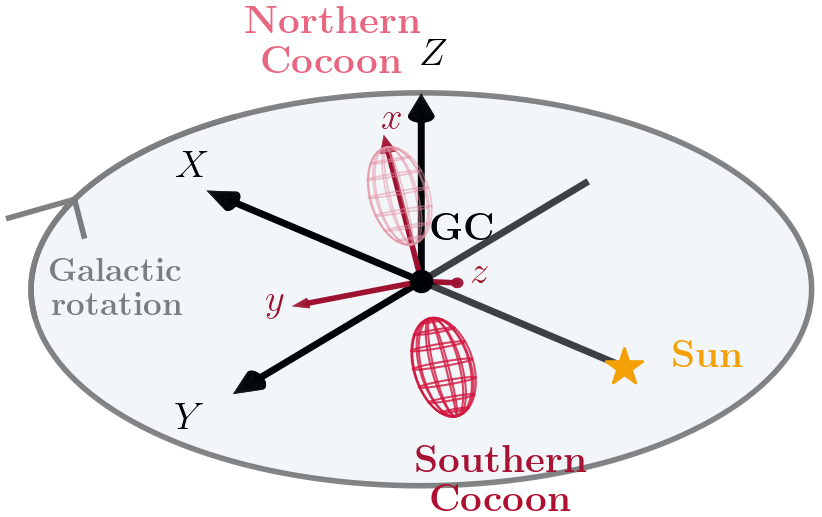}
    \caption{3D model of the Southern Cocoon (red) and the Northern Cocoon (pink) within the \emph{Fermi} Bubbles under the galactocentric coordinate $X$-$Y$-$Z$. $x$-$y$-$z$ represents the rotated principal axes of the Cocoons. See text for details.}
    \label{fig:3D_cocoons}
\end{figure}

To model the Cocoons, we adopt a galactocentric Cartesian coordinate system with its origin at the GC. The $X$-axis points from the Sun toward the GC, the $Z$-axis is perpendicular to the Galactic plane and points toward the North Galactic Pole, and the $Y$-axis lies in the Galactic plane such that $X$-$Y$-$Z$ forms a right-handed system; see~\figref{3D_cocoons}. In this system, the Sun is located at approximately $\bm{x}_\odot = (-8.5, 0, 0)\,\mathrm{kpc}$.

We model the Cocoons as a coaxial pair of identical prolate ellipsoids, each with semi-axes $a \geq b = c$, so that the two transverse axes are equal. Their major axes lie on a common line through the GC, and their centers lie on this line at a distance $z_0 > 0$ on either side of the GC. To specify the orientation of the pair in three dimensions, we start from a reference configuration and apply rigid rotations to it; in general this requires three rotations (yaw, pitch, and roll). Here we take the reference configuration to be the one in which the common major axis coincides with the $Z$-axis. Because each ellipsoid is axisymmetric about the common major axis, only two angles are needed: a pitch angle $\theta$ about the $Y$-axis and a yaw angle $\phi$ about the $Z$-axis, with $0 \leq \theta \leq 90^\circ$ and $-180^\circ < \phi \leq 180^\circ$ (for $\theta = 0$ the yaw angle is immaterial). More details on the 3D rotations can be found in~\cite{Hu:2026apn}.

Since $\theta \leq 90^\circ$, the ellipsoid whose center lies on the positive $Z$-axis in the reference configuration stays north of the Galactic plane after the rotation, and its partner stays south of it. We call them the Northern and Southern Cocoons, respectively. The two Cocoons are allowed to differ in their gamma-ray emissivity  
by a fixed ratio $\varepsilon_{\textrm{S}}/\varepsilon_{\textrm{N}}$; because this ratio is free, restricting $\theta$ to non-negative values costs no generality. Within each ellipsoid, the emissivity is taken to be uniform, and the same spectral shape is assumed throughout both Cocoons. The spectral shape is therefore identical in every pixel. The observed flux depends on the morphology, both in the model templates and in the fit to the data. The predicted intensity in a given direction is the common spectral shape multiplied by the line-of-sight integral of the emissivity through the pair, as seen from $\bm{x}_\odot$. That is then rescaled from the fit to the \textit{Fermi} data.

We describe the bounding surfaces of the ellipsoid pair by,
\beq
\left(\bm{R}\,\bm{x} \mp z_0\,\hat{\bm{e}}_z\right)^\intercal
\begin{pmatrix}
b^{-2} & 0 & 0 \\
0 & b^{-2} & 0 \\
0 & 0 & a^{-2} \\
\end{pmatrix}
\left(\bm{R}\,\bm{x} \mp z_0\,\hat{\bm{e}}_z\right) = 1,
\label{eq:cocoonpair}
\eeq
where the upper (lower) sign refers to the Northern (Southern) Cocoon.
Here $\bm{x} = (x, y, z)^\intercal$ is the galactocentric position vector and $\hat{\bm{e}}_z = (0, 0, 1)^\intercal$. The 3D rotation matrix
\begin{equation}
\bm{R} = \begin{pmatrix}
\cos \theta & 0 & -\sin \theta \\
0 & 1 & 0 \\
\sin \theta & 0 & \cos \theta\\
\end{pmatrix}
\begin{pmatrix}
\cos \phi & \sin \phi & 0 \\
-\sin \phi & \cos \phi & 0 \\
0 & 0 & 1\\
\end{pmatrix},
\label{eq:tilted}
\end{equation}
is composed in the \emph{extrinsic} pitch-then-yaw order, i.e., the sequence $Y[\theta] \to Z[\phi]$ of~\cite{Hu:2026apn}. Positive values of $\theta$ and $\phi$ correspond to right-handed active rotations of the ellipsoid pair. For $\theta = \phi = 0$, the configuration is untilted: the major axes of both ellipsoids lie along $Z$, the two equal transverse axes lie along $X$ and $Y$, and the centers of the Northern and Southern Cocoons sit at $+z_0\,\hat{\bm{e}}_z$ and $-z_0\,\hat{\bm{e}}_z$, respectively. After the rotation, the common major axis points along $\hat{\bm{n}} = \bm{R}^\intercal \hat{\bm{e}}_z = (\sin\theta\cos\phi,\ \sin\theta\sin\phi,\ \cos\theta)^\intercal$, and the two centers sit at $\pm z_0\,\hat{\bm{n}}$.

By adding the Cocoons as an additional template to the existing emission templates of: Pi0+Bremss, ICS, Isotropic, HLE, Loop I Bright, Loop I Dim and Flat Bubbles, we perform an eight-parameter fit, as described in Table~\ref{tab:fit_schemes_2}, that presents the fitting schemes that we used in probing the properties of the Bubbles and Cocoons.

In Table~\ref{tab:CocoonsProperties}, we give the best-fit properties for the Cocoons. 
We have tested these properties for two combinations of an eight-parameter template fit. In the first case in addition to the Pi0+Bremss, ICS, Isotropic, Loop I Bright, and Loop I Dim we use the Flat Bubbles, the HLE (with NFW) and the Cocoons. In the second case which is our final option in addition to the first five templates we use the Flat Bubbles, the Inner Bubbles and the Cocoons. The significance of the Cocoons is large. Compared to the case where we performed a seven-parameter fit using the first five templates, the Flat Bubbles and the HLE, we get a log-likelihood improvement of $2\Delta \! \ln (\mathcal L) =3998$. Most of that significance comes from the Southern Cocoon as due to the rotations is closer to us, but also it has a significantly larger emissivity. 

\begin{table}[t]
\caption{The best-fit properties of the Cocoons. The log-likelihood improvement provided is evaluated using the SL\,Z6\,R20\,T100000\,C5 galactic diffuse emission model for the Pi0+Bremss and the ICS templates. The parameter $z_{0, \, \textrm{cocoon}}$ is fixed to 3.5 kpc.}
\begin{center}
\begin{tabular}{lr}
\hline
$\varepsilon_\text{S, cocoon}/\varepsilon_\text{N, cocoon}$ & 10\\
$z_{0, \, \textrm{cocoon}}$ & 3.5 kpc\\
$a_{\textrm{cocoon}}$ & 2.0 kpc\\
$b_{\textrm{cocoon}}$ & 1.0 kpc\\
$\theta_{\textrm{cocoon}}$ &$35^{\circ}$\\
$\phi_{\textrm{cocoon}}$ & $-29^{\circ}$\\
\hline
$-2\Delta \! \ln (\mathcal L)$ & $-3998$\\ 
\hline
\end{tabular}
\end{center}
\label{tab:CocoonsProperties}
\end{table}

\begin{figure*}[t]
\hspace{-0.07in}
\includegraphics[width=2.32in,angle=0]{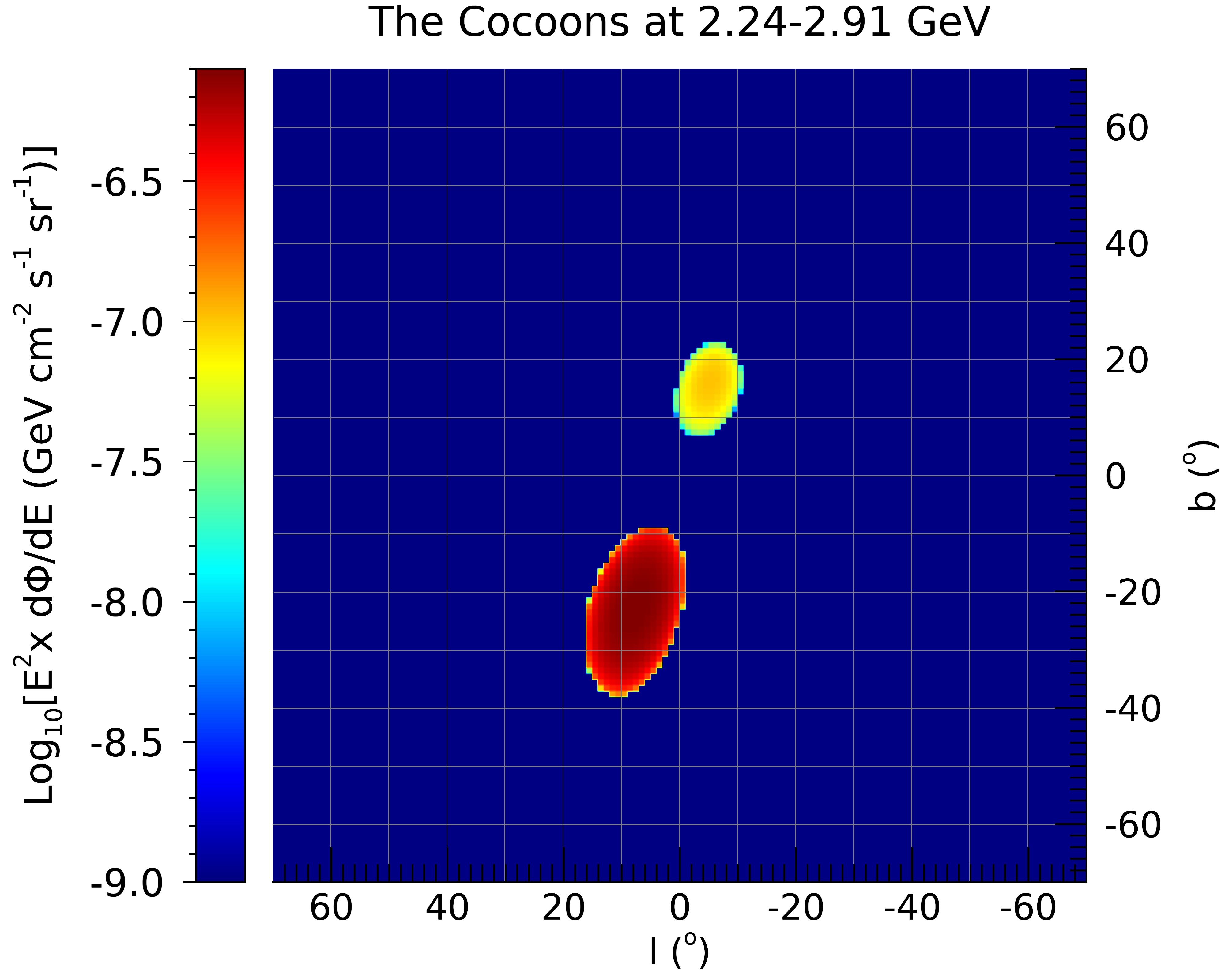} 
\hspace{-0.03in}
\includegraphics[width=2.32in,angle=0]{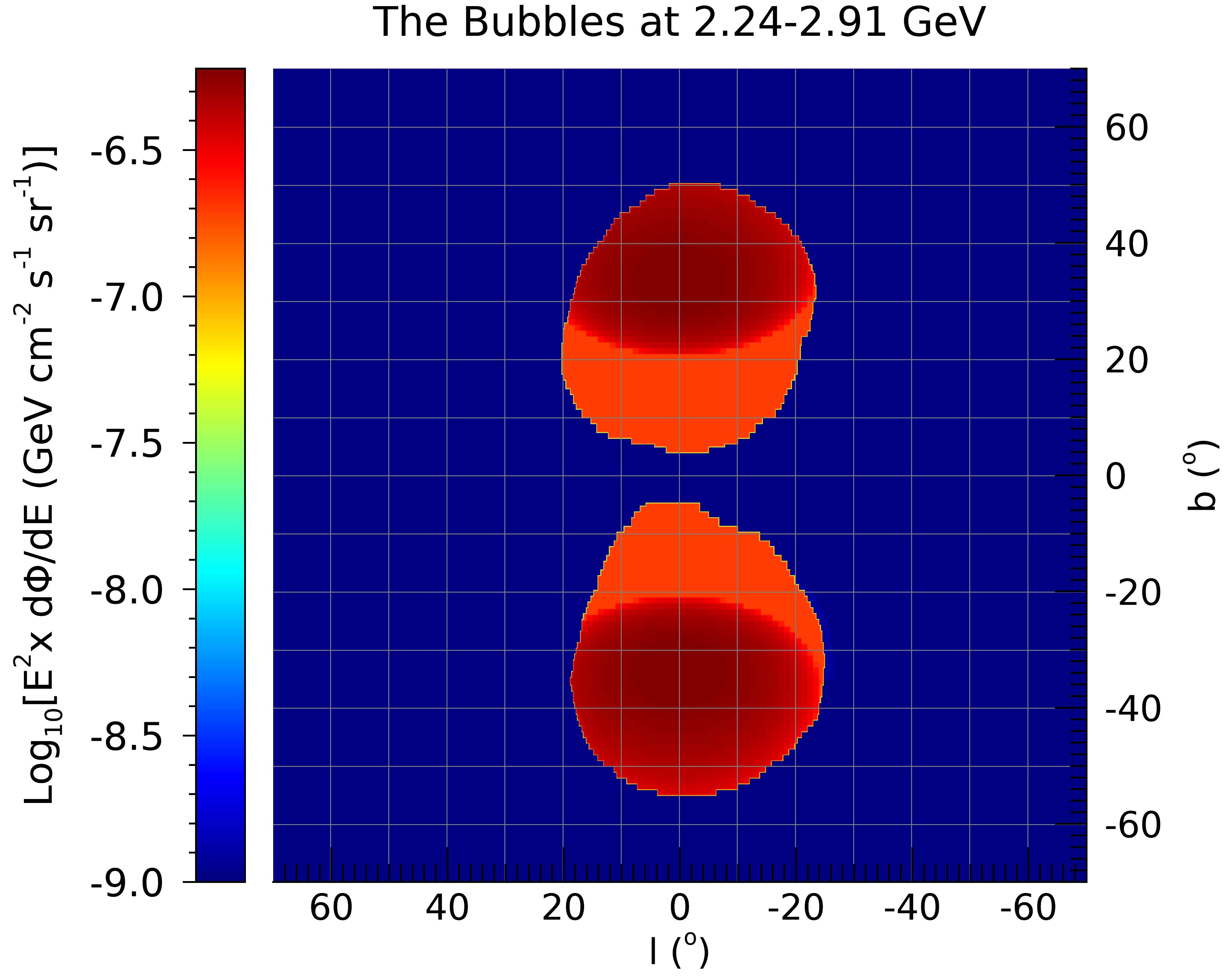} 
\hspace{-0.03in}
\includegraphics[width=2.32in,angle=0]{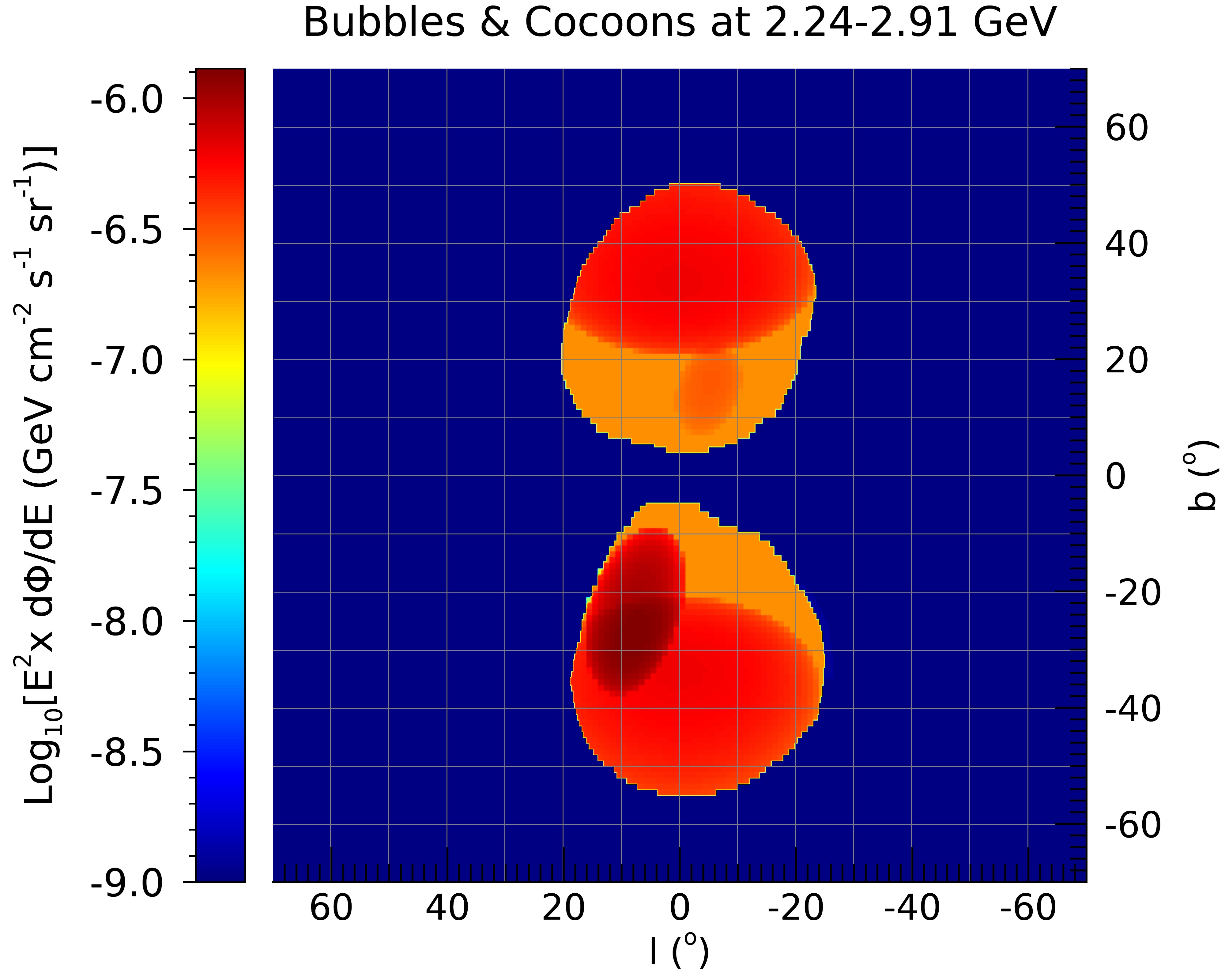}
\vspace{0.00in}
\caption{The morphology of the Cocoons and main \textit{Fermi} Bubbles for energies 
of 2.24--2.91 GeV shown over a region of $140^{\circ} \times 140^{\circ}$. \textit{Left}, the Cocoons, \textit{middle}, the Combination of the Flat Bubbles $+$ Inner Bubbles (i.e. the main Bubbles) 
and \textit{right}, the combination of the Cocoons and the main Bubbles. We use a resolution of 
$0.25^{\circ} \times 0.25^{\circ}$ pixels in a Cartesian grid. We do not include in these maps 
the PSF smoothing of the \textit{Fermi}-LAT instrument. This emphasizes the differences 
in the morphology between the various emission components. We show the differential flux multiplied 
by the energy squared ($E^{2} \times d\Phi/dE$). We add a flux on these maps of $1.0\times 10^{-9} \; \textrm{GeV} \, \textrm{cm}^{-2} \, \textrm{s}^{-1} \textrm{sr}^{-1}$ to make clear where the boundaries of these templates lie.}
\label{fig:Bubbles_and_Cocoons_Templates}
\end{figure*}

In Fig.~\ref{fig:Bubbles_and_Cocoons_Templates}, we show in the first panel the diffuse emission flux from the Cocoons with their best-fit properties and best-fit flux, for the final configuration for the \textit{Fermi} Bubbles $\&$ Cocoons. We show those templates for the energy bin of $2.24\leq E \leq 2.91\,\gev$, at which the averaged flux over the entire ROI \textbf{c}, from the Cocoons: the Inner Bubbles: the Flat Bubbles is roughly 2:5:5. Given that the Cocoons template spans a smaller area on the sky than the other two components, the brightness especially of the Southern Cocoon is pretty prominent; which can be seen in the combined map shown in the right panel of Fig.~\ref{fig:Bubbles_and_Cocoons_Templates}.

\begin{figure}[t]
    \centering
\includegraphics[width=1.0\linewidth]{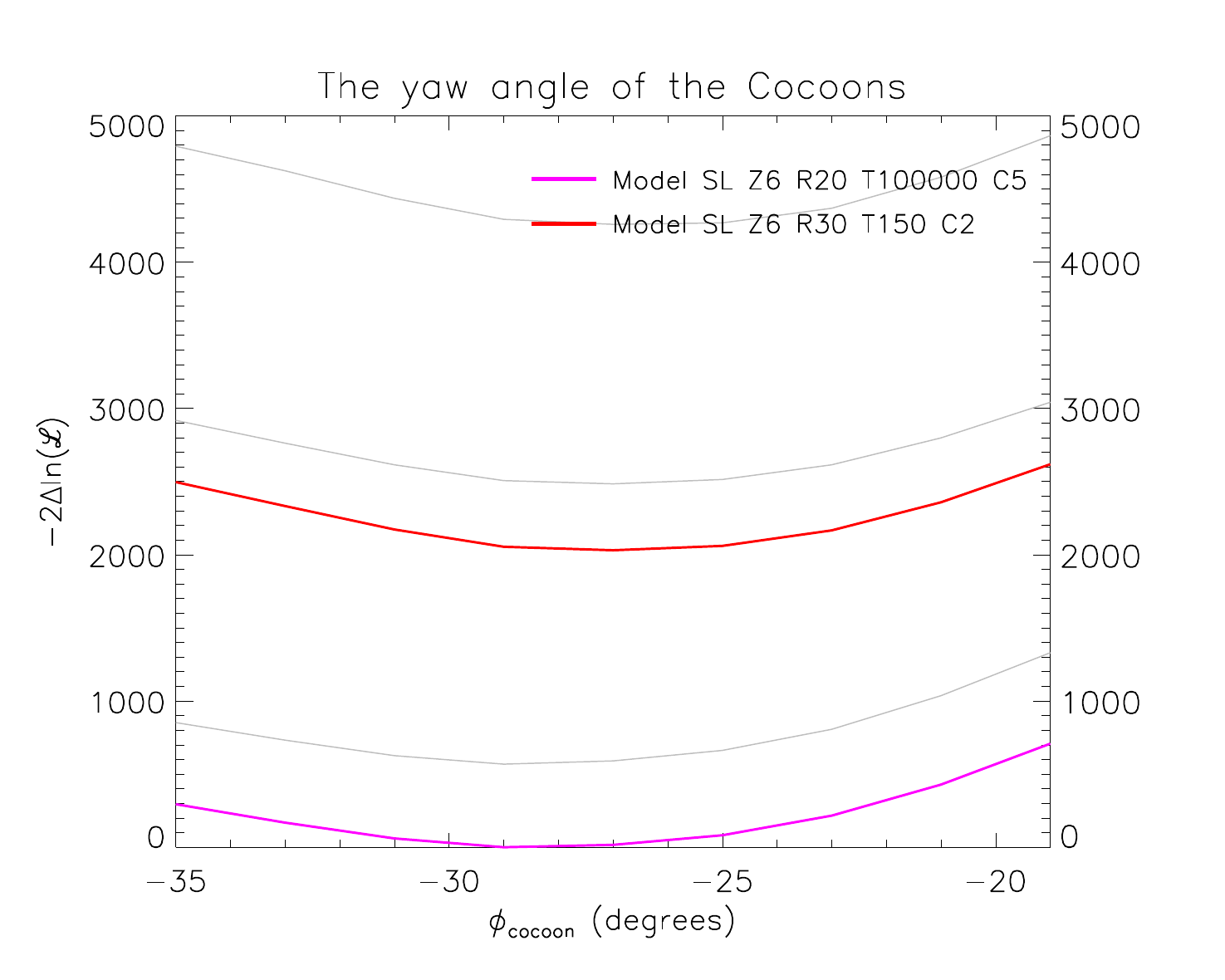}\\
    \vspace{-0.4cm}
    \includegraphics[width=1.0\linewidth]{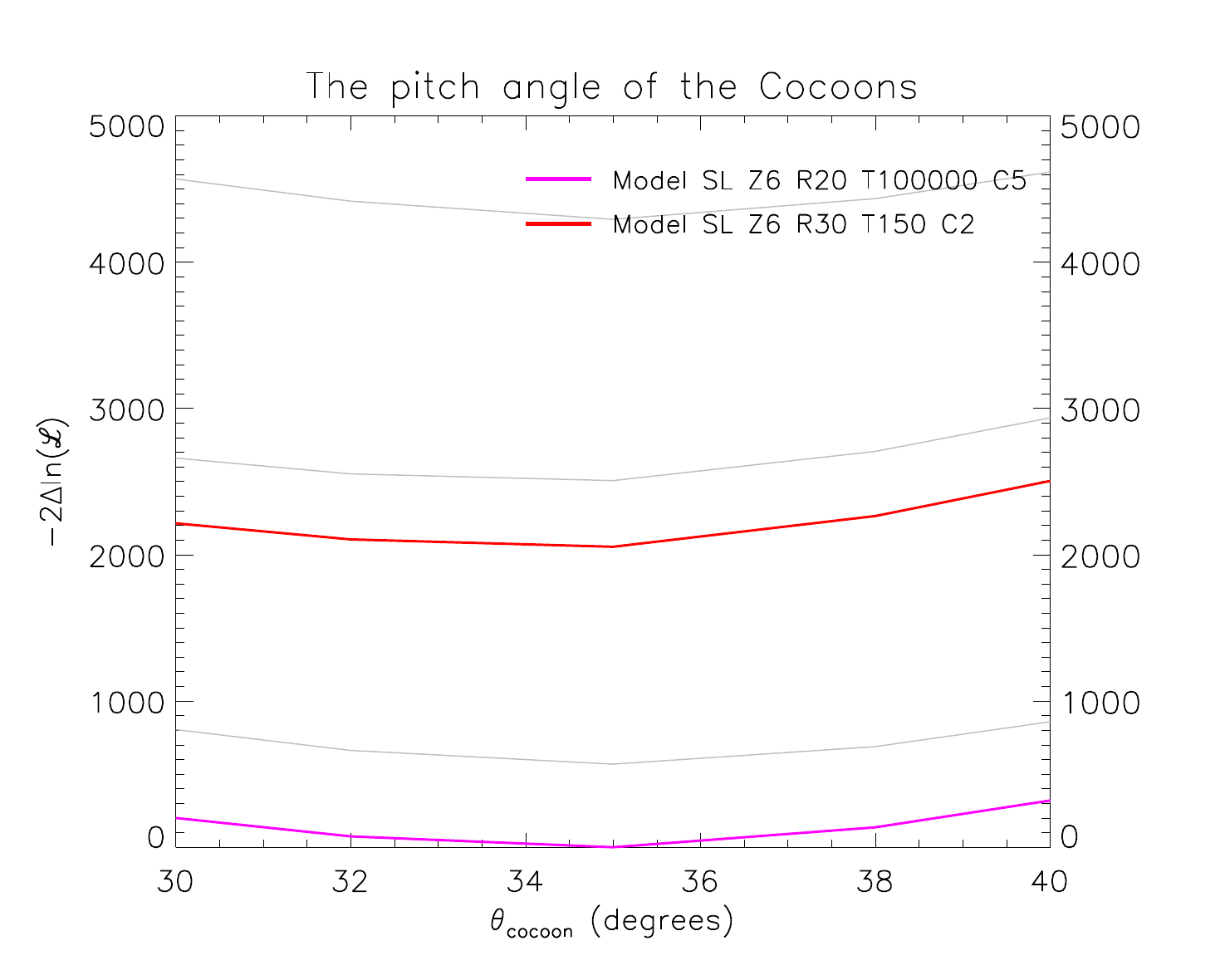}\\
    \caption{\textit{Top}: The yaw of the Cocoons for  $-35^{\circ} \leq \phi_{\textrm{cocoon}} \leq -19^{\circ}$. 
    \textit{Bottom}: The pitch of the Cocoons for  $30^{\circ} \leq \theta_{\textrm{cocoon}} \leq 40^{\circ}$. 
    Our $y$-axis gives the difference in the fit between choices for the Cocoons' rotation, defined as $-2 \ln \mathcal{L}$ for the alternative model minus $-2 \ln \mathcal{L}$ of the best fit choice. The magenta (lowest) line showcases the results for the best-fit galactic diffuse emission model SL\,Z6\,R20\,T100000\,C5, the red line the results for SL\,Z6\,R30\,T150\,C2, while the gray lines show the results for the Cocoons' under three more galactic diffuse emission models.}
    \label{fig:CocoonsYaw_ans_Pitch}
\end{figure}

In Fig.~\ref{fig:CocoonsYaw_ans_Pitch}, we provide the log-likelihood difference, $-2\Delta \! \ln (\mathcal L)$, for five galactic diffuse emission models including model SL\,Z6\,R20\,T100000\,C5 (in magenta line), as a function of the Cocoons'  pitch $\theta_{\textrm{cocoon}}$ and yaw $\phi_{\textrm{cocoon}}$ angles. Lower points along the y-axis represent a better fit.
In Appendix~\ref{app:CocoonFitProperties}, we provide further tests of log-likelihood differences to show the impact on the quality of the fit by varying the other properties of the Cocoons for the same five alternative galactic diffuse emission models. In every case, we vary the value of one parameter for the Cocoons, keeping the others fixed to their best-fit values of Table~\ref{tab:CocoonsProperties}. 

We have checked that adding the Cocoons template as a separate eighth component affects only somewhat the ranking of the templates used for the evaluation of the main Bubbles' edges, making option ``B'' the less preferred alternative than without the Cocoons.
Alternative edges ``E'' remains the best option among all choices, including the original edges of Ref.~\cite{Su:2010qj}.
Comparing several combinations of the Cocoons' properties and alternative templates for the boundaries of the Bubbles, we find that the $-2\Delta \! \ln (\mathcal L)$ are approximately as much as the $-2\Delta \! \ln (\mathcal L)$ found between option E and Ref.~\cite{Su:2010qj}, without having the Cocoons present (i.e. from our seven-parameter fit). Similarly, the optimization of the Cocoons' properties led to the same results irrespective of the fine details on the edges of the main Bubbles.
The properties of the edges of the Bubbles and the properties of the Cocoons are clearly separable in the fit, as only one boundary of the Southern Cocoon marginally overlaps with the boundaries of the main Bubbles (see again Fig.~\ref{fig:Bubbles_and_Cocoons_Templates}). 

\begin{figure}[t]
    \centering
    \includegraphics[width=0.98\linewidth]{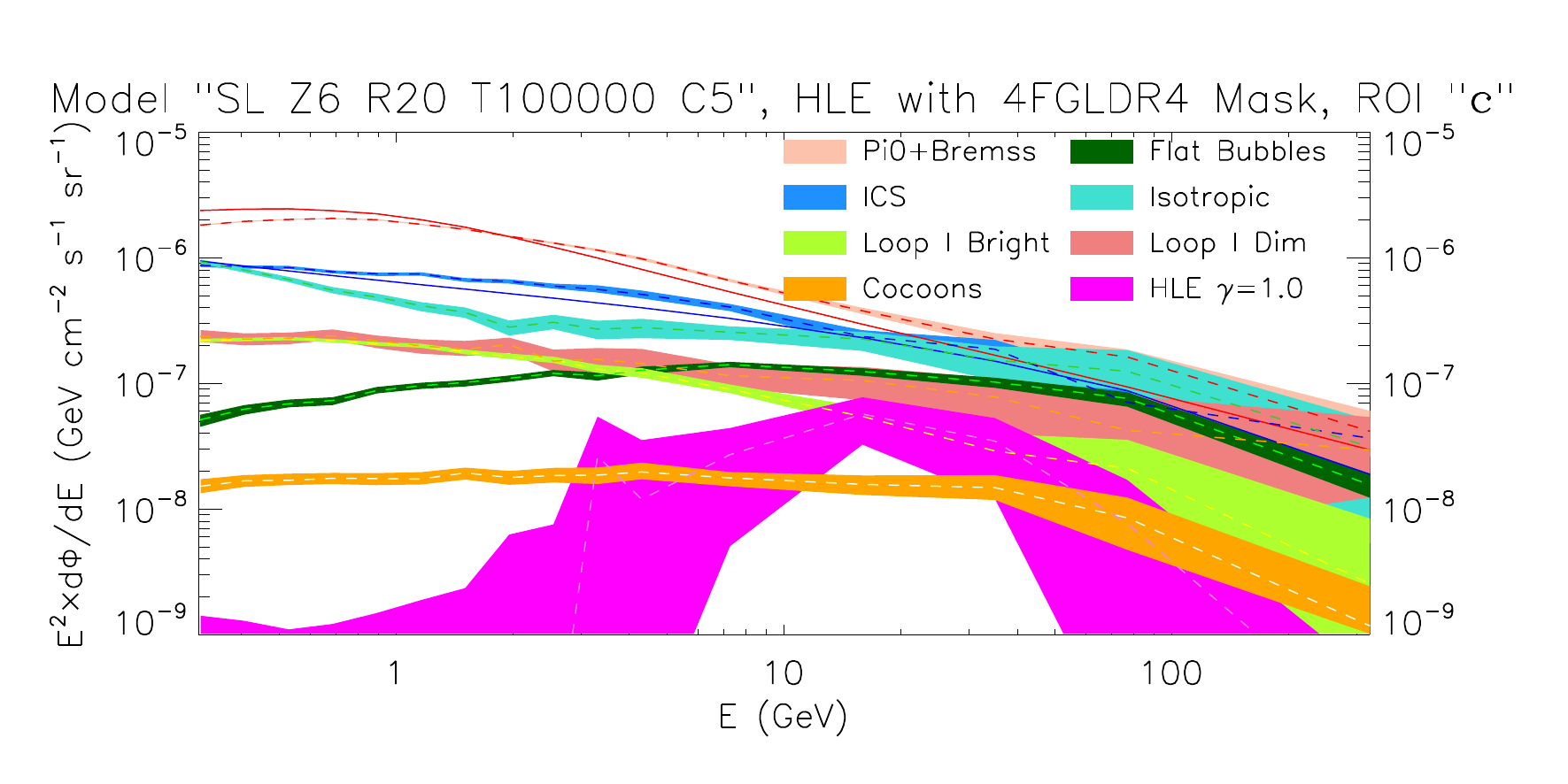}
    \vspace{-0.3cm}
    \caption{The galactic diffuse emission spectra for the ROI \textbf{c}, with the 4FGLDR4 \textit{Fermi} point and extended sources masked out in the fit. For simplicity in evaluating the averaged spectra from ROI \textbf{c}, we include the \textit{modeled} diffuse emission flux from the directions of the point and extended sources' masks, but exclude the diffuse emission from $|b|\leq 10^{\circ}$.
    For the Pi0, Bremss and ICS templates we use the predictions from model SL\,Z6\,R20\,T100000\,C5. Adding the Cocoons, significantly suppresses the flux associated with the HLE and improves the fit.}
    \label{fig:BubblesE_Cocoons_and_HLE}
\end{figure}

In Fig.~\ref{fig:BubblesE_Cocoons_and_HLE}, we provide the galactic diffuse emission spectra from ROI \textbf{c}, for the best-fit combination of edges for the Flat Bubble component and the Cocoons. In this fit, we retain the HLE template as a separate component assuming a NFW profile for the underlying dark matter distribution. However, compared to the result of Fig.~\ref{fig:HLE_120x120_WO_ResMaps}, for the same ROI \textbf{c}, the associated flux of the HLE gets suppressed by a factor of at least 2 (at 20 GeV) and in fact is effectively zero for energies less than 2 GeV and above 100 GeV. The Cocoons absorb a significant part of the HLE's emission.
In fact, having found the best-fit combination for the Flat Bubbles edges (model ``E'') and the Cocoons' properties, the further improvement in the log-likelihood that we get by having the HLE as an additional template is only $2\Delta \! \ln (\mathcal L) \simeq 120$.

We further note that the ICS flux is unaffected by the HLE and that the ratio of fluxes from the Loop I Bright to the Loop I Dim templates is about one, which is closer to our expectations from radio observations. These are conceptual improvements compared to results from Section~\ref{sec:Results} (especially those of Fig~\ref{fig:HLE_120x120_WITH_ResMaps}).
Finally, the fitted fluxes from all the templates remain in good agreement between alternative assumptions on the Bubble edges and the properties of the Cocoons, assuming we get roughly right the Cocoons' orientation and brightness.

\subsection{The Inner morphology of the \textit{Fermi} Bubbles}
\label{subsec:InnerBubbles}

We model the Inner Bubbles (IB) in the same way as the Cocoons: as a coaxial pair of identical ellipsoids whose symmetry axes (the semi-axes $a_\text{IB}$) lie along the $Z$-axis, with centers placed symmetrically about the GC at $\pm z_{0,\,\textrm{IB}}\,\hat{\bm{e}}_z$. The Inner Bubbles differ from the Cocoons in four respects. (1) We keep the pair in the reference configuration and apply no 3D rotation, i.e., $\theta_\text{IB} = \phi_\text{IB} = 0$; their bounding surfaces are therefore given by~\eqref{cocoonpair} with $\bm{R} = \mathbf{1}$. (2) To construct the template, we project the Inner Bubbles with their best-fit parameters along the line of sight and restrict the projected map to the region enclosed by the Flat Bubbles with their optimal boundaries, i.e., we multiply it by a mask that is unity inside those boundaries and zero outside. No such mask is applied to the Cocoon template. (3) We set the emissivities per volume of the two Inner Bubbles equal, $\varepsilon_\text{S, IB}/\varepsilon_\text{N, IB}=1$. (4) The fit prefers flattened ellipsoids with $a_\text{IB} < b_\text{IB} = c_\text{IB}$: whereas the Cocoons are prolate, the Inner Bubbles are oblate. The middle panel of Fig.~\ref{fig:Bubbles_and_Cocoons_Templates}, which combines the best-fit Inner Bubbles with the Flat Bubbles at their optimal boundaries, shows the resulting template; the contrast between the oblate Inner Bubbles and the prolate Cocoons (of the left panel) is clearest there.

In Table~\ref{tab:InnerBubblesProperties}, we give these parameters describing the Inner Bubbles. 
In Appendix~\ref{app:Inner_Bubbles_Properties}, we show log-likelihood profiles as we did with the Cocoons, where we change the values of the modeling parameters around the best-fit combination for the Inner Bubbles and show the effect different choices have on the log-likelihood. 
\begin{table}[t]
\caption{The best-fit properties of the Inner Bubbles. The log-likelihood improvement provided is evaluated using the SL\,Z6\,R20\,T100000\,C5 galactic diffuse emission model for the Pi0+Bremss and the ICS templates, and refers to the statistical improvement of replacing the HLE of $\gamma=1.0$ with the Inner Bubbles. In comparing the log-likelihood improvement of replacing the HLE template with the Inner Bubbles template, we use in both fits the best-fit properties for the location, morphology and orientation of the Cocoons described in Section~\ref{subsec:Cocoons} and Table~\ref{tab:CocoonsProperties}, and alternative Bubble edges E.}
\begin{center}
\begin{tabular}{lr}
\hline
$\varepsilon_\text{S, IB}/\varepsilon_\text{N, IB}$ & 1\\
$z_{0, \, \textrm{IB}}$ & 6.2 kpc \\
$a_{\textrm{IB}}$  & 2.5 kpc\\
$b_{\textrm{IB}}$  & 4.3 kpc\\
$\theta_{\textrm{IB}}$ &$0$\\
$\phi_{\textrm{IB}}$ & $0$\\
\hline
$-2\Delta \! \ln (\mathcal L)$ & $-5018$ \\
\hline
\end{tabular}
\end{center}
\label{tab:InnerBubblesProperties}
\end{table}

\begin{figure}[t]
    \centering
    \includegraphics[width=0.98\linewidth]{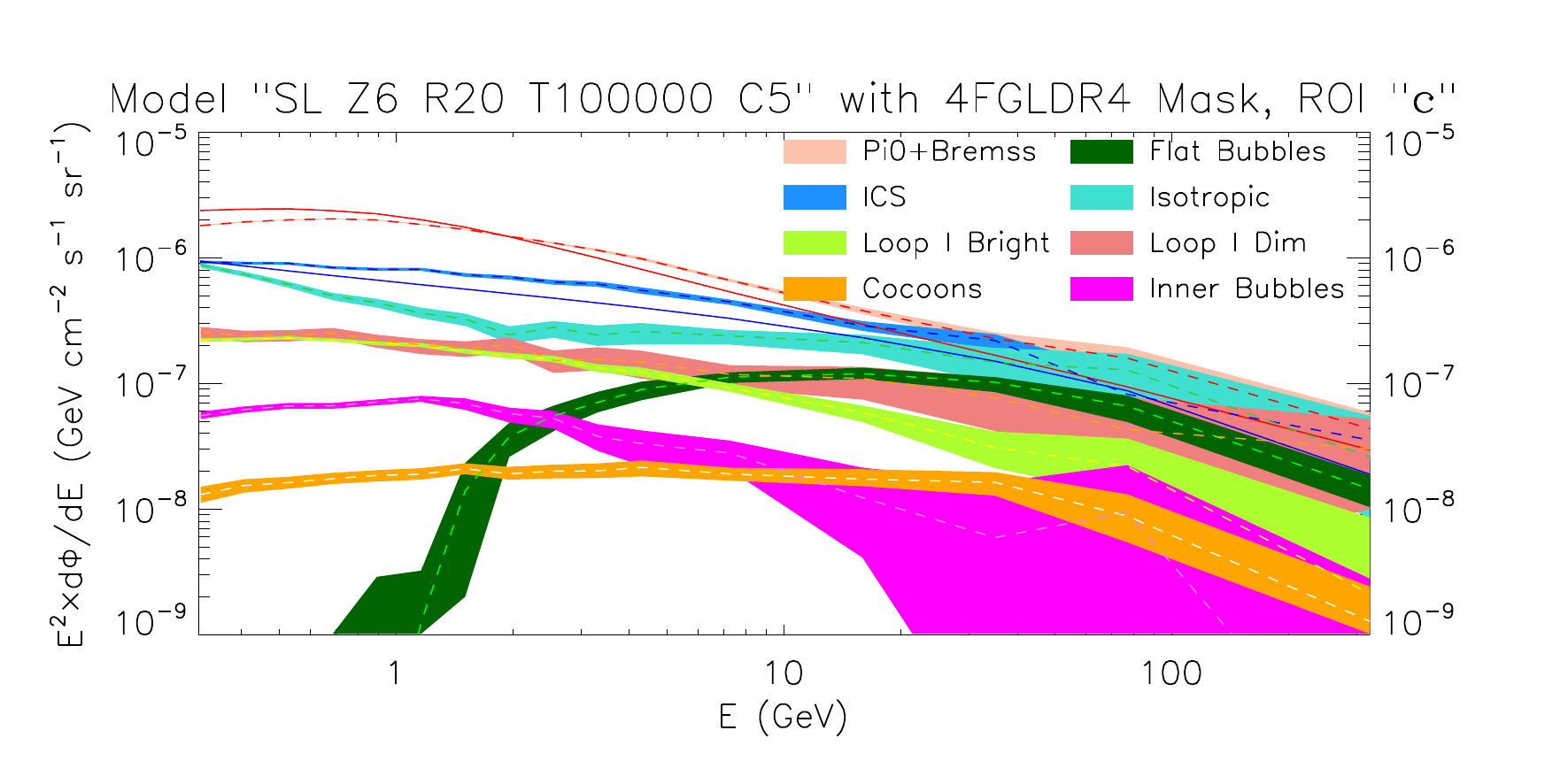}
    \vspace{-0.3cm}
    \caption{The galactic diffuse emission spectra for the ROI \textbf{c}, with the 4FGLDR4 \textit{Fermi} point and extended sources masked out in the fit. Like in Fig.~\ref{fig:BubblesE_Cocoons_and_HLE}, in evaluating the averaged spectra from ROI \textbf{c}, we include the \textit{modeled} diffuse emission flux from the directions of the point and extended sources' masks, and just exclude the emission from $|b|\leq 10^{\circ}$. For the Pi0, Bremss and ICS templates we use the predictions from model SL\,Z6\,R20\,T100000\,C5. Replacing the HLE with the Inner Bubbles template significantly improves the fit to the result presented in Fig.~\ref{fig:BubblesE_Cocoons_and_HLE}. The Inner Bubbles spectrum is softer than that of the Flat Bubbles, suggesting a slightly softer population of underlying cosmic-ray electrons inside the volume of the Inner Bubbles (see text for more details).}
    \label{fig:FinalBubblesSLZ6R20T100000C5}
\end{figure}

In Fig.~\ref{fig:FinalBubblesSLZ6R20T100000C5}, we give the galactic diffuse emission spectra for the ROI \textbf{c}, for all eight templates used. The HLE has been replaced by the Inner Bubbles that provide a much better fit by a $2\Delta \! \ln (\mathcal L) = 5018$. This is statistically an even more dramatic improvement than the one of adding the Cocoons. 
In this figure, we use the templates for the Pi0, Bremss and ICS from the SL\,Z6\,R20\,T100000\,C5 galactic diffuse emission model. 
We note that the Inner Bubbles template is statistically significant up to $\simeq 20$ GeV, being more important at low energies, 
while the Flat Bubbles is more important at energies above 2 GeV and statistically significant up to $\simeq 1$ TeV. 
More importantly, the spectral shapes of the Flat Bubbles and the Inner Bubbles are not identical, with the Inner Bubbles having a softer spectrum and the Flat Bubbles' flux being suppressed below 2 GeV. 
Both templates given their smoothness and lack of correlation with the ISM gas, are likely the result of ICS from high-energy electrons and positrons. In short, we refer to both the cosmic-ray electrons and potential positrons as electrons.  

These cosmic-ray electrons have a spectrum and a spatial distribution. In fact, their spectrum does not need to be the same everywhere within the Bubbles' volume as defined by their boundaries/edges.

The Flat Bubbles effectively assumes that the cosmic-ray electrons responsible for them have a higher density close to the bounding surfaces (edges) of their three-dimensional volume. 
The fact that the Flat Bubbles is suppressed compared to the Inner Bubbles template at the lower gamma-ray energies and more dominant at the higher gamma-ray energies further suggests that the volume closer to the boundaries/edges of the Bubbles has more energetic cosmic-ray electrons, i.e. our fits suggest that the higher cosmic-ray electrons are spatially more concentrated closer to the boundaries/edges of the bubbles region and at the lower latitudes.

The Inner Bubbles template, instead, assumes a homogeneous emissivity within the ellipsoids, effectively absorbing in the fit better the ICS signal from the slightly lower-energy cosmic-ray electrons. Given the shape of the Inner Bubbles these cosmic-ray electrons are separated from the galactic center by several kiloparsecs possibly rising that far from the galactic center due to some buoyancy effect with a shape that resembles the top of a mushroom. 

We note that in Fig.~\ref{fig:FinalBubblesSLZ6R20T100000C5}, while the emission from the Bubbles and the Cocoons is likely of ICS origin, there is very little degeneracy in the fit between these emission components and the modeled galactic diffuse ICS emission component. In fact, both the Pi0+Bremss and the ICS components after the fit (depicted in the figure by the dashed lines), are very close to the spectral shapes of the original model prediction (depicted by the solid lines), and within $\mathcal O(10\%)$ in their normalization relative to the original model flux prediction. Furthermore, especially the spectrum of the Flat Bubbles and to a great extent the Cocoons are harder than the galactic diffuse ICS. These provide a certain level of soundness checks on our results.  

\begin{figure}[t]
    \centering
    \includegraphics[width=0.98\linewidth]{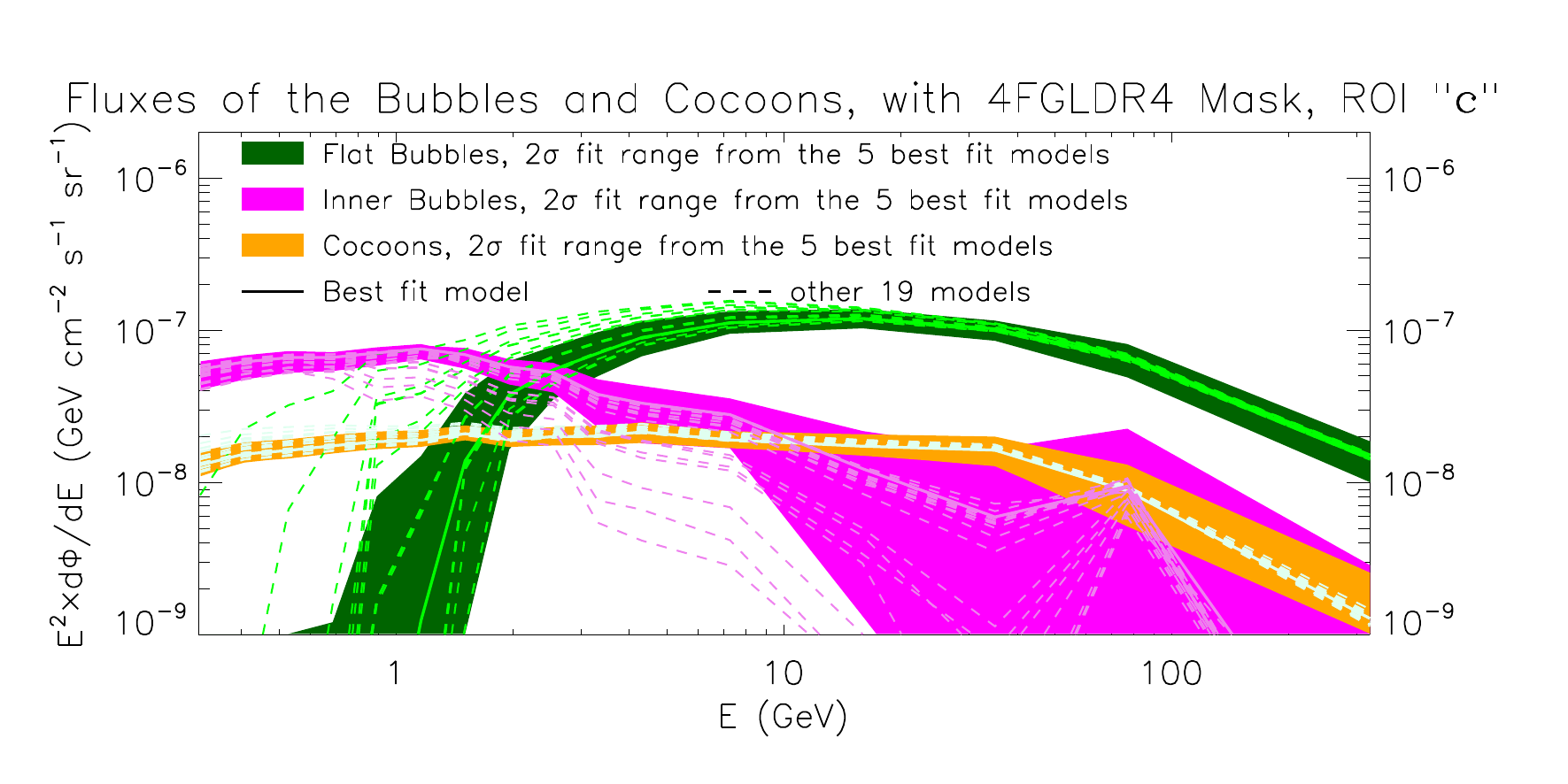}
    \vspace{-0.3cm}
    \caption{The fitted emission spectra within ROI \textbf{c}, for the Flat Bubbles (light green lines and dark green band), the Inner Bubbles (magenta lines and band) and the Cocoons (light cyan lines and orange band). In the solid lines we show the best-fit spectrum for each component, using the best-fit galactic diffuse emission model. The dashed lines give the derived best-fit spectra from the remaining 19 galactic diffuse emission models of Ref.~\cite{Fermi-LAT:2012edv}. For each component, its band gives the combined 2-$\sigma$ fit bands from the five galactic diffuse emission models that provide the best fit to the \textit{Fermi} data among the 20 (see text for more details).}
\label{fig:FinalBubbles5emissionModels}
\end{figure}

To further test the robustness of our results on the spectral properties of the \textit{Fermi} Bubbles and the Cocoons, in Fig.~\ref{fig:FinalBubbles5emissionModels}, we show the spectra associated with the Flat Bubbles, the Inner Bubbles and the Cocoons templates from the fit using all 20 diffuse emission models from Ref.~\cite{Fermi-LAT:2012edv}.
As before the galactic diffuse emission model that provides the best fit is SL\,Z6\,R20\,T100000\,C5, followed by SY\,Z6\,R30\,T100000\,C5, SL\,Z6\,R30\,T100000\,C2, SL\,Z8\,R30\,T150\,C5, and SY\,Z10\,R30\,T150\,C5. The Bubbles and Cocoons spectra derived using the fit SL\,Z6\,R20\,T100000\,C5 galactic diffuse emission model are shown in the solid lines, while the results from the other 19 galactic diffuse emission models are shown with dashed lines.

We notice that the spectrum of the Cocoons is consistently the same irrespective of the galactic diffuse emission model, roughly being $d\Phi/dE \propto E^{-2}$ up to $\simeq 50$ GeV softening at higher energies.  Furthermore, within ROI \textbf{c}, the Flat Bubbles averaged spectrum is always higher than the averaged spectrum from the Inner Bubbles at energies above 3 GeV, with the flux from the Inner Bubbles being higher for almost all models below 1 GeV. If Combined the Flat Bubbles and the Inner Bubbles will give $d\Phi/dE \propto E^{-2}$ up to gamma-ray energies of $\simeq 50$ GeV. However, these results show that at low energies there is little emission from the Flat Bubbles and that at high energies there is little emission associated with the Inner Bubbles. Thus, their relative brightness significantly changes with gamma-ray energy.

\begin{table*}[t]
\caption{The spectra from the \textit{Fermi} Flat Bubbles, Inner Bubbles and Cocoons, averaged over ROI \textbf{c}. The fluxes are given in units of $E^{2}d\Phi/dE$.}
\begin{center}
\begin{tabular}{ccccc}
\hline
Energy Range & Energy & Flat Bubbles & Inner Bubbles & Cocoons \\
(GeV) & (GeV) & $\times10^{-8}$ (GeV cm$^{-2}$ s$^{-1}$ sr$^{-1}$) & $\times10^{-8}$ (GeV cm$^{-2}$ s$^{-1}$ sr$^{-1}$) & $\times10^{-8}$ (GeV cm$^{-2}$ s$^{-1}$ sr$^{-1}$) \\
\hline \hline
0.275 -- 0.357 & 0.313  & $0^{+0.08}$ & $5.53^{+0.70}_{-1.47}$ &  $1.30^{+0.30}_{-0.25}$  \\ 
0.357 -- 0.464 & 0.407  & $0^{+0.09}$  & $6.18^{+0.65}_{-1.44}$ & $1.53^{+0.30}_{-0.26}$  \\  
0.464 -- 0.603 & 0.529  & $0^{+0.09}$  & $6.60^{+0.68}_{-1.35}$  & $1.62^{+0.30}_{-0.27}$  \\ 
0.603 -- 0.784 & 0.688  & $0^{+0.12}$  & $6.61^{+0.55}_{-1.23}$ & $1.74^{+0.31}_{-0.27}$  \\
0.784 -- 1.02  & 0.894  & $0^{+0.81}$  & $7.08^{+0.62}_{-1.25}$ & $1.85^{+0.31}_{-0.24}$ \\
1.02 -- 1.32   & 1.16   & $0.12^{+1.37}_{-0.12}$  & $7.53^{+0.57}_{-1.11}$ & $1.89^{+0.31}_{-0.24}$ \\
1.32 -- 1.72   & 1.51   & $1.38^{+2.47}_{-1.38}$  & $6.89^{+0.74}_{-1.31}$  & $2.09^{+0.28}_{-0.22}$  \\
1.72 -- 2.24   & 1.96   & $3.76^{+2.50}_{-2.11}$  & $5.73^{+0.69}_{-1.35}$ & $1.90^{+0.29}_{-0.21}$  \\
2.24 -- 2.91   & 2.55   & $5.49^{+2.35}_{-1.98}$  & $5.26^{+0.83}_{-1.37}$  & $1.99^{+0.31}_{-0.25}$  \\
2.91 -- 3.78   & 3.31   & $7.03^{+2.66}_{-2.06}$  & $3.78^{+0.97}_{-1.48}$ & $2.01^{+0.35}_{-0.25}$  \\
3.78 -- 4.91   & 4.31   & $8.93^{+2.57}_{-2.21}$  & $3.31^{+0.99}_{-1.51}$  &  $2.14^{+0.34}_{-0.33}$ \\
4.91 -- 10.8   & 7.28   & $11.19^{+2.08}_{-1.72}$  & $2.78^{+0.77}_{-1.11}$  &  $1.90^{+0.27}_{-0.21}$ \\
10.8 -- 23.7   & 16.0   & $11.92^{+1.85}_{-1.65}$  & $1.23^{+0.94}_{-1.23}$ & $1.73^{+0.34}_{-0.23}$  \\
23.7 -- 51.9   & 35.1   & $10.20^{+1.35}_{-1.69}$  & $0.59^{+1.07}_{-0.59}$  & $1.63^{+0.37}_{-0.35}$  \\
51.9 -- 114.0  & 76.9   & $6.58^{+1.51}_{-1.69}$  &  $0.94^{+1.31}_{-0.94}$ & $0.88^{+0.43}_{-0.36}$  \\
114.0 -- 928.0 & 325.3  & $1.44^{+0.42}_{-0.44}$  & $0.01^{+0.28}_{-0.01}$  & $0.13^{+0.13}_{-0.13}$  \\
\hline \hline 
\end{tabular}
\end{center}
\label{tab:BubblesAndCocoonsSpectra}
\end{table*}

In Table~\ref{tab:BubblesAndCocoonsSpectra}, we provide the fitted spectra for the Flat Bubbles, the Inner Bubbles and the Cocoons averaged over ROI \textbf{c}, for the 16 energy bins that we use in our analysis. 
The central values of these three spectra come from their best-fit values when using the galactic diffuse emission model SL\,Z6\,R20\,T100000\,C5, that in combination with the Bubbles $\&$ Cocoons, the Isotropic and the Loop I Bright and Dim components, provides the best fit to the total gamma-ray data in that ROI.  
The 2-$\sigma$ reported bands come from combining the 2-$\sigma$-fit bands of each of these components from the five best-fit diffuse emission models out of the 20. The combined 2-$\sigma$-fit bands are also shown in Fig.~\ref{fig:FinalBubbles5emissionModels}. 
This is done to account for both the statistical uncertainties related to the number of photons associated with each of these emission components, and the systematic uncertainties associated with the modeling of the galactic diffuse emission. 

\section{Conclusions and Discussion}
\label{sec:conclusions}

In this paper, we revisit a claim made by Ref.~\cite{Totani:2025fxx}, on the presence of an extended gamma-ray excess emission from galactic latitudes of $10^{\circ} \leq |b| \leq 60^{\circ}$ and galactic longitudes of $|\ell|\leq 60^{\circ}$, that resembles the signal of an annihilating dark matter particle. We refer to that excess emission as Halo-like emission (HLE), which is beyond the expectations from the combination of galactic diffuse gamma-ray emission, the isotropic gamma-ray emission, the emission from the Loop I region and from the \textit{Fermi} Bubbles. We describe the modeling of each one of these known emissions in Section~\ref{sec:gammaraysky}. 

Analyzing the \textit{Fermi} gamma-ray data in energies of 0.3 GeV to 900 GeV, from 2008 to 2025 (described in Section~\ref{sec:analysis}), we perform a sequence of tests to assess the robustness of the HLE as an excess gamma-ray emission, its morphological characteristics and its spectrum. 
We test the original region of interest (ROI) of Ref.~\cite{Totani:2025fxx}, as well as break it into many smaller ROIs, as we describe in Section~\ref{subsec:roi}, while also testing alternative masks for the exclusion/inclusion of the emission from the detected gamma-ray point and extended sources (discussed in Sections~\ref{sec:gammaraysky} and~\ref{subsec:mask}). 
We also use an expanded library of galactic diffuse gamma-ray emission models. We expand the number of utilized models from the original six of Ref.~\cite{Totani:2025fxx}, to 100, relying on published libraries of such models from Refs.~\cite{Fermi-LAT:2012edv} and~\cite{Cholis:2021rpp}. This allows us to more extensively account for the modeling uncertainties of the galactic diffuse emission (see 
Section~\ref{sec:gammaraysky}). We perform a template fit to the gamma-ray sky at different energies (described in Section~\ref{subsec:fitting_procedure}) and vary the number of free normalization parameters to test the benefits of additional modeling complexity and its effect on the HLE properties. 

We present the results of these tests in Section~\ref{sec:Results}. We confirm that the HLE is indeed an emission that is not accounted for in the pre-existing literature, and that it is not related to the galactic center excess (the GCE) that has long been detected and characterized. However, our results suggest that above latitudes of $|b| \geq 20^{\circ}$, the HLE has the wrong normalization over different ROIs, i.e. morphology, for a signal of dark matter annihilation. Instead, it absorbs/fits \emph{a new component that is not a real dark matter annihilation signal, but some other emission connected to the \textit{Fermi} Bubbles}. The claim that it is a dark matter-like signal is the result of degeneracies in the fit between the HLE, and the combination of the emission from the \textit{Fermi} Bubbles, the galactic ICS and the emission from Loop I. Those degeneracies and the wrong normalization properties of the HLE as a signal of dark matter annihilation are presented in Figs.~\ref{fig:CornerPlotTotaniMethod},~\ref{fig:HLE_120x120_WITH_ResMaps},~\ref{fig:HLE_120x120_WO_ResMaps},~\ref{fig:SLZ6R30T150C2_spectra_6parameters},~\ref{fig:HLE_spectrum_AllModels_AltROIs} and~\ref{fig:DMprofile_for_ROIs}.
To some extent, there is also some dependency of the HLE in the manner in which the original ROI is studied (see Fig.~\ref{fig:HLE_TotaniMask}). 

However, the HLE is a bright excess emission \textit{irrespective of the galactic diffuse emission model or the template fitting scheme used} as shown in Figs.~\ref{fig:CornerPlotTotaniMethod}-\ref{fig:SLZ6R30T150C2_spectra_6parameters}, and mostly in~Figs.~\ref{fig:HLE_spectrum_FermiModels_ROIe} and~\ref{fig:HLE_spectrum_AllModels_AltROIs}. The question we then address is what its true origin is. 
In Section~\ref{sec:NewFermiBubbles}, we study the hypothesis that the HLE absorbs part of the emission from the \textit{Fermi} Bubbles, not accounted for/modeled in the literature as such. We show that this is the case, and thus provide a \textit{new model} for the \textit{Fermi} Bubbles including the Cocoons in the part of the sky the Bubbles are at.  

Our model of \textit{Fermi} Bubbles considers the emission from the Bubbles as the sum of three (template) components, i) a flat in brightness template within the boundaries of the Bubbles on the galactic sky (our Flat Bubbles), ii) a template for the emission from the Cocoons (which, as a component, has been previously claimed) and iii) an entirely new template component, the Inner Bubbles. Those improvements in fact replace/are responsible for the HLE.    

We provide an update on the Bubbles' boundaries/edges, shown in Fig.~\ref{fig:alt_flat_bubbles} (see also discussion in Section~\ref{subsec:BubblesBoundaries}). Those edges differ from the original result of Ref.~\cite{Su:2010qj}, by up to a couple of degrees.
There is a robust statistical preference for these updated boundaries as shown in Table~\ref{tab:Bubbles_Edges}.
We provide a digitized version of those boundaries in appendix~\ref{app:FermiBubblesEdgesE}.

Both the Cocoons and the Inner Bubbles, described in Sections~\ref{subsec:Cocoons} and~\ref{subsec:InnerBubbles} respectively, are modeled as a coaxial pair of ellipsoids, each with a pair of semi-axes $a$ and $b$, with their centers separated from the galactic center by $z_{0}$. 
For the Cocoons that are schematically shown in Fig.~\ref{fig:3D_cocoons}. Within those ellipsoids, we assume a constant gamma-ray emissivity, making both the Cocoons and the Inner Bubbles brighter in the directions on the sky with the greater line-of-sight integration within their volume and along the parts of that volume that are closer to us, the observers from the location of the Solar System.
While the Inner Bubbles have the same emissivity in the Northern and Southern Galactic Hemispheres, the Cocoons are found to have a significantly higher emissivity from the Cocoon located in the Southern Hemisphere. Furthermore, for the coaxial ellipsoid used for the Cocoons to explain the \textit{Fermi} observations, it needs to undergo a 3D rotation, consisting of a yaw and a pitch (described in Section~\ref{subsec:Cocoons}), resulting in the Southern Cocoon's volume being closer to the Solar System, making it appear even bigger in size. Finally, while the Cocoons are prolate ellipsoids, the Inner Bubbles are oblate ellipsoids instead. Our Fig.~\ref{fig:Bubbles_and_Cocoons_Templates}, and Tables~\ref{tab:CocoonsProperties} and~\ref{tab:InnerBubblesProperties}, provide the best-fit morphological properties for these structures. Some of the parameters describing the properties of these structures change when changing the galactic diffuse emission model as is shown in Fig.~\ref{fig:CocoonsYaw_ans_Pitch} and in the Figures of Appendices~\ref{app:CocoonFitProperties} and~\ref{app:Inner_Bubbles_Properties}.
Yet, for the galactic diffuse emission models that provide the best fits to the \textit{Fermi} data, the morphological characteristics of the Cocoons, the Inner Bubbles, as well as the location on the sky of the Bubbles' edges, are quite reliably derived.   

The addition of Cocoons and the Inner Bubbles improves the fit to the gamma-ray data by a very significant amount as those structures are responsible for an appreciable flux of photons from latitudes $10^{\circ} \leq |b| \leq 60^{\circ}$ as we show in Figs.~\ref{fig:BubblesE_Cocoons_and_HLE} and~\ref{fig:FinalBubblesSLZ6R20T100000C5}.
By testing a variety of alternative galactic diffuse emission models we derive the averaged gamma-ray spectra from the Flat Bubbles, the Inner Bubbles and the Cocoons, estimating the uncertainty on those fluxes. We use the sky window of $10^{\circ} \leq |b| \leq 60^{\circ}$, $|\ell| \leq 60^{\circ}$, from where their properties can be reliably derived. We show those spectra in Figure~\ref{fig:FinalBubbles5emissionModels} and Table~\ref{tab:BubblesAndCocoonsSpectra}. 
We find that the Flat Bubbles spectrum is harder than that of the Inner Bubbles, resulting in the Inner Bubbles being the brightest structures 
at low gamma-ray energies while the Flat Bubbles dominate the Bubbles' emission at high
energies. 
The spectrum of the Cocoons scales roughly as  $d\Phi/dE \propto E^{-2}$ up to $\simeq 50$ GeV, softening beyond that energy. Given the smooth morphology of all three structures, we think the most likely mechanism for their emission is ICS. 

\begin{figure}[t]
    \centering
    \includegraphics[width=0.98\linewidth]{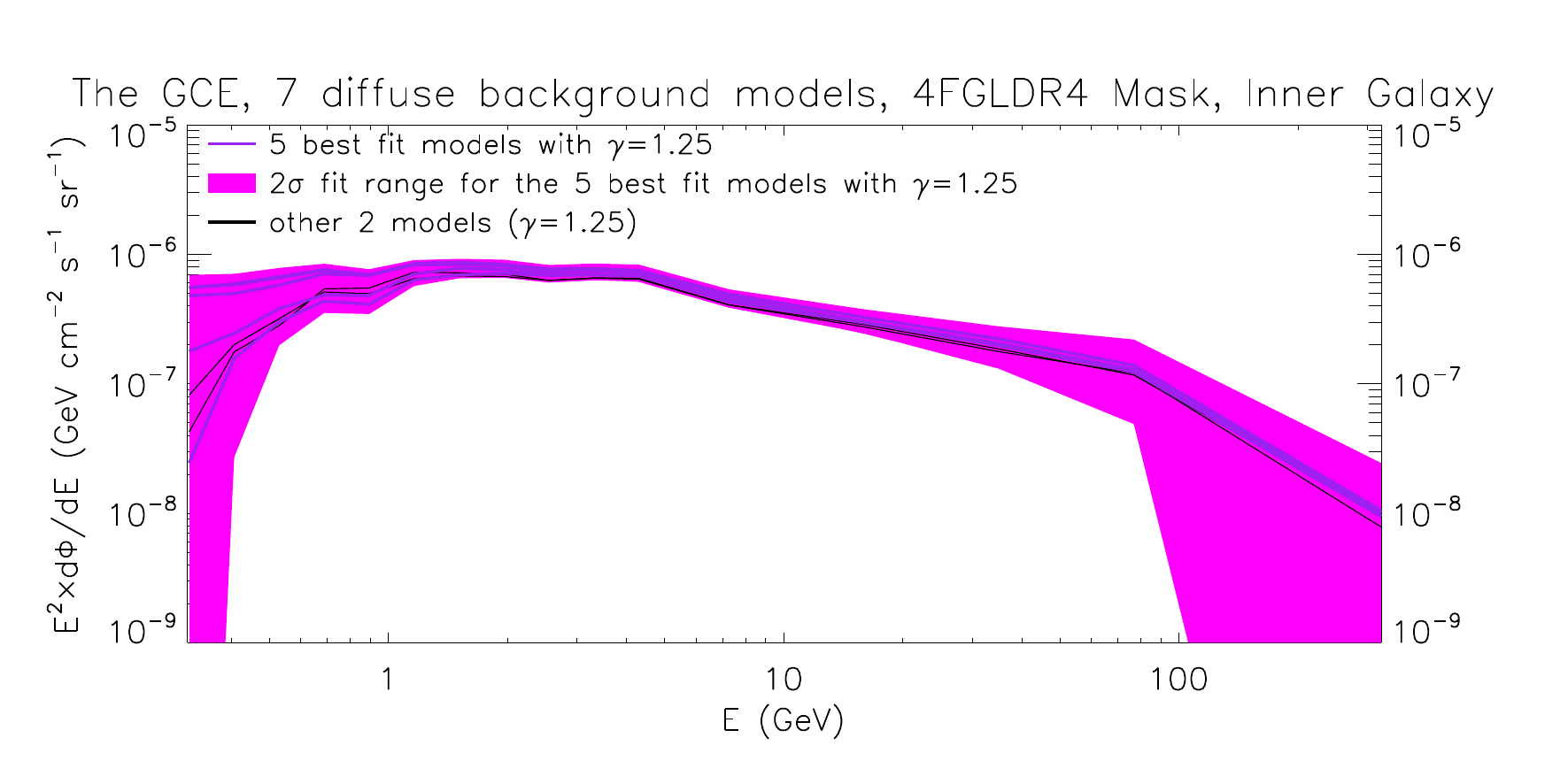}
    \vspace{-0.3cm}
    \caption{The GCE from fitting the Inner Galaxy i.e. the $40^{\circ} \times 40^{\circ}$ ROI excluding $|b| \leq 2^{\circ}$ and masking in the fit the 4FGLDR4 point and extended sources. We use the new model for the \textit{Fermi} Bubbles and Cocoons. We show results from seven galactic diffuse emission models: X, XV, XLVIII, XLIX and LIII from Ref.~\cite{Cholis:2021rpp} (that give the best fit to the \textit{Fermi} data in that ROI), and models SL\,Z6\,R20\,T100000\,C5 and SL\,Z6\,R30\,T150\,C2 from \cite{Fermi-LAT:2012edv}. The derived GCE spectrum using the updated model for the \textit{Fermi} Bubbles and Cocoons is in full agreement with earlier results from the Inner Galaxy.}
    \label{fig:GCE_with_new_Bubbles_model}
\end{figure}

While a much dimmer dark matter signal may exist at high latitudes, the need for a dark matter-like HLE component with the magnitude claimed by Ref.~\cite{Totani:2025fxx}, vanishes with this new model for the Bubbles. 
As a final result, we study the Inner Galaxy ROI of $2^{\circ} \leq |b| \leq 20^{\circ}$ and $|\ell| \leq 20^{\circ}$, masking the 4FGLDR4 point and extended gamma-ray sources. We use seven alternative galactic diffuse emission models from Refs.~\cite{Cholis:2021rpp} and~\cite{Fermi-LAT:2012edv}. We find that even with this new model for the \textit{Fermi} Bubbles the spectrum and amplitude of the GCE are in full agreement with previous analyses of the Inner Galaxy \cite{Calore:2014xka, Cholis:2021rpp, DiMauro:2021raz, Zhong:2024vyi, Hu:2026apn, DiMauro:2026fnp}. We show the GCE spectrum from this analysis in Fig.~\ref{fig:GCE_with_new_Bubbles_model}. 

Previous hydrodynamic and magnetohydrodynamic simulations, as in \cite{Guo:2011eg, Guo:2011ip, Yang:2012fy}, have discussed the origin of the \textit{Fermi} Bubbles as the result of AGN jet activity. 
The updated morphological characteristics of the \textit{Fermi} Bubbles including the substructure of Cocoons and Inner Bubbles, allow us to shed further light on the history of the environment around the supermassive black hole of our Galaxy over the last few million years, and on the conditions of the ISM far away from the galactic center, where cosmic rays interact forming the Bubbles. 

Given the different orientation of the main Bubbles being perpendicular to the galactic disk and the Cocoons having an angle significantly different than $90^{\circ}$ from the disk, and being smaller in size, it's reasonable to expect that these two features are separated in age with the Cocoons likely coming from a later event. 
The main Bubbles are expected to be a few Myr old, with a relatively short AGN injection timescale, to explain the fact that they have an axis ratio close to one as observed by us (i.e. they are relatively ``fat'') \cite{Guo:2011eg}. Instead the Cocoons are more elongated, suggesting that the ratio of their age to the timescale of the AGN injection phase was smaller. 
The further confirmation of the sharpness of the edges of the Flat Bubbles, with the 17-year \textit{Fermi} data, shows a suppression of cosmic-ray diffusion along their boundary, as also suggested in \cite{Guo:2011eg}. Moreover, the
fact that the Flat Bubbles are still responsible for most of the observed gamma-ray flux (power), and they have a harder spectrum than the less powerful Inner Bubbles, suggests that the higher energy cosmic rays are localized close to the edges of the Bubbles volume, an effect that can be explained from stochastic 2nd-order Fermi acceleration around the location of the edges \cite{Mertsch:2011es} (see also~\cite{Fujita:2022wny} for a study where long AGN activity is needed to explain the Bubbles). 

Given that the ICS emission from the Flat Bubbles clearly dominates above 3 GeV the ICS emission from the Inner Bubbles,
there are more cosmic-ray electrons with $E \gtrsim 20$ GeV within the volume occupied by the Flat Bubbles than the volume occupied by the Inner Bubbles. Subsequently, the expected synchrotron radiation from the same energy cosmic-ray electrons at the observed \textit{WMAP} and \textit{Planck} (microwave) wavelengths \cite{Finkbeiner:2003im, Dobler:2007wv, Dobler:2012ef, Planck:2012opn}, is dominated morphologically by the Flat Bubbles.
In order for synchrotron radiation from the sub-10 GeV-energy cosmic-ray electrons within the Inner Bubbles to be clearly identified, observations at larger wavelengths are required.

We note that there are clear indications that the emission from the Southern Cocoon in the \textit{Planck} microwaves has already been observed \cite{Planck:2012opn}. However, given how dim the Northern Cocoon is in gamma rays, and how difficult it is to remove the microwave foreground emission from the location on the galactic sky that is at, we expect it will be very challenging to detect it in microwaves.  

We make publicly available the templates that model the emission from Loop I and the templates that compose the new model of the \textit{Fermi} Bubbles and Cocoons at \texttt{https://zenodo.org/records/23199004}.

\acknowledgements
We thank Tomonori Totani for kindly providing positive and negative residual templates of his work and for his valuable comments. We thank Tim Linden, Nicholas Rodd and Deheng Song for useful discussions.
I.C. acknowledges that this material is based upon work supported by the U.S. Department of Energy, Office of Science, 
Office of High Energy Physics, under Award No. DE-SC0022352.
L.H. and Y.Z. are supported by the GRF Grants No.
11302824, No. 11310925, and the France/Hong Kong
Joint Research Scheme F-CityU106/25 from the Research Grants Council, University Grants Committee, and the Grants No. 9610645 and No. 7020130 from the City University of Hong Kong. Y.Z. acknowledges Aspen Center for Physics for their hospitality and support by NSF Grant No. PHY-2210452 and a grant from the Simons Foundation (1161654, Troyer). Part of this work was carried out using the computational facilities, CityUHK Burgundy, managed and provided by the Computing Services Center at City University of Hong Kong.

\appendix

\section{The models from Ref.~\cite{Fermi-LAT:2012edv}}
\label{app:FermiModels}

In Table~\ref{tab:FemriCollModels}, we list the 20 galactic diffuse emission models we used from Ref.~\cite{Fermi-LAT:2012edv}. In short, the first two letters of each model provide some information on the assumptions on the spatial distribution of cosmic-ray sources in the Milky Way. The ``Z4"-``Z10" provide information on the scale height of the simulated cylindrical diffusion zone, while the ``R20'' or ``R30'' of this zone's assumed radius. The ``T150'', ``T100000'' describe the underlying spin temperature relevant for the modeling of the ISM distribution of the neutral hydrogen atoms. The final two characters ``C2'' or ``C5'' describe alternative magnitude cuts (of 2 or 5) on the E(B-V) reddening
maps of Ref.~\cite{Schlegel:1997yv}. All these models have been developed around the \textit{Fermi} gamma-ray observations. See Ref.~\cite{Fermi-LAT:2012edv}, for a full discussion of their underlying modeling assumptions.

\begin{table}[t]
\caption{The list of the names of the 20 galactic diffuse emission models used from Ref.~\cite{Fermi-LAT:2012edv}.}
\begin{center}
 \begin{tabular}{c}
         \hline
Name of galactic diffuse emission model from Ref.~\cite{Fermi-LAT:2012edv} \\
            \hline \hline
             SL\,Z4\,R30\,T150\,C5 \\   
             SL\,Z6\,R20\,T150\,C2 \\
             SL\,Z6\,R20\,T100000\,C5 \\
             SL\,Z6\,R30\,T150\,C2 \\
             SL\,Z6\,R30\,T100000\,C2 \\
             SL\,Z6\,R30\,T150\,C5 \\
             SL\,Z8\,R30\,T150\,C5 \\
             SL\,Z10\,R30\,T150\,C2 \\
             SL\,Z10\,R30\,T150\,C5 \\
             SO\,Z4\,R30\,T150\,C5 \\
             SO\,Z6\,R30\,T150\,C2 \\
             SO\,Z10\,R30\,T150\,C5 \\
             SS\,Z6\,R20\,T150\,C5 \\
             SS\,Z6\,R30\,T150\,C2 \\
             SS\,Z8\,R20\,T150\,C5 \\
             SS\,Z10\,R30\,T150\,C5 \\
             SY\,Z6\,R20\,T100000\,C2 \\
             SY\,Z6\,R30\,T150\,C2 \\
             SY\,Z6\,R30\,T100000\,C5 \\
             SY\,Z10\,R30\,T150\,C5 \\
             \hline \hline 
\end{tabular}
\end{center}
\label{tab:FemriCollModels}
\end{table}

\section{The new boundaries of the \textit{Fermi} Bubbles}
\label{app:FermiBubblesEdgesE}

In Table~\ref{tab:AlternativeEdgesE}, we give the boundaries for alternative edges ``E'' that provide the best-fit to the \textit{Fermi} data. 
Those boundaries are used for both the Flat Bubbles and the Inner Bubbles templates (see main text for more details).
\begin{table*}[t]
\caption{The coordinates of the boundaries for alternative edges ``E'' for the \textit{Fermi} Bubbles.}
\begin{center}
\small
\begin{tabular}{cccc}
\hline
Northern Bubble Edge \; &  Northern Bubble Edge \; & Southern Bubble Edge  \; & Southern Bubble Edge  \; \\
$b>0$, $\ell > 0$ &  $b > 0$, $\ell < 0$  & $b < 0$, $\ell > 0$  & $b < 0$, $\ell < 0$  \; \\
\hline \hline
\, $4.0^{\circ} \leq b \leq 5.0^{\circ}$ \,: $\ell =2.0^{\circ}$ \, &  \, $4.0^{\circ} \leq b \leq 5.0^{\circ}$ \,: $\ell =-5.0^{\circ}$ \,  &  & \\   
\, $5.0^{\circ} < b \leq 5.5^{\circ}$\, : $\ell =4.0^{\circ}$ \, &  \, $5.0^{\circ} < b \leq 5.5^{\circ}$\, : $\ell =-7.9^{\circ}$ \,  & \, $-5.5^{\circ} \leq b \leq -5.0^{\circ}$\, : $\ell =5.5^{\circ}$ \, & \, $-5.5^{\circ} \leq b \leq -5.0^{\circ}$\, : $\ell =-3.5^{\circ}$ \, \\  
\, $5.5^{\circ} < b \leq 6.5^{\circ}$\, : $\ell =8.0^{\circ}$ \, & \, $5.5^{\circ} < b \leq 6.5^{\circ}$\, : $\ell =-10.2^{\circ}$ & \, $-6.5^{\circ} < b \leq -5.5^{\circ}$\, : $\ell =6.5^{\circ}$ \, &  \, $-6.5^{\circ} < b \leq -5.5^{\circ}$\, : $\ell =-3.5^{\circ}$ \, \\   
\, $6.5^{\circ} < b \leq 7.5^{\circ}$\, : $\ell =12.0^{\circ}$ & \, $6.5^{\circ} < b \leq 7.5^{\circ}$\, : $\ell =-12.0^{\circ}$ &  \, $-7.5^{\circ} < b \leq -6.5^{\circ}$\, : $\ell =7.5^{\circ}$ \, & \, $-7.5^{\circ} < b \leq -6.5^{\circ}$\, : $\ell =-5.0^{\circ}$ \, \\  
\, $7.5^{\circ} < b \leq 9.0^{\circ}$ \,: $\ell =14.0^{\circ}$ & \, $7.5^{\circ} < b \leq 9.0^{\circ}$ \,: $\ell =-13.2^{\circ}$ & \, $-9.0^{\circ} < b \leq -7.5^{\circ}$ \,: $\ell =8.1^{\circ}$ \, & \, $-9.0^{\circ} < b \leq -7.5^{\circ}$ \,: $\ell =-6.9^{\circ}$ \, \\  
\, $9.0^{\circ}  < b \leq 10.0^{\circ}$: $\ell =15.2^{\circ}$ & \, $9.0^{\circ}  < b \leq 10.0^{\circ}$: $\ell =-14.4^{\circ}$ & $-10.0^{\circ} < b \leq -9.0^{\circ}$ \,: $\ell =9.4^{\circ}$ \, &  $-10.0^{\circ}  < b \leq -9.0^{\circ}$ \,: $\ell =-10.0^{\circ}$ \\
$10.0^{\circ} < b \leq 11.5^{\circ}$: $\ell =16.5^{\circ}$ & $10.0^{\circ} < b \leq 11.5^{\circ}$: $\ell =-16.5^{\circ}$ & $-11.5^{\circ} < b \leq -10.0^{\circ}$: $\ell =10.7^{\circ}$ & $-11.5^{\circ} < b \leq -10.0^{\circ}$: $\ell =-13.8^{\circ}$ \\
$11.5^{\circ} < b \leq 12.5^{\circ}$: $\ell =17.5^{\circ}$ & $11.5^{\circ} < b \leq 12.5^{\circ}$: $\ell =-17.5^{\circ}$ & $-12.5^{\circ} < b \leq -11.5^{\circ}$: $\ell =11.2^{\circ}$ & $-12.5^{\circ} < b \leq -11.5^{\circ}$: $\ell =-15.5^{\circ}$ \\
$12.5^{\circ} < b \leq 14.0^{\circ}$: $\ell =18.1^{\circ}$ & $12.5^{\circ} < b \leq 14.0^{\circ}$: $\ell =-18.1^{\circ}$ & $-14.0^{\circ} < b \leq -12.5^{\circ}$: $\ell =11.8^{\circ}$ & $-14.0^{\circ} < b \leq -12.5^{\circ}$: $\ell =-16.3^{\circ}$ \\     
$14.0^{\circ} < b \leq 15.0^{\circ}$: $\ell =18.7^{\circ}$ & $14.0^{\circ} < b \leq 15.0^{\circ}$: $\ell =-18.7^{\circ}$ & $-15.0^{\circ} < b \leq -14.0^{\circ}$: $\ell =12.3^{\circ}$ & $-15.0^{\circ} < b \leq -14.0^{\circ}$: $\ell =-17.5^{\circ}$ \\
$15.0^{\circ} < b \leq 16.5^{\circ}$: $\ell =19.3^{\circ}$ & $15.0^{\circ} < b \leq 16.5^{\circ}$: $\ell =-19.3^{\circ}$ & $-16.5^{\circ} < b \leq -15.0^{\circ}$: $\ell =12.9^{\circ}$ & $-16.5^{\circ} < b \leq -15.0^{\circ}$: $\ell =-18.5^{\circ}$ \\  
$16.5^{\circ} < b \leq 17.5^{\circ}$: $\ell =19.9^{\circ}$ & $16.5^{\circ} < b \leq 17.5^{\circ}$: $\ell =-19.9^{\circ}$ & $-17.5^{\circ} < b \leq -16.5^{\circ}$: $\ell =13.4^{\circ}$ & $-17.5^{\circ} < b \leq -16.5^{\circ}$: $\ell =-19.0^{\circ}$ \\   
$17.5^{\circ} < b \leq 20.0^{\circ}$: $\ell =20.0^{\circ}$ & $17.5^{\circ} < b \leq 20.0^{\circ}$: $\ell =-20.4^{\circ}$ & $-20.0^{\circ} < b \leq -17.5^{\circ}$: $\ell =14.0^{\circ}$ & $-19.0^{\circ} < b \leq -17.5^{\circ}$: $\ell =-19.8^{\circ}$ \\       
$20.0^{\circ} < b \leq 22.5^{\circ}$: $\ell =20.1^{\circ}$ & $20.0^{\circ} < b \leq 22.5^{\circ}$: $\ell =-20.9^{\circ}$ & $-21.0^{\circ} < b \leq -20.0^{\circ}$: $\ell =14.5^{\circ}$ & $-20.0^{\circ} < b \leq -19.0^{\circ}$: $\ell =-20.5^{\circ}$ \\   
$22.5^{\circ} < b \leq 24.0^{\circ}$: $\ell =19.9^{\circ}$ & $22.5^{\circ} < b \leq 24.0^{\circ}$: $\ell =-21.2^{\circ}$ & $-22.0^{\circ} < b \leq -21.0^{\circ}$: $\ell =15.1^{\circ}$ & $-21.0^{\circ} < b \leq -20.0^{\circ}$: $\ell =-21.5^{\circ}$ \\   
$24.0^{\circ} < b \leq 25.0^{\circ}$: $\ell =19.8^{\circ}$ & $24.0^{\circ} < b \leq 25.0^{\circ}$: $\ell =-22.0^{\circ}$ & $-23.0^{\circ} < b \leq -22.0^{\circ}$: $\ell =15.5^{\circ}$ & $-22.0^{\circ} < b \leq -21.0^{\circ}$: $\ell =-22.0^{\circ}$ \\
$25.0^{\circ} < b \leq 26.0^{\circ}$: $\ell =19.5^{\circ}$ & $25.0^{\circ} < b \leq 26.0^{\circ}$: $\ell =-22.5^{\circ}$ & $-24.0^{\circ} < b \leq -23.0^{\circ}$: $\ell =15.9^{\circ}$ & $-23.0^{\circ} < b \leq -22.0^{\circ}$: $\ell =-22.5^{\circ}$ \\      
$26.0^{\circ} < b \leq 27.0^{\circ}$: $\ell =19.2^{\circ}$ & $26.0^{\circ} < b \leq 27.0^{\circ}$: $\ell =-22.7^{\circ}$ & $-25.0^{\circ} < b \leq -24.0^{\circ}$: $\ell =16.3^{\circ}$ & $-24.0^{\circ} < b \leq -23.0^{\circ}$: $\ell =-23.0^{\circ}$ \\   
$27.0^{\circ} < b \leq 28.0^{\circ}$: $\ell =18.9^{\circ}$ & $27.0^{\circ} < b \leq 28.0^{\circ}$: $\ell =-22.9^{\circ}$ & $-27.0^{\circ} < b \leq -25.0^{\circ}$: $\ell =16.6^{\circ}$ & $-25.0^{\circ} < b \leq -24.0^{\circ}$: $\ell =-23.5^{\circ}$ \\   
$28.0^{\circ} < b \leq 29.0^{\circ}$: $\ell =18.6^{\circ}$ & $28.0^{\circ} < b \leq 29.0^{\circ}$: $\ell =-23.0^{\circ}$ & $-29.0^{\circ} < b \leq -27.0^{\circ}$: $\ell =16.9^{\circ}$ & $-26.0^{\circ} < b \leq -25.0^{\circ}$: $\ell =-24.0^{\circ}$ \\   
$29.0^{\circ} < b \leq 30.0^{\circ}$: $\ell =18.5^{\circ}$ & $29.0^{\circ} < b \leq 30.0^{\circ}$: $\ell =-23.2^{\circ}$ & $-30.0^{\circ} < b \leq -29.0^{\circ}$: $\ell =17.3^{\circ}$ & $-27.0^{\circ} < b \leq -26.0^{\circ}$: $\ell =-24.3^{\circ}$ \\   
$30.0^{\circ} < b \leq 31.0^{\circ}$: $\ell =18.2^{\circ}$ & $30.0^{\circ} < b \leq 30.5^{\circ}$: $\ell =-23.4^{\circ}$ & $-31.0^{\circ} < b \leq -30.0^{\circ}$: $\ell =17.5^{\circ}$ & $-28.0^{\circ} < b \leq -27.0^{\circ}$: $\ell =-24.5^{\circ}$ \\   
$31.0^{\circ} < b \leq 32.0^{\circ}$: $\ell =17.9^{\circ}$ & $30.5^{\circ} < b \leq 32.5^{\circ}$: $\ell =-23.5^{\circ}$ & $-32.0^{\circ} < b \leq -31.0^{\circ}$: $\ell =17.8^{\circ}$ & $-29.0^{\circ} < b \leq -28.0^{\circ}$: $\ell =-24.7^{\circ}$ \\   
$32.0^{\circ} < b \leq 33.0^{\circ}$: $\ell =17.6^{\circ}$ & $32.5^{\circ} < b \leq 33.5^{\circ}$: $\ell =-23.4^{\circ}$ & $-33.0^{\circ} < b \leq -32.0^{\circ}$: $\ell =18.0^{\circ}$ & $-30.0^{\circ} < b \leq -29.0^{\circ}$: $\ell =-24.8^{\circ}$ \\   
$33.0^{\circ} < b \leq 34.0^{\circ}$: $\ell =17.3^{\circ}$ & $33.5^{\circ} < b \leq 34.5^{\circ}$: $\ell =-23.3^{\circ}$ & $-34.0^{\circ} < b \leq -33.0^{\circ}$: $\ell =18.2^{\circ}$ & $-31.0^{\circ} < b \leq -30.0^{\circ}$: $\ell =-24.9^{\circ}$ \\   
$34.0^{\circ} < b \leq 35.0^{\circ}$: $\ell =16.9^{\circ}$ & $34.5^{\circ} < b \leq 35.0^{\circ}$: $\ell =-23.1^{\circ}$ & $-35.0^{\circ} < b \leq -34.0^{\circ}$: $\ell =18.4^{\circ}$ & $-33.0^{\circ} < b \leq -31.0^{\circ}$: $\ell =-25.0^{\circ}$ \\   
$35.0^{\circ} < b \leq 36.0^{\circ}$: $\ell =16.5^{\circ}$ & $35.0^{\circ} < b \leq 36.0^{\circ}$: $\ell =-22.5^{\circ}$ & $-36.0^{\circ} < b \leq -35.0^{\circ}$: $\ell =18.5^{\circ}$ & $-35.0^{\circ} < b \leq -33.0^{\circ}$: $\ell =-24.9^{\circ}$ \\   
$36.0^{\circ} < b \leq 37.0^{\circ}$: $\ell =15.9^{\circ}$ & $36.0^{\circ} < b \leq 37.0^{\circ}$: $\ell =-22.3^{\circ}$ & $-37.0^{\circ} < b \leq -36.0^{\circ}$: $\ell =18.4^{\circ}$ & $-36.0^{\circ} < b \leq -35.0^{\circ}$: $\ell =-24.8^{\circ}$ \\   
$37.0^{\circ} < b \leq 38.0^{\circ}$: $\ell =15.3^{\circ}$ & $37.0^{\circ} < b \leq 38.0^{\circ}$: $\ell =-22.0^{\circ}$ & $-38.0^{\circ} < b \leq -37.0^{\circ}$: $\ell =18.2^{\circ}$ & $-37.0^{\circ} < b \leq -36.0^{\circ}$: $\ell =-24.7^{\circ}$ \\   
$38.0^{\circ} < b \leq 39.0^{\circ}$: $\ell =14.5^{\circ}$ & $38.0^{\circ} < b \leq 39.0^{\circ}$: $\ell =-21.7^{\circ}$ & $-39.0^{\circ} < b \leq -38.0^{\circ}$: $\ell =18.0^{\circ}$ & $-39.0^{\circ} < b \leq -37.0^{\circ}$: $\ell =-24.4^{\circ}$ \\   
$39.0^{\circ} < b \leq 40.0^{\circ}$: $\ell =13.8^{\circ}$ & $39.0^{\circ} < b \leq 40.0^{\circ}$: $\ell =-21.1^{\circ}$ & $-40.0^{\circ} < b \leq -39.0^{\circ}$: $\ell =17.8^{\circ}$ & $-40.0^{\circ} < b \leq -39.0^{\circ}$: $\ell =-24.2^{\circ}$ \\   
$40.0^{\circ} < b \leq 41.0^{\circ}$: $\ell =12.9^{\circ}$ & $40.0^{\circ} < b \leq 41.0^{\circ}$: $\ell =-20.5^{\circ}$ & $-41.0^{\circ} < b \leq -40.0^{\circ}$: $\ell =17.5^{\circ}$ & $-41.0^{\circ} < b \leq -40.0^{\circ}$: $\ell =-24.0^{\circ}$ \\   
$41.0^{\circ} < b \leq 42.0^{\circ}$: $\ell =12.3^{\circ}$ & $41.0^{\circ} < b \leq 42.0^{\circ}$: $\ell =-19.4^{\circ}$ & $-42.0^{\circ} < b \leq -41.0^{\circ}$: $\ell =17.3^{\circ}$ & $-42.0^{\circ} < b \leq -41.0^{\circ}$: $\ell =-23.8^{\circ}$ \\  
$42.0^{\circ} < b \leq 43.0^{\circ}$: $\ell =11.5^{\circ}$ & $42.0^{\circ} < b \leq 43.0^{\circ}$: $\ell =-18.8^{\circ}$ & $-43.0^{\circ} < b \leq -42.0^{\circ}$: $\ell =17.0^{\circ}$ & $-43.0^{\circ} < b \leq -42.0^{\circ}$: $\ell =-23.2^{\circ}$ \\    
$43.0^{\circ} < b \leq 44.0^{\circ}$: $\ell =11.1^{\circ}$ & $43.0^{\circ} < b \leq 44.0^{\circ}$: $\ell =-17.7^{\circ}$ & $-44.0^{\circ} < b \leq -43.0^{\circ}$: $\ell =16.7^{\circ}$ & $-44.0^{\circ} < b \leq -43.0^{\circ}$: $\ell =-22.0^{\circ}$ \\   
$44.0^{\circ} < b \leq 45.0^{\circ}$: $\ell =10.2^{\circ}$ & $44.0^{\circ} < b \leq 45.0^{\circ}$: $\ell =-16.5^{\circ}$ & $-45.0^{\circ} < b \leq -44.0^{\circ}$: $\ell =16.3^{\circ}$ & $-45.0^{\circ} < b \leq -44.0^{\circ}$: $\ell =-21.0^{\circ}$ \\   
$45.0^{\circ} < b \leq 46.0^{\circ}$: $\ell =8.3^{\circ}$ \, & $45.0^{\circ} < b \leq 46.0^{\circ}$: $\ell =-14.8^{\circ}$ & $-46.0^{\circ} < b \leq -45.0^{\circ}$: $\ell =15.8^{\circ}$ & $-46.0^{\circ} < b \leq -45.0^{\circ}$: $\ell =-20.4^{\circ}$ \\  
$46.0^{\circ} < b \leq 47.0^{\circ}$: $\ell =6.7^{\circ}$ \, & $46.0^{\circ} < b \leq 47.0^{\circ}$: $\ell =-13.2^{\circ}$ & $-47.0^{\circ} < b \leq -46.0^{\circ}$: $\ell =15.3^{\circ}$ & $-47.0^{\circ} < b \leq -46.0^{\circ}$: $\ell =-19.8^{\circ}$ \\    
$47.0^{\circ} < b \leq 48.0^{\circ}$: $\ell =5.5^{\circ}$ \, & $47.0^{\circ} < b \leq 48.0^{\circ}$: $\ell =-12.0^{\circ}$ & $-48.0^{\circ} < b \leq -47.0^{\circ}$: $\ell =14.6^{\circ}$ & $-48.0^{\circ} < b \leq -47.0^{\circ}$: $\ell =-18.8^{\circ}$ \\   
$48.0^{\circ} < b \leq 49.0^{\circ}$: $\ell =4.0^{\circ}$ \, & $48.0^{\circ} < b \leq 49.0^{\circ}$: $\ell =-10.0^{\circ}$ & $-49.0^{\circ} < b \leq -48.0^{\circ}$: $\ell =13.9^{\circ}$ & $-49.0^{\circ} < b \leq -48.0^{\circ}$: $\ell =-17.8^{\circ}$ \\   
$49.0^{\circ} < b \leq 50.0^{\circ}$: $\ell =1.5^{\circ}$ \, & $49.0^{\circ} < b \leq 50.0^{\circ}$: $\ell =-7.0^{\circ}$ \, & $-50.0^{\circ} < b \leq -49.0^{\circ}$: $\ell =12.8^{\circ}$ & $-50.0^{\circ} < b \leq -49.0^{\circ}$: $\ell =-15.8^{\circ}$ \\      &  & $-51.0^{\circ} < b \leq -50.0^{\circ}$: $\ell =11.2^{\circ}$ & $-51.0^{\circ} < b \leq -50.0^{\circ}$: $\ell =-14.8^{\circ}$ \\
&  & $-52.0^{\circ} < b \leq -51.0^{\circ}$: $\ell =10.6^{\circ}$ & $-52.0^{\circ} < b \leq -51.0^{\circ}$: $\ell =-13.8^{\circ}$ \\
&  & $-53.0^{\circ} < b \leq -52.0^{\circ}$: $\ell =8.9^{\circ}$ \, & $-53.0^{\circ} < b \leq -52.0^{\circ}$: $\ell =-12.0^{\circ}$ \\  
&  & $-54.0^{\circ} < b \leq -53.0^{\circ}$: $\ell =7.2^{\circ}$ \, & $-54.0^{\circ} < b \leq -53.0^{\circ}$: $\ell =-10.1^{\circ}$ \\ 
&  & $-55.0^{\circ} < b \leq -54.0^{\circ}$: $\ell =3.7^{\circ}$ \, & $-55.0^{\circ} < b \leq -54.0^{\circ}$: $\ell =-6.3^{\circ}$ \, \\                                      
\hline \hline 
\end{tabular}
\end{center}
\label{tab:AlternativeEdgesE}
\end{table*}

\section{Fit-results on the properties of the Cocoons}
\label{app:CocoonFitProperties}

\begin{figure*}[t]
    \centering
    \includegraphics[width=0.34\linewidth]{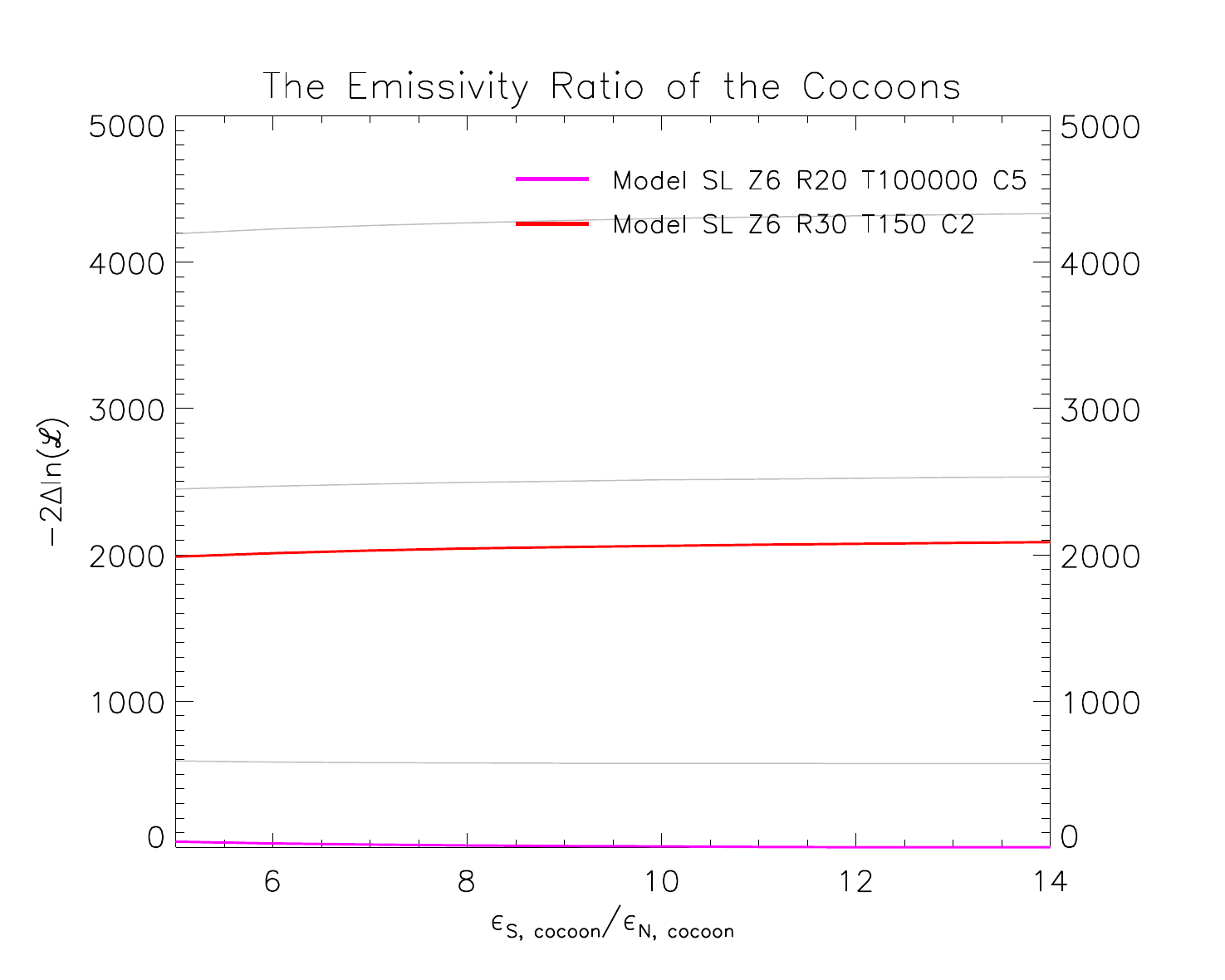}
    \hspace{-0.5cm}
    \includegraphics[width=0.34\linewidth]{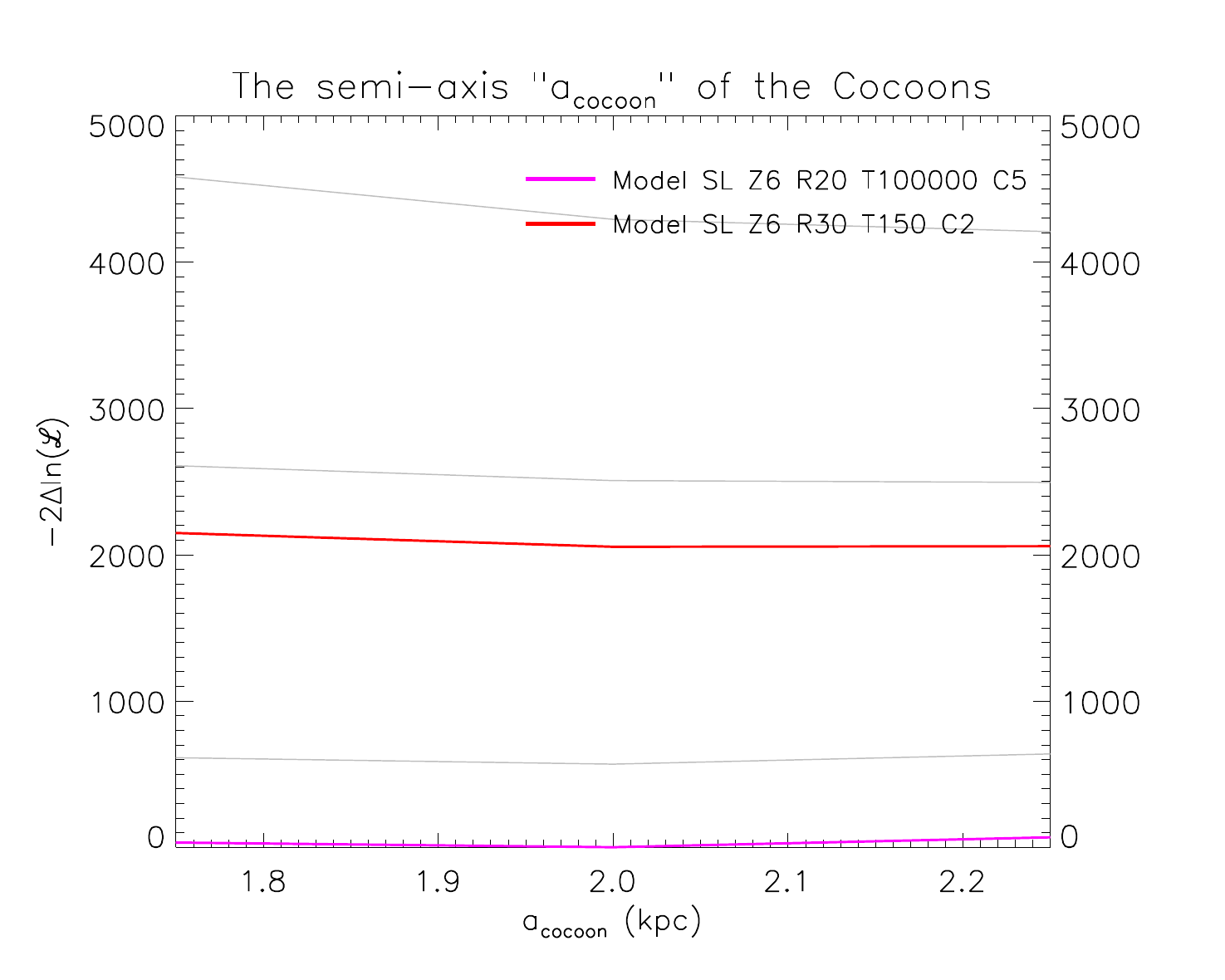}
    \hspace{-0.5cm}
    \includegraphics[width=0.34\linewidth]{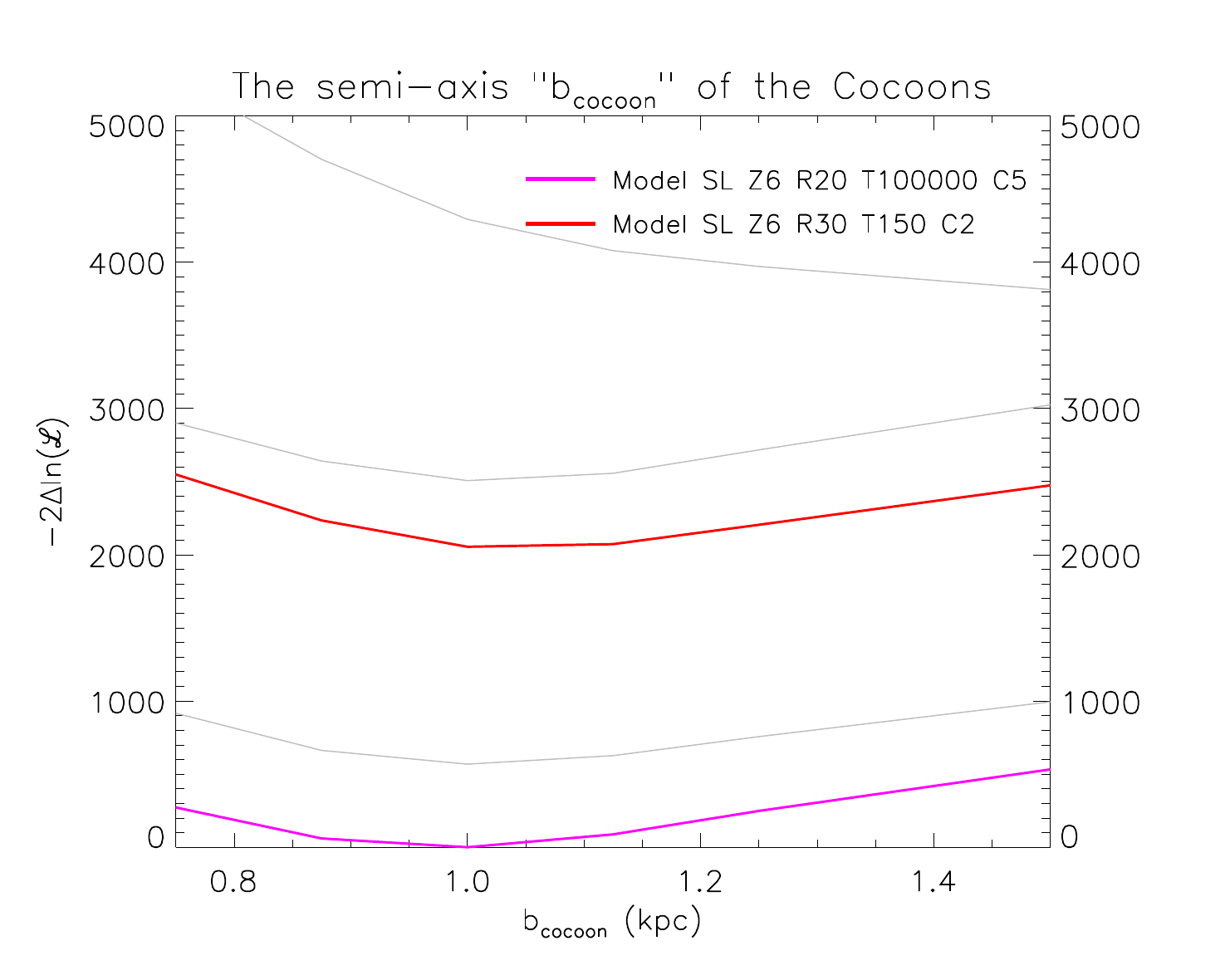}
    \vspace{-0.3cm}
    \caption{The change in log-likelihood $\mathcal L$ by varying the properties of the Cocoons. \textit{Left}: the effect on the value of the emissivity ratio $\varepsilon_\text{S, cocoon}/\varepsilon_\text{N, cocoon}$. \textit{Center}: the effect on the value of the semi-axis $a_{\textrm{cocoon}}$ of the ellipsoids. \textit{Right}: the effect on the value of the semi-axis $b_{\textrm{cocoon}}$ of the ellipsoids. We also use the best-fit edges (version ``E'') and the best-fit properties for the Inner Bubbles (see main text and Table~\ref{tab:InnerBubblesProperties}). We show the $-2\Delta \! \ln (\mathcal L)$ profiles for five galactic diffuse emission models, highlighting the results for SL\,Z6\,R20\,T100000\,C5 and SL\,Z6\,R30\,T150\,C2.}
    \label{fig:LogLikelihood_Profile_for_Cocoons_remaining_properties}
\end{figure*}

In Fig.~\ref{fig:LogLikelihood_Profile_for_Cocoons_remaining_properties}, we give the log-likelihood profiles for the displacement of the Cocoon's the emissivity ratio $\varepsilon_\text{S, cocoon}/\varepsilon_\text{N, cocoon}$, and the values of their semi-axes $a_{\textrm{cocoon}}$ and $b_{\textrm{cocoon}}$. This figure is in addition to the results on the Cocoon's yaw and pitch angles presented in Fig.~\ref{fig:CocoonsYaw_ans_Pitch} of the main text. 
In these fits we use the best-fit Bubbles' edges and the best-fit version of the Inner Bubbles.
In every case we vary only one parameter of the Cocoons fixing the remaining to their best-fit choice. We always keep $z_{0, \, \textrm{cocoon}} = 3.5$ kpc. 
For the galactic diffuse emission models that give the better fits (lower lines along the $-2\Delta \! \ln (\mathcal L)$ axis) the properties of the Cocoons are in agreement with the results presented in Table~\ref{tab:CocoonsProperties}, of the main text, while for galactic models with significantly worse fits, the morphology of the Cocoons can deviate from the results of Table~\ref{tab:CocoonsProperties}.

\section{Fit-results on the properties of the Inner Bubbles}
\label{app:Inner_Bubbles_Properties}

\begin{figure*}[t]
    \centering
    \includegraphics[width=0.34\linewidth]{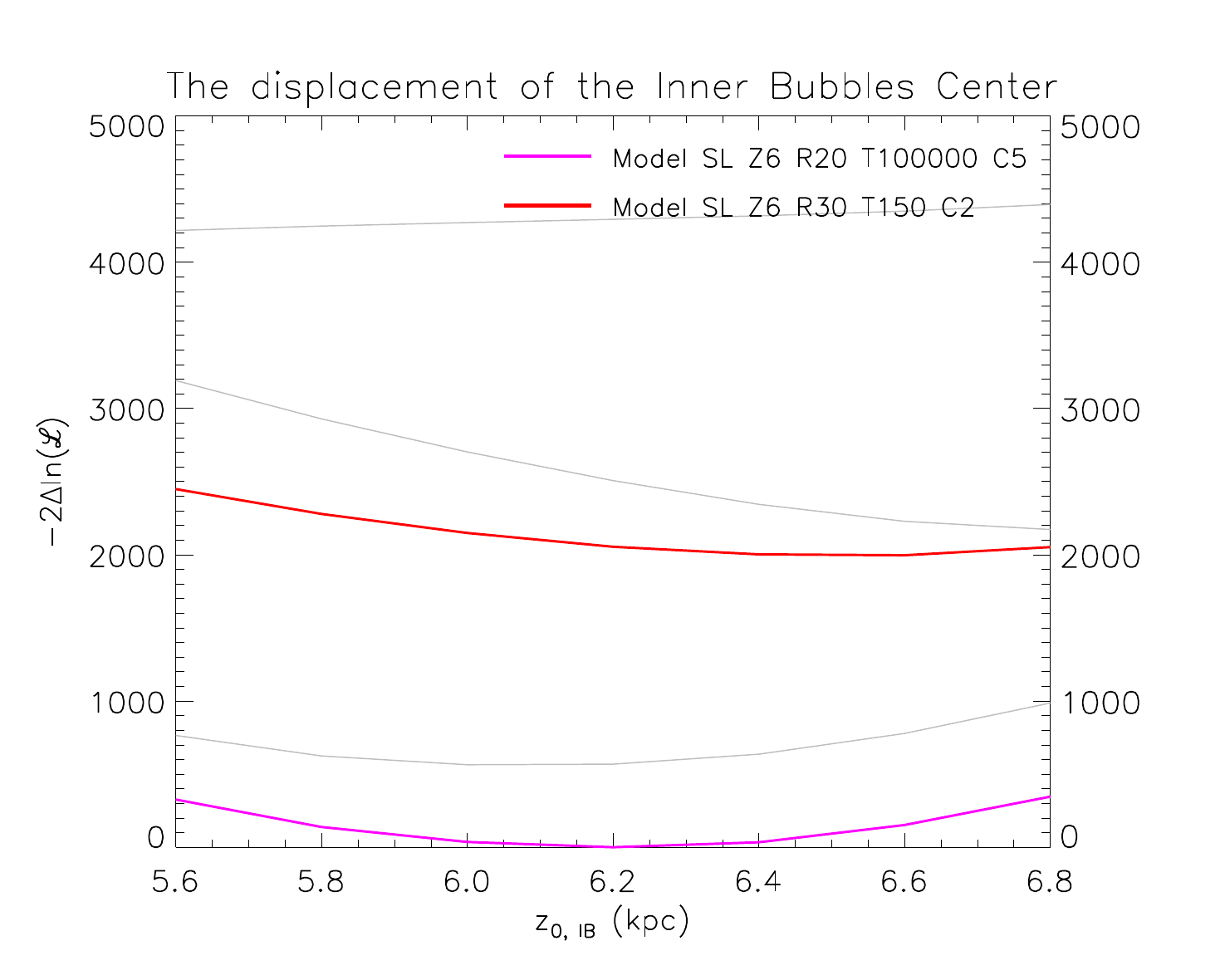}
    \hspace{-0.5cm}
    \includegraphics[width=0.34\linewidth]{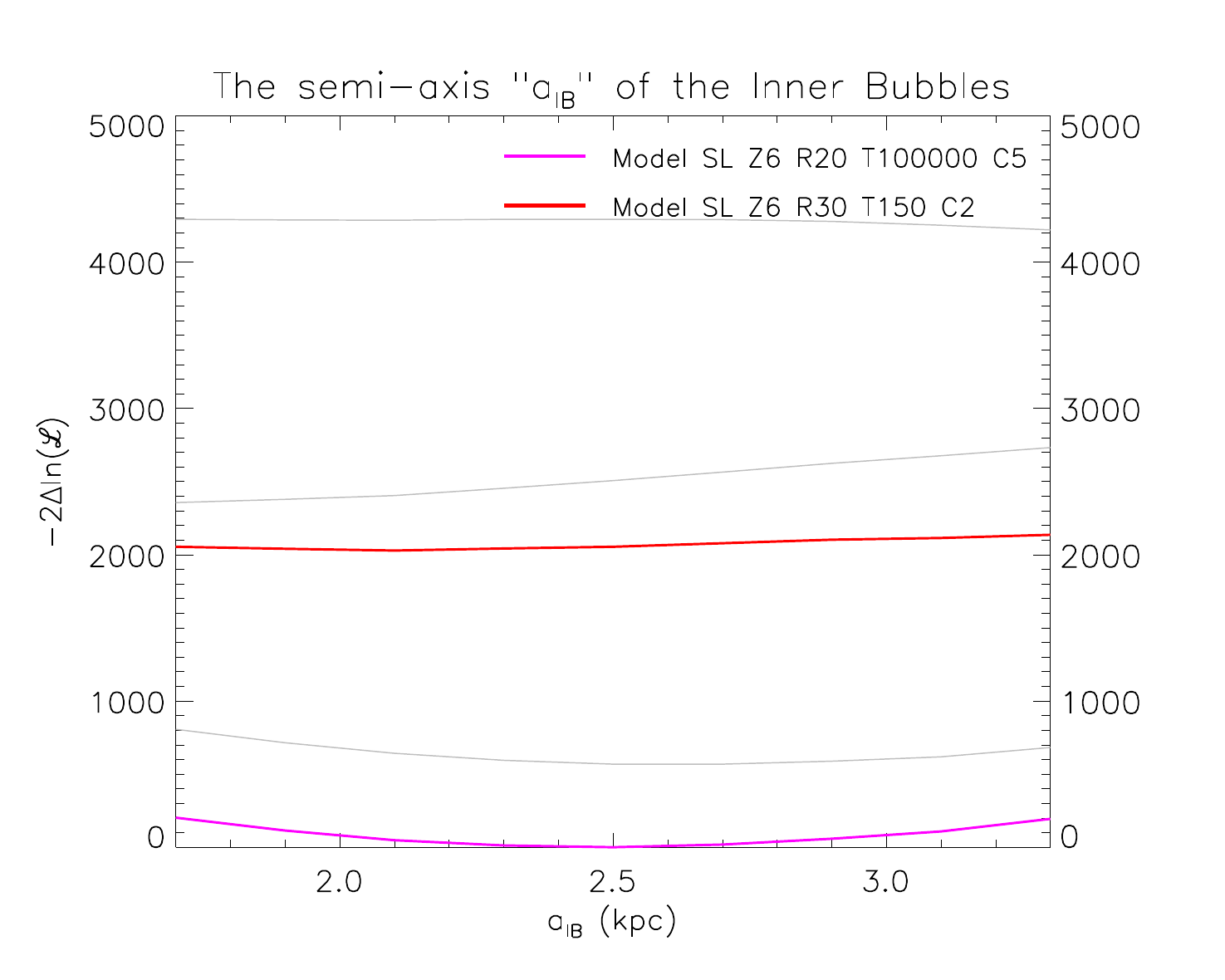}
    \hspace{-0.5cm}
    \includegraphics[width=0.34\linewidth]{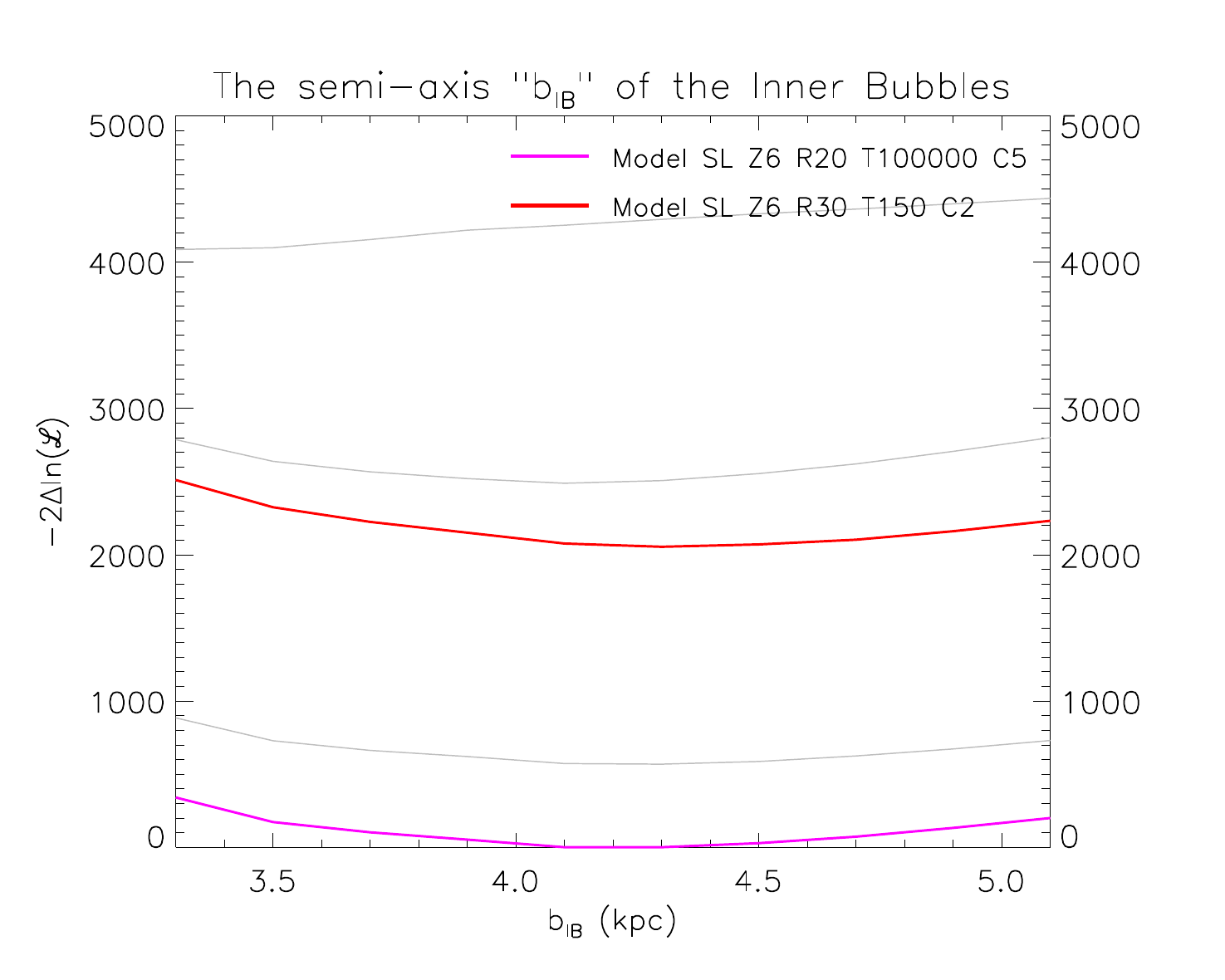}
    \vspace{-0.3cm}
    \caption{The change in log-likelihood $\mathcal L$ by varying the properties of the Inner Bubbles. \textit{Left}: the effect on the value of the displacement $z_{0, \, \textrm{IB}}$ of the co-axial ellipsoids' centers from the  center of the Galaxy. \textit{Center}: the effect on the value of the semi-axis $a_{\textrm{IB}}$ of the ellipsoids. \textit{Right}: the effect on the value of the semi-axis $b_{\textrm{IB}}$ of the ellipsoids. We keep fixed the emissivity ratio $\varepsilon_\text{S, IB}/\varepsilon_\text{N, IB} = 1$, with $\theta_{\textrm{IB}} =0$ and $\phi_{\textrm{IB}}=0$. We also use the best-fit edges (version E) and the best-fit properties for the Cocoons (see main text and Table~\ref{tab:CocoonsProperties}). We show the $-2\Delta \! \ln (\mathcal L)$ profiles for five galactic diffuse emission models, highlighting the results for SL\,Z6\,R20\,T100000\,C5 and SL\,Z6\,R30\,T150\,C2.}
    \label{fig:LogLikelihood_Profile_for_Inner_Bubbles}
\end{figure*}

In Fig.~\ref{fig:LogLikelihood_Profile_for_Inner_Bubbles}, we give the log-likelihood profiles for the three non-trivial properties of the Inner Bubbles: the displacement of the co-axial ellipsoids' centers from the center of the Milky Way $z_{0, \, \textrm{IB}}$, and the values of their semi-axes $a_{\textrm{IB}}$ and $b_{\textrm{IB}}$. In these fits we use the best-fit Bubbles' edges and the best-fit version of the Cocoons. 
As in Fig.~\ref{fig:LogLikelihood_Profile_for_Cocoons_remaining_properties}, in every case we vary only one parameter of the Inner Bubbles fixing the remaining to their best-fit choice.
For the galactic diffuse emission models that give the better fits (lower lines along the $-2\Delta \! \ln (\mathcal L)$ axis) the properties of the Inner Bubbles are in agreement with the results presented in Table~\ref{tab:InnerBubblesProperties}, of the main text.

\newpage

\bibliography{reference}
\bibliographystyle{h-physrev.bst}

\end{document}